\documentclass[12pt]{article}
\usepackage[british]{babel}
\def\feynVersion{n}

\def\AnswerYes{y}
\ifx\feynVersion\AnswerYes
   \usepackage{feynmp}                    %%% for Feynman diagrams; a few
\fi

\usepackage[dvips]{color}                 %%% for colours in the graphs
\usepackage{graphicx}                     %%% for pictures and figures
\usepackage{amssymb}
\usepackage{amsmath}
\usepackage{xspace}                       %%% to get the correct spacing of
\newcommand{\3}{\ss}
\newcommand{\absatz}{\vspace{2ex}\noindent}

\newcommand{\blue}[1]{#1}

\newcommand{\journal}[4]{\textit{#1}\xspace\textbf{#2}, #4 (#3)}
\newcommand{\PR}{\textnormal{Phys.\ Rev.\ }}

\newcommand{\PRC}{\PR\textnormal{C}~}
\newcommand{\PRD}{\PR\textnormal{D}~}
\newcommand{\PRL}{\PR\textnormal{Lett.\ }}

\newcommand{\dis}{\displaystyle}

\newcommand{\fs}{\scriptstyle} % adjusts size of labels e.g. in feynmf-diagrams
\newcommand{\non}{\nonumber}
\newcommand{\hq}{\hspace{0.5em}}
\newcommand{\hqm}{\hspace*{-0.25em}}

\newcommand{\half}{\frac{1}{2}}
\newcommand{\e}{\mathrm{e}}
\newcommand{\ii}{\mathrm{i}}
\newcommand{\dd}{\mathrm{d}}
\newcommand{\tr}{\mathrm{tr}}

\newcommand{\deint}[2]{\dd^{#1}\! #2\;}
\newcommand{\deintdim}[2]{\frac{\dd^{#1}\! #2}{(2\pi)^{#1}}\;}

\newcommand{\kv}{\vec{k}}
\newcommand{\pv}{\vec{p}}%{\vec{\,\!p}\!\:{}}
\newcommand{\qv}{\vec{q}}%{\vec{\,\!q}\!\:{}}

\newcommand{\Folgt}{\Longrightarrow}

\newcommand{\mpi}{\ensuremath{m_\pi}}

\newcommand{\MeV}{\ensuremath{\mathrm{MeV}}}
\newcommand{\fm}{\ensuremath{\mathrm{fm}}}
\newcommand{\EFTNoPion}{EFT(${\pi\hskip-0.55em /}$)\xspace}

\newcommand{\NXLO}[1]{N\ensuremath{{}^{#1}}LO\xspace}

\newcommand{\wave}[3]{\ensuremath{{}^{#1}\mathrm{#2}_{#3}}}
\newcommand{\oneS}{\wave{1}{S}{0}}
\newcommand{\twoS}{\wave{2}{S}{\half}}
\newcommand{\threeS}{\wave{3}{S}{1}}
\newcommand{\fourS}{\wave{4}{S}{\frac{3}{2}}}

\newcommand{\LambdaNoPion}{\ensuremath{\Lambda_{\pi\hskip-0.4em /}}}

\newcommand{\de}{\partial}
\newcommand{\dev}{\vec{\de}}

\newcommand{\calA}{\mathcal{A}} \newcommand{\calD}{\mathcal{D}}
\newcommand{\calH}{\mathcal{H}} \newcommand{\calK}{\mathcal{K}}
\newcommand{\calL}{\mathcal{L}} 
 \newcommand{\calO}{\mathcal{O}}
\newcommand{\calP}{\mathcal{P}} 
\newcommand{\calZ}{\mathcal{Z}}

\newcommand{\mytitle}[1]{\begin{center}\LARGE{\textbf{#1}}\end{center}}
\newcommand{\myauthor}[1]{\textbf{#1}}
\newcommand{\myaddress}[1]{\textit{#1}}
\newcommand{\mypreprint}[1]{\begin{flushright}#1\end{flushright}}

\begin{document}
%%%%%%%%%%%%%%%%%%%%%%%%%%%%%%%%%%%%%%%%%%%%%%%%%%%%%%%%%%%%%%%%%%%%%%%%%%%%%%%
% The following is used only if the Feynman diagrams are to be generated
% explicitly:
\ifx\feynVersion\AnswerYes
\begin{fmffile}{zparfeyn}
  
  \fmfset{curly_len}{2mm} \fmfset{dash_len}{1.5mm} \fmfset{wiggly_len}{3mm}
% the following sets the galleries on which the external verices lie as 
% straight lines, not curved. 
  \fmfstraight

% invisible frame around fmfgraph*-environment:
% #1: x-size of fmfgraph* in \unitlength
% #2: y-size of fmfgraph* in \unitlength
% #3: additional space on LEFT of graph in \unitlength  
% #4: additional space on TOP of graph in \unitlength  
% #5: additional space on RIGHT of graph in \unitlength  
% #6: additional space on BOTTOM of graph in \unitlength  
%
% USAGE: \feyngraph{xsize}{ysize}{left}{top}{right}{bottom}{graph}  
%  
% If second and next-to-last line are commented out, a box is put around the
% invisible frame: Good for proof-reading.
%
  \newlength{\feyngraphlength}
\newcommand{\feyngraph}[7]{%
  \setlength{\feyngraphlength}{#1\unitlength}%
  \addtolength{\feyngraphlength}{#3\unitlength}%
  \addtolength{\feyngraphlength}{#5\unitlength}%
  \parbox[c]{\feyngraphlength}{%
%    \framebox{%%%%%
      \fmfframe(#3,#4)(#5,#6){%
        \begin{fmfgraph*}(#1,#2)%
          #7%
        \end{fmfgraph*}%
        }%
      }%
%    }%%%%%
    }
%  The following produces an ellipse along a line given by two vertices.
%  Application: \fmf{ellipse,foreground=red,rubout=1}{v1,v2}
%               draws ellipse with major semi-axis 0.6*(distance v1-v2) and
%                                  minor semi-axis 0.3*(distance v1-v2). 
%               around centre of {v1,v2}
    \fmfcmd{vardef ellipseraw (expr p, ang) = save radx; numeric radx;
      radx=6/10 length p; save rady; numeric rady; rady=3/10 length p; pair
      center; center:=point 1/2 length(p) of p; save t; transform t;
      t:=identity xscaled (2*radx*h) yscaled (2*rady*h) rotated (ang + angle
      direction length(p)/2 of p) shifted center; fullcircle transformed t
      enddef;
      style_def ellipse expr p= shadedraw ellipseraw (p,0); enddef; }
%  The following produces an ellipse along a line given by two vertices.
%  Application: \fmf{newellipse,foreground=red,rubout=1}{v1,v2}
%               draws ellipse with major semi-axis 0.6*(distance v1-v2) and
%                                  minor semi-axis 0.3*(distance v1-v2)
%               around centre of {v1,v2}
    \fmfcmd{vardef newellipseraw (expr p, ang) = save radx; numeric radx;
      radx=6/10 length p; save rady; numeric rady; rady=3/10 length p; pair
      center; center:=point 1/2 length(p) of p; save t; transform t;
                t:=identity xscaled (radx*h) %here the difference to ellipse
                yscaled (rady*h)             %here the difference to ellipse
                rotated (ang + angle direction length(p)/2 of p) shifted
                center; fullcircle transformed t enddef;
                style_def newellipse expr p= shadedraw newellipseraw (p,0);
                enddef; }
%  The following produces a circle along a line given by two vertices.
%  Application: \fmf{mycircle,foreground=red,rubout=1}{v1,v2}
%               draws circle with radius 0.1*(distance v1-v2)
%               around centre of {v1,v2}
              \fmfcmd{vardef mycircleraw (expr p, ang) = save radx; numeric
                radx; radx=1/10 length p; save rady; numeric rady; rady=1/10
                length p; pair center; center:=point 1/2 length(p) of p; save
                t; transform t; t:=identity xscaled (2*radx*h) yscaled
                (2*rady*h) rotated (ang + angle direction length(p)/2 of p)
                shifted center; fullcircle transformed t enddef;
                style_def mycircle expr p= shadedraw mycircleraw (p,0);
                enddef; }
%  The following produces a semi-ellipse along a line given by two vertices.
%  Application: \fmf{(left/right)semicircle,foreground=red,rubout=1}{v1,v2}
%               draws semi-circle with major semi-axis 0.6*(distance v1-v2) and
%                                  minor semi-axis 0.3*(distance v1-v2)
%               around centre of {v1,v2}
%  leftsemicircle: decresecnt moon; rightsemicircle: crescent moon.
              \fmfcmd{vardef mysemicircleraw (expr p, ang) = save radx;
                numeric radx; radx=6/10 length p; save rady; numeric rady;
                rady=4/10 length p; pair center; center:=point 1/2 length(p)
                of p; save t; transform t; t:=identity xscaled (2*radx*h)
                yscaled (2*rady*h) rotated (ang + angle direction length(p)/2
                of p) shifted center; halfcircle transformed t enddef;
                style_def leftsemicircle expr p= shadedraw mysemicircleraw
                (p,0)--cycle enddef; style_def rightsemicircle expr p=
                shadedraw mysemicircleraw (p,180)--cycle enddef; }
%  The following produces a triple line. Improvable.
              \fmfcmd{style_def triple expr p = save oldpen; pen oldpen;
                oldpen := currentpen; pickup oldpen scaled 5; ccutdraw p;
                pickup oldpen scaled 3; cullit; draw p withcolor white;
                cullit; pickup oldpen scaled 1; cullit; ccutdraw p; cullit;
                enddef;}
              \fi
% End of part which is active when Feynman graphs to be drawn explicitely.
%%%%%%%%%%%%%%%%%%%%%%%%%%%%%%%%%%%%%%%%%%%%%%%%%%%%%%%%%%%%%%%%%%%%%%%%%%%%%%%

%%%%%%%%%%%%%%%%%%%%%%%%%%%%%%%%%%%%%%%%%%%%%%%%%%%%%%%%%%%%%%%%%%%%%%%%%%%%%%%
%%%%%%%%%%%%%%%%%%%%%%%%%%%%%%%%%%%%%%%%%%%%%%%%%%%%%%%%%%%%%%%%%%%%%%%%%%%%%%%
% This is a nice title page including abstract ....
%

\begin{titlepage}
  \setcounter{page}{0}
  
  \mypreprint{
    %%%%%%%%%%%%%%%%%%%%%%%%%%%%%%%%%%%%%
    %\textbf{Draft version \today} \hfill
    %%%%%%%%%%%%%%%%%%%%%%%%%%%%%%%%%%%%%
    nucl-th/0404073\\
    TUM-T39-04-05\\
    23rd April 2004\\
    Revised version 9th July 2004\\
    Re-revised version 7th September 2004\\
    Accepted for publication by Nuclear Physics \textbf{A}}
  
  %\vspace*{0.5cm}
  \vspace*{0.1cm}
  
  \mytitle{Improved Convergence in the Three-Nucleon System at Very Low
    Energies}
  
  \vspace*{0.5cm}

\begin{center}
  \myauthor{Harald W.\ Grie\3hammer$^{a,b,}$}\footnote{Email:
    hgrie@physik.tu-muenchen.de;
    permanent address: a}\\[2ex]
  
  \vspace*{0.5cm}
  
  \myaddress{$^a$
    Institut f{\"u}r Theoretische Physik (T39), Physik-Department,\\
    Technische Universit{\"a}t M{\"u}nchen, D-85747 Garching, Germany}
  \\[2ex]
  \myaddress{$^b$ ECT*, Villa Tambosi, I-38050 Villazzano (Trento), Italy}
  
  \vspace*{0.2cm}

\end{center}

\vspace*{0.5cm}

\begin{abstract}
  Neutron-deuteron scattering in the context of ``pion-less'' Effective Field
  Theory at very low energies is investigated to next-to-next-to-leading
  order. Convergence is improved by fitting the two-nucleon contact
  interactions to the tail of the deuteron wave-function, a procedure known as
  Z-parameterisation and extended here to the three-nucleon system. The
  improvement is particularly striking in the doublet-$\mathrm{S}$ wave
  (triton) channel, where better agreement to potential-model calculations and
  better convergence from order to order in the power counting is achieved for
  momenta as high as $\sim 120\;\MeV$. Investigating the cut-off dependence of
  the phase-shifts, one confirms numerically the analytical finding that the
  first momentum-dependent three-body force enters at \NXLO{2}. The other
  partial waves converge also substantially faster. Effective-range parameters
  of the $nd$-system are determined, e.g.~for the quartet-$\mathrm{S}$-wave
  scattering length $a_q=[6.35\pm0.02]\;\fm$, which compares favourably both
  in magnitude and uncertainty with recent high-precision potential-model
  determinations. Differential cross-sections up to
  $E_\mathrm{lab}\approx15\;\MeV$ agree with data.
\end{abstract}
\vskip 1.0cm
\noindent
\begin{tabular}{rl}
Suggested PACS numbers:& %\begin{minipage}[t]{11cm}
                    11.80.Jy, 13.75.Cs, 14.20.Dh, 21.30.-x, 25.40.Dn, 27.10.+h 
                    %\end{minipage}
                    \\[1ex]
Suggested Keywords: &\begin{minipage}[t]{11cm}
                    Effective Field Theory, three-body system, three-body
                    force, Faddeev equation, partial waves. 
                    \end{minipage}
\end{tabular}

\vskip 1.0cm

\end{titlepage}

\setcounter{footnote}{0}

\newpage

%%%%%%%%%%%%%%%%%%%%%%%%%%%%%%%%%%%%%%%%%%%%%%%%%%%%%%%%%%%%%%%%%%%%%%%%%%%%%%%
%%%%%%%%%%%%%%%%%%%%%%%%%%%%%%%%%%%%%%%%%%%%%%%%%%%%%%%%%%%%%%%%%%%%%%%%%%%%%%%
%%%%%%%%%%%%%%%%%%%%%%%%%%%%%%%%%%%%%%%%%%%%%%%%%%%%%%%%%%%%%%%%%%%%%%%%%%%%%%%
% Main Body
%

%%%%%%%%%%%%%%%%%%%%%%%%%%%%%%%%%%%%%%%%%%%%%%%%%%%%%%%%%%%%%%%%%%%%
\section{Introduction}
\setcounter{equation}{0}
\label{sec:introduction}

The few-nucleon system at very low energies is an important tool to understand
key physical questions.  On the one hand, a new generation of high-precision
experiments at the lower end of the energy spectrum use both polarised targets
and beams, and employ neutrino detectors or radioactive-beam facilities to
extract e.g.~neutron properties, neutrino masses and reactions relevant for
nuclear astro-physics. On the other hand, a plethora of pivotal physical
processes is hard to access directly in experiments, like reaction rates in
Big-Bang nucleo-synthesis. In both cases, it is mandatory that binding effects
in light nuclei are taken into account with as few bias as possible towards a
particular model of the few-nucleon system. Theory must provide such
model-independent extraction and calibration methods or predictions at very
low energies, often well below $20\;\MeV$.

While \emph{precision} -- the numerical stability of computations -- can be
controlled by combining sophisticated algorithms with Moore's law~\cite{Moore}
and is hence (albeit sometimes formidable) not a fundamental problem, the
necessary \emph{accuracy} -- namely an estimate of the systematic
uncertainties of a theoretical ansatz -- is harder to win and mandates
understanding the physical system at hand. In the three-nucleon sector, this
calls in particular for a systematic understanding of the r\^ole three-nucleon
forces play. Traditionally, these were often introduced \emph{a posteriori} to
cure discrepancies between data and calculations, e.g.~for the triton binding
energy, but such a path is of course untenable when data are scarce or absent.
There is also a host of deviations between experiment and theory for which a
suitable three-body force could not yet be constructed, like the famed
$A_y$-problem~\cite{Huber:1998hu}.

The so-called ``pion-less'' version of Effective Field Theory in Nuclear
Physics (\EFTNoPion) aspires to provide just such a systematic classification
of all forces. At its heart lies the tenet that Physics at those very low
energies can be described by point-like interactions between nucleons only:
One cannot identify pions as the lightest exchange-particles between nucleons
as long as the typical external momentum $p_\mathrm{typ}$ in a reaction is
below the pion mass $\mpi$ because the Compton wave-lengths are not small
enough to resolve the nuclear forces as originating in part from one-pion
exchange. With all particles but the nucleons thus ``integrated out'' as
heavy, one can identify a small, dimension-less parameter
$Q=\frac{p_\mathrm{typ}}{\LambdaNoPion}\ll 1$, where $\LambdaNoPion\sim\mpi$
is the typical momentum scale at which the one-pion exchange is resolved and
\EFTNoPion must break down. The resultant power-counting orders each process
according to the power in $Q$ at which it starts to contribute and establishes
therefore an ordering scheme which is used for systematically improvable,
rigorous error estimates. This and the systematic, gauge invariant inclusion
of external electro-weak currents and relativistic effects, see
e.g.~\cite{Chen:1999tn}, distinguishes \EFTNoPion from its historical roots,
the Effective Range Expansion~\cite{Bethe} and the model-independent approach
to three-body physics~\cite{efimovI}.

\EFTNoPion is a mathematically well-defined, systematic low-energy theory of
QCD. It is computationally considerably simpler than potential models or the
``pion-ful'' version of EFT, which attempts to extend Chiral Perturbation
Theory to the few-nucleon
system~\cite{Weinberg,Ordonez:1995rz,Epelbaum:2003we}. Conceptually, many
problems which are also found when formulating a fully consistent ``pion-ful''
EFT are encountered in a simpler setting. It was used quite successfully to
provide model-independent results for a cornucopia of two-nucleon processes
with external electro-weak probes, see the
reviews~\cite{bira_review,seattle_review,bedaque_bira_review} and references
therein, also for a sketch of its development. Bedaque et
al.~\cite{3stooges_boson,3stooges_doublet} showed that a momentum-independent
three-body force must be LO for the triton and sketched the path to include
further three-body forces systematically at higher orders. Recently, a rigorous
power-counting to all orders was developed for the three-nucleon forces in
this approach, opening the path to new high-accuracy extractions and
predictions of nucleonic and nuclear properties~\cite{4stooges}. Barford and
Birse confirmed this in an analysis of the renormalisation group flow of the
position-space version of the problem~\cite{BarfordBirse}. Now, convergence
issues become interesting. They are the subject of this article.

\absatz In \EFTNoPion, the strengthes of the contact interactions between the
nucleons can be determined from low-energy observables in various ways, which
differ in principle only by higher-order effects. Still, it is standard
practise to to improve the speed of convergence by physical considerations.
One would for example not try to describe deuteron properties by starting from
the Effective Range Expansion (ERE) in the \threeS-channel of $NN$-scattering
around zero momentum, in which the correct position of the deuteron pole at
$B_t=2.2246\dots\;\MeV$ is reached only perturbatively.  Instead, it is
prudent to put the deuteron pole-position in the right place, and give the
pole its correct strength.  The deuteron wave-function has then the proper
asymptotic fall-off and normalisation.  Phillips et al.~\cite{Phillips:1999hh}
showed that this \emph{Z-parameterisation}
%~\footnote{Pronounced
%  ``Zed-parameterisation''.}
is an effective way to sum up the dominant effective-range contributions. It
both simplifies calculations of deuteron properties and improves convergence.

As the \threeS-channel of $NN$-scattering is an important sub-cluster of the
three-nucleon system, it is natural to consider the implications of this
choice for the triton and other partial waves.  With a slight extension, it
can be included into the standard EFT treatment of the three-nucleon system in
which an auxiliary two-nucleon field is introduced to simplify computations.
This article will show that not only is the convergence both from order to
order in the expansion and to experimental phase-shifts greatly improved; its
results support also the power counting for the three-body forces pertinent to
the triton -- mathematically rigorously proven to all orders
in~\cite{4stooges} -- by an error analysis of the \twoS-channel.

\absatz This article is organised as follows: The next Section merges
Z-parameterisation with the auxiliary-field method. After a brief review of
the formalism to compute $Nd$ scattering in \EFTNoPion and the occurrence of
three-body forces in the triton channel at the beginning of Section
\ref{sec:results}, the results for the \twoS-channel, triton properties and
the Phillips line are examined in Sub-Sect.~\ref{sec:tritonresults}. I touch
upon the pros and cons of including the effective range as leading-order in
Sect.~\ref{sec:allorders}.  Finally, the effect of Z-parameterisation on the
other partial waves of the $Nd$-system is discussed in
Sub-Sect.~\ref{sec:higherpws}.  Various effective-range parameters of low
partial waves are also listed, together with comparisons between experimental
and theoretical cross-sections. The Conclusions in Sect.~\ref{sec:conclusions}
are followed by an Appendix on the derivation of the three-body equations in
\EFTNoPion, defining also the spin-isospin projection operators in the
three-nucleon system.

%%%%%%%%%%%%%%%%%%%%%%%%%%%%%%%%%%%%%%%%%%%%%%%%%%%%%%%%%%%%%%%%%%%%
\section{Z-Parameterisation and Auxiliary-Field-Formalism in the Two-Nucleon
  System} \setcounter{equation}{0}
\label{sec:Zparam+aux}

%%%%%%%%%%%%%%%%%%%%%%%%%%%%%%%%%%
\subsection{Merging the Auxiliary Field Formalism \dots}
\label{sec:auxiliaryfield}

It is a standard technique to simplify calculations in \EFTNoPion both in the
two-and three-nucleon system by considering not directly contact interactions
between four nucleons, but to introduce auxiliary fields with the quantum
numbers of the two-nucleon real and virtual bound states, coupling to two
nucleons~\footnote{As by-product, I attempt to unify the notational cornucopia
  of which the present author is not completely innocent.  Throughout, the
  sub-script $t$ ($s$) denotes quantities in the spin-triplet (singlet)
  channel of $NN$ scattering.}, see
e.g.~\cite{Kaplan:1996nv,2stooges_quartet,3stooges_quartet,pbhg,Beane:2000fi}:
\begin{eqnarray}
  \label{eq:threeSlagrangean}
  \calL_{2N,t}&=&
  -y\left[ d_t^{i \dagger} (N^T P^i_t N) +\mathrm{H.c.}\right]\\ 
  &&
  +d_t^{i\dagger}\left[\Delta_t
  -c_{0t}\left(\ii\partial_0+\frac{\dev^2}{4M}+\frac{\gamma_t^2}{M}\right)
  -\sum\limits_{n=1}^\infty c_{nt}
  \left(\ii\partial_0+\frac{\dev^2}{4M}+\frac{\gamma_t^2}{M}\right)^{n+1} 
  \right]
  d_t^i\non
\end{eqnarray}
Here, $N={p \choose n}$ is the nucleon iso-doublet and
$P^i_t=\frac{1}{\sqrt{8}}\tau_2 \sigma_2\sigma^i$ the projector onto the
spin-triplet iso-spin-singlet state with vector index $i=1,2,3$. $\sigma^i$
($\tau^A$) are the spin (iso-spin) Pauli matrices, $A=1,2,3$ the iso-vector
indices. The auxiliary field $d_t$ represents the deuteron with binding
momentum $\gamma_t=\sqrt{MB_t}=45.7025\;\MeV$. Strictly speaking, one should
replace the (iso-scalar) nucleon mass $M\to M-B_t/2$ to obtain the correct
deuteron mass, but this effect is negligible in what follows. Only the
\threeS-channel of $NN$ scattering is considered here; the \oneS-channel is
discussed in Sect.~\ref{sec:oneS}. The choice of sign for $c_{nt}$ is
traditional~\cite{Kaplan:1996nv}.

That this Lagrangean is on-shell equivalent to the one containing only nucleon
fields was formally shown in~\cite{pbhg} by a Gau\3'ian integration over
$d_t$, followed by a field-redefinition and disregarding terms with more than
four nucleon fields. The advantage of this scheme -- stressed repeatedly,
e.g.~in~\cite{4stooges,2stooges_quartet,3stooges_quartet,pbhg,chickenpaper} --
is that the parameters follow na\"ive dimensional analysis: As again discussed
in Sect.~\ref{sec:zparameterisation}, $\Delta_t\sim Q$, with mass-dimension 1,
is LO; the dimension-less parameter $c_{0t}\sim Q^0$ first appears at NLO
since it comes with two powers of momentum, $c_{0t}\pv^2\sim Q^2$; while a
dimension-ful operator proportional to $c_{nt}\sim Q^0$ enters at \NXLO{2n+1}
because it is accompanied by $2n+2$ powers of the typical momentum.

All interactions which are not pure $\mathrm{S}$-wave -- like
$\mathrm{SD}$-mixing, $\mathrm{P}$-wave scattering between two nucleons
etc.~-- are added either as interactions between the deuteron $d_t$ and two
nucleons, or between four nucleons~\cite{Beane:2000fi,chickenpaper}. They are
not listed as they are of higher order than necessary in the following -- with
the exception of $\mathrm{SD}$ mixing, whose effect will be neglected in the
three-nucleon system, see the discussion in Sect.~\ref{sec:formalism}.  The
only additional term at \NXLO{2} is the kinetic energy of the nucleon, which
up to relativistic corrections is:
\begin{eqnarray}
  \label{eq:Nlagrangean}
  \calL_{1N}&=&
  N^\dagger\left(\ii\partial_0+\frac{\dev^2}{2M}\right)N
\end{eqnarray}

The bare deuteron propagator $\ii/\Delta_t$ is thus dressed at LO by all
interactions proportional to $y$ with an arbitrary number of loops, see
Fig.~\ref{fig:dprop}. At NLO, one perturbative insertion proportional to
$c_{0t}$ is included, followed by two at \NXLO{2}, and in general $n$ at
\NXLO{n}. In addition, one insertion of the operator proportional to $c_{nt}$
enters at \NXLO{2n+1}, etc.

\begin{figure}[!htb]
\begin{center}
  \ifx\feynVersion\AnswerYes
      %%% if Feynman graphs explicit
  \feyngraph{30}{40}{0}{0}{0}{0}{ \fmfleft{i} \fmfright{o}
    \fmf{double,width=thin}{i,o} } \hq $=$ \hq \feyngraph{30}{40}{0}{0}{0}{0}{
    \fmfleft{i} \fmfright{o}
    \fmf{vanilla,width=3thin,foreground=(0.5,,0.5,,0.5),
      label=$\blue{\fs\frac{\ii}{\Delta_t}}$,
      label.side=right,label.dist=0.15h}{i,o} } \hq $+$ \hq
  \feyngraph{30}{40}{0}{0}{0}{0}{ \fmfleft{i} \fmfright{o}
    \fmf{vanilla,width=3thin,tension=6, foreground=(0.5,,0.5,,0.5)}{i,v1}
    \fmf{vanilla,width=thin,left=0.65}{v1,v2}
    \fmf{vanilla,width=thin,left=0.65}{v2,v1}
    \fmf{vanilla,width=3thin,tension=6, foreground=(0.5,,0.5,,0.5)}{v2,o} }
  \hq $+$ \hq \feyngraph{54}{40}{0}{0}{0}{0}{ \fmfleft{i} \fmfright{o}
    \fmf{vanilla,width=3thin,tension=6, foreground=(0.5,,0.5,,0.5)}{i,v1}
    \fmf{vanilla,width=thin,left=0.65}{v1,v2}
    \fmf{vanilla,width=thin,left=0.65}{v2,v1}
    \fmf{vanilla,width=3thin,tension=6, foreground=(0.5,,0.5,,0.5)}{v2,v3}
    \fmf{vanilla,width=thin,left=0.65}{v3,v4}
    \fmf{vanilla,width=thin,left=0.65}{v4,v3}
    \fmf{vanilla,width=3thin,tension=6, foreground=(0.5,,0.5,,0.5)}{v4,o} }
  \hq $+$ \hq \feyngraph{78}{40}{0}{0}{0}{0}{ \fmfleft{i} \fmfright{o}
    \fmf{vanilla,width=3thin,tension=6, foreground=(0.5,,0.5,,0.5)}{i,v1}
    \fmf{vanilla,width=thin,left=0.65}{v1,v2}
    \fmf{vanilla,width=thin,left=0.65}{v2,v1}
    \fmf{vanilla,width=3thin,tension=6, foreground=(0.5,,0.5,,0.5)}{v2,v3}
    \fmf{vanilla,width=thin,left=0.65}{v3,v4}
    \fmf{vanilla,width=thin,left=0.65}{v4,v3}
    \fmf{vanilla,width=3thin,tension=6, foreground=(0.5,,0.5,,0.5)}{v4,v5}
    \fmf{vanilla,width=thin,left=0.65}{v5,v6}
    \fmf{vanilla,width=thin,left=0.65}{v6,v5}
    \fmf{vanilla,width=3thin,tension=6, foreground=(0.5,,0.5,,0.5)}{v6,o} }
  \hq $+$ \hq \dots
  \\
  \feyngraph{60}{40}{0}{0}{0}{0}{ \fmfleft{i} \fmfright{o}
    \fmf{double,width=thin}{i,v,o} \fmfv{decor.shape=cross,foreground=red,
      label=$\fs-\ii c_{0t}\big(p_0-\frac{\pv^2}{4M}
      +\frac{\gamma_t^2}{M}\big)$, label.angle=-90,label.dist=0.2h}{v} }
  \hq\hq \hq\hq \hq\hq \hq\hq \feyngraph{60}{40}{0}{0}{0}{0}{ \fmfleft{i}
    \fmfright{o} \fmf{double,width=thin}{i,v1,v2,o}
    \fmfv{decor.shape=cross,foreground=red}{v1}
    \fmfv{decor.shape=cross,foreground=red}{v2} } \hq\hq \hq\hq \hq\hq \hq\hq
  \feyngraph{80}{40}{0}{0}{0}{0}{ \fmfleft{i} \fmfright{o}
    \fmf{double,width=thin}{i,v1,v2,v3,o}
    \fmfv{decor.shape=cross,foreground=red}{v1}
    \fmfv{decor.shape=cross,foreground=red}{v2}
    \fmfv{decor.shape=cross,foreground=red}{v3} } \hq,\hq
  \feyngraph{60}{40}{0}{0}{0}{0}{ \fmfleft{i} \fmfright{o}
    \fmf{double,width=thin}{i,v,o} \fmfv{decor.shape=hexacross,foreground=red,
      label=$\fs-\ii c_{1t}\big(p_0-\frac{\pv^2}{4M}
      +\frac{\gamma_t^2}{M}\big)^2$, label.angle=-90,label.dist=0.2h}{v} }
  \else
      %%% if Feynman graphs as .eps file
  \includegraphics*[width=0.82\textwidth]{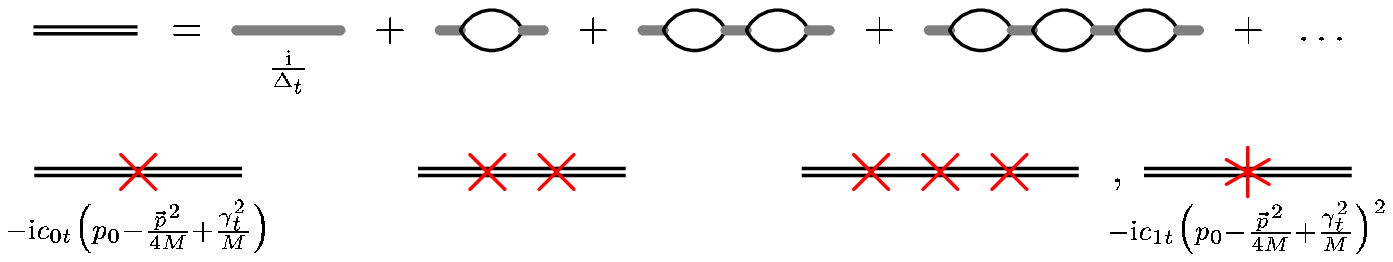} \fi
\caption{Top: Re-summation of the bare deuteron propagator (thick gray line)
  into the dressed deuteron propagator (double line) at LO by dressing with
  two-nucleon bubbles. Bottom: The NLO, \NXLO{2} and \NXLO{3} corrections to
  the deuteron propagator. The cross denotes an insertion of the deuteron
  kinetic-energy operator proportional to $C_{0t}$, the star one proportional
  to $c_{1t}$.}
\label{fig:dprop}
\end{center}
\end{figure}

Choosing for the $d_tNN$ coupling constant without loss of generality
\begin{equation}
  \label{eq:ychoice}
  y^2=\frac{4\pi}{M}\;\;,
\end{equation}
the auxiliary-field propagator with kinetic energy $p_0$ and momentum $\pv$
becomes particularly simple in the low-energy expansion:
\begin{eqnarray}
  \label{eq:dprop}
  &&\hqm\hqm\hqm\hqm\hqm\hqm
    \ii\;D_t(p_0,\pv)\non\\
  &&\hqm\hqm\hqm\hqm\hqm\hqm
    =\frac{\ii}{\Delta_t+\mu-\sqrt{\frac{\pv^2}{4}-Mp_0-\ii\epsilon}}\;
    \sum\limits_{m=0}^\infty 
    \left[\frac{c_{0t}(p_0-\frac{\pv^2}{4M}+\frac{\gamma_t^2}{M})
     +\sum\limits_{n=1}^\infty c_{nt}\left(p_0-\frac{\pv^2}{4M}
       +\frac{\gamma_t^2}{M}\right)^{n+1}}
    {\Delta_t+\mu-\sqrt{\frac{\pv^2}{4}-Mp_0-\ii\epsilon}}\right]^m\\
  &&\hqm\hqm\hqm\hqm\hqm\hqm
    \to\frac{\ii}{\Delta_t+\mu-\sqrt{\frac{\pv^2}{4}-Mp_0-\ii\epsilon}
    -c_{0t}(p_0-\frac{\pv^2}{4M}+\frac{\gamma_t^2}{M})
    -\sum\limits_{n=1}^\infty c_{nt}
    \left(p_0-\frac{\pv^2}{4M}+\frac{\gamma_t^2}{M}\right)^{n+1}}\non
\end{eqnarray}
Here, $\mu$ is the regulator of the linear divergence in the nucleon loop,
regularised using dimensional regularisation with the PDS subtraction
scheme~\cite{Kaplan:1998we}. The scattering amplitude between two
non-relativistic nucleons with relative centre-of-mass (cm) momentum $k$ in
the \threeS-channel is obtained by multiplying with $-y^2=-4\pi/M$ and setting
the nucleons on-shell ($p_0=k^2/M$, $\pv=\vec{0}$):
\begin{eqnarray}
  \label{eq:NNscatteringEFT}
  \calA_{NN}(k) &=&-\frac{4\pi}{M}\;\frac{1}{\Delta_t+\mu+\ii k}\;
  \sum\limits_{m=0}^\infty 
  \left[\frac{c_{0t}(\frac{k^2}{M}+\frac{\gamma_t^2}{M})
     +\sum\limits_{n=1}^\infty c_{nt}\left(\frac{k^2}{M}
       +\frac{\gamma_t^2}{M}\right)^{n+1}}
   {\Delta_t+\mu+\ii k}\right]^m\\
&\to&-\frac{4\pi}{M}\;\frac{1}{\Delta_t+\mu+\ii k
  -c_{0t}(\frac{k^2}{M}+\frac{\gamma_t^2}{M})
  -\sum\limits_{n=1}^\infty c_{nt}
  \left(\frac{k^2}{M}+\frac{\gamma_t^2}{M}\right)^{n+1}}\non
\end{eqnarray}
To sum all effective-range corrections to all orders in the second line of
(\ref{eq:dprop}) and (\ref{eq:NNscatteringEFT}) has no advantage but to
shorten the following determination of the low-energy coefficients; actual
calculations involve only a finite number of $c_{nt}$'s, in which case the
difference between the two lines is formally of higher order.  Problems with
re-summing the effective-range corrections ``to all orders'' are discussed in
Sect.~\ref{sec:allorders}.

%%%%%%%%%%%%%%%%%%%%%%%%%%%%%%%%%%
\subsection{\dots with Z-Parameterisation}
\label{sec:zparameterisation}

In Effective Range Expansion (ERE) around the deuteron pole~\cite{Bethe}, the
scattering amplitude of two non-relativistic nucleons with relative cm
momentum $k$ in the \threeS-channel is
\begin{equation}
  \label{eq:NNscatteringERE}
  \calA_{NN}(\threeS,k)=
  -\frac{M}{4\pi}\;\frac{1}{\gamma_t-\frac{\rho_{0t}}{2}\;(\gamma_t^2+k^2)- 
    \sum\limits_{n=1}^{\infty}\rho_{nt}(\gamma_t^2+k^2)^{n+1}+\ii\;k}\;\;,
\end{equation}
where $\rho_{0t}=1.764\;\fm$ is the effective range, $\rho_{1t}=0.389\;\fm^3$
the shape parameter etc.  Matching (\ref{eq:NNscatteringEFT}) to the
ERE-result (\ref{eq:NNscatteringERE}), one obtains the
\emph{ERE-parameterisation} of \EFTNoPion:
\begin{equation}
  \label{eq:parametersERE}
  \Delta_{t,\mathrm{ERE}}+\mu_\mathrm{ERE}=\gamma_t\;\;,\;\;
  c_{0t,\mathrm{ERE}}=\frac{\rho_{0t}}{2}\;M\;\;,\;\;
  c_{nt,\mathrm{ERE}}=\rho_{nt}\;M^{n+1}\;\;\forall n\ge 1
\end{equation}
While $\gamma_t$ is clearly a typical low-momentum scale in \EFTNoPion,
$\gamma_t\ll\LambdaNoPion\sim\mpi$, the other parameters encode Physics beyond
the breakdown scale and hence one expects $\rho_{nt}\sim\LambdaNoPion^{-2n-1}$
from dimensional analysis. Indeed, contributions to the scattering amplitude
for example from the shape-parameter are small, with the expansion parameter
$Q$ estimated as $Q\sim\gamma_t \rho_{1t}^{1/3}\approx 0.17$. Therefore, the
amplitude can be expanded in powers of $Q\sim\gamma_t^{1/(2n+1)}\rho_{nt}$,
and an error estimate follows, rendering Effective Range Theory useful:
\begin{equation}
  \label{eq:NNscatteringEREexpanded}
  \calA_{NN}(\threeS,k)=
  -\frac{M}{4\pi}\;\frac{1}{\gamma_t+\ii\;k}
  \left[1+\frac{\rho_{0t}}{2}\;\frac{\gamma_t^2+k^2}{\gamma_t+\ii\;k}+
  \left(\frac{\rho_{0t}}{2}\;\frac{\gamma_t^2+k^2}{\gamma_t+\ii\;k}\right)^2
  +\calO(Q^3) \right]
\end{equation}
%Such an expansion is also prudent in view of the fact that the re-summed form
%shows $2n+2$ poles when all coefficients up to $c_{nt}$ are kept. The
%additional $2n+1$ spurious poles cause complications, as discussed in more
%detail in Sect.~\ref{sec:allorders}.
A problem with the expanded version (\ref{eq:NNscatteringEREexpanded}) arises
however because the effective range is numerically somewhat larger than
na\"ively expected, $\rho_0>\mpi^{-1}$ or
\begin{equation}
  \label{eq:expansion}
  \gamma_t \rho_{0t}\approx 0.41\;\;.
\end{equation}
While the deuteron binding energy is in ERE-parameterisation reproduced
immediately at LO, the residue of the scattering amplitude at the deuteron
pole is now only expanded in this numerically not-so-small parameter,
\begin{eqnarray}
  \label{eq:residue}
  \mathrm{Res}[\calA_{NN}(\threeS,k=\ii\gamma_t)]&&=-\frac{4\pi}{M}\; Z_t \;\;
    \mbox{ with }\nonumber\\
  Z_t=\frac{1}{1-\gamma_t\rho_{0t}}&&=1.690(3)\\
  =1+\gamma_t\rho_{0t}+(\gamma_t\rho_{0t})^2+(\gamma_t\rho_{0t})^3+\dots&&=
  1+0.409+0.167+0.068+\dots\;\;,\nonumber
\end{eqnarray}
so that even including \NXLO{2}-effects, seven percent are missed to close the
gap to the exact value. In contrast, not considering the \NXLO{3}
shape-parameter effects in $\calA$ leads only to a deviation on the order of
$\gamma_t^3\rho_{1t}=0.5\%$.

Clearly, the residue is an important characteristic of the two-body system as
it determines the asymptotic normalisation of the deuteron wave-function at
large distances $r$:
\begin{equation}
  \label{eq:deuteronwavefu}
  \Psi_\mathrm{deuteron}(r\to\infty)=\sqrt{\frac{\gamma Z_t}{2\pi}}\;
  \frac{\e^{-\gamma r}}{r}\;\;.
\end{equation}
Phillips et al.~\cite{Phillips:1999hh} re-summed therefore the series in
$\gamma_t\rho_{0t}$ partially by expressing $\rho_{0t}$ via $Z_t$. To discuss
its advantage, re-write (\ref{eq:NNscatteringERE}) in a superficially more
complicated way as
\begin{eqnarray}
  \label{eq:effrangeZ}
  &&\frac{1+(Z_t-1)}{\gamma_t+\ii k}\;
  \frac{1}{1+\frac{Z_t-1}{2}\;\left(1+\frac{\ii k}{\gamma_t}\right)-
  Z_t\sum\limits_{n=1}^\infty \rho_{nt}\;(\gamma_t-\ii k)
  \;(\gamma_t^2+k^2)^n}\\
 &&
 =\frac{1}{\gamma_t+\ii k}\;
 \left[1+(Z_t-1)\right]
 \left[1-\frac{Z_t-1}{2}\left(1+\frac{\ii k}{\gamma_t}\right)+
   \left(\frac{Z_t-1}{2}\left(1+\frac{\ii k}{\gamma_t}\right)\right)^2+
   \calO(Q^3)
 \right]
  \;\;.\nonumber
\end{eqnarray}
Formally, $Z_t-1\approx 0.69$ is still treated as a small expansion parameter
in the resulting \emph{Z-parameterisation}, but the deuteron residue is now
restored already at NLO with no corrections from higher orders as the second
denominator in (\ref{eq:effrangeZ}) has no residue at $\gamma_t=-\ii k$:
\begin{equation}
  Z_t=\underbrace{\rule[-1ex]{0ex}{0ex}1}_\mathrm{LO}+
  \underbrace{\rule[-1ex]{0ex}{0ex}(Z_t-1)}_\mathrm{NLO} +
  \underbrace{\rule[-1ex]{0ex}{0ex}0}_\text{\NXLO{n}}\;\;,\;\;n\ge2\;.
\end{equation}
Outside the deuteron pole, the additional terms from $Z_t$ also converge
faster because the second term in the second denominator is a power series in
$(Z_t-1)/2\approx0.3$. A slight dis-advantage is that the expansion parameter
for the higher-order correction is now bigger by $70\%$:
$Z_t\;\rho_{nt}\gamma_t^{2n+1}$. Thus, the \NXLO{3}-correction from the shape
parameter is now $\sim 0.8\%$. At this level of accuracy, however, other
corrections (e.g.~from $\mathrm{P}$-wave interactions) must be considered,
whose scale is also set by the pion mass and which are estimated to be
stronger because $(\gamma_t/(\LambdaNoPion\sim\mpi))^3\sim 3\%$.

In Z-parameterisation, the expanded version of the \EFTNoPion-amplitude
(\ref{eq:NNscatteringEFT}) is therefore first matched to reproduce the correct
deuteron pole position: $\Delta_t+\mu=\gamma_t$. Then, the residue is found
starting at NLO as
\begin{equation}
  \label{eq:residuematching}
  Z_t\stackrel{!}{=}1+\sum\limits_{n=1}^\infty
  \left(c_{0t}\;\frac{2\gamma_t}{M}\right)^n \;\;.
\end{equation}
At LO, no additional free parameter exists in EFT and the residue is one. At
higher orders, the residue is set equal to $Z_t$ when $c_{0t}$ is suitably
chosen: With the expansion truncated at some finite order,
$c_{0t}=\sum\limits_{n=0}^{n_\text{max}} c_{0t}^{(n)}$ contains now
contributions from higher orders in the $Q$-expansion, $c_{0t}^{(n)}$ being
$\calO(Q^n)$. To summarise, the parameters in Z-parameterisation are
\begin{eqnarray}
  \label{eq:parametersZ}
  y^2&=&\frac{4\pi}{M}\sim Q^0\\
  \Delta_t+\mu&=&\gamma_t\sim Q\\
  c_{0t}^{(n)}&=&(-)^n\;(Z_t-1)^{n+1}\;\frac{M}{2\gamma_t}\sim Q^n\\
  c_{nt}&=&\rho_{nt}\;M^{n+1}\sim Q^0\;\;\forall n\ge 1\;\;.
\end{eqnarray}
In contradistinction, $c_{0t}$ receives in ERE-parameterisation only a
contribution at order $Q^0$, and none at higher orders
(\ref{eq:parametersERE}).  However, such an expansion is with our choice of
Lagrangean (\ref{eq:threeSlagrangean}) only encountered for one parameter, and
not -- as previously -- for
all~\cite{4stooges,3stooges_quartet,pbhg,chickenpaper}.

The auxiliary-field propagator to \NXLO{2},
\begin{equation}
  \label{eq:dpropexpanded}
  \begin{split}
    &D_t(p_0,\pv)=\frac{1}{\gamma_t-\sqrt{\frac{\pv^2}{4}-Mp_0-\ii\epsilon}}\\
    &\times \left[\underbrace{\rule[-3ex]{0ex}{0ex}1}_{\mathrm{LO}}+
      \underbrace{\rule[-3ex]{0ex}{0ex}\frac{Z_t-1}{2\gamma_t}\;
        \left(\gamma_t+\sqrt{\frac{\pv^2}{4}-Mp_0-\ii\epsilon}\right)
      }_{\mathrm{NLO}}+ \underbrace{\rule[-3ex]{0ex}{0ex}
        \left(\frac{Z_t-1}{2\gamma_t}\right)^2\;
        \left(\frac{\pv^2}{4}-Mp_0-\gamma_t^2\right)}_{\text{\NXLO{2}}}+
      \calO(Q^3) \right]\;\;,
  \end{split}
\end{equation}
is thus a faster-converging alternative to the usual ERE of the
$NN$-scattering amplitude in the \threeS-channel. The wave-function
renormalisation, i.e.~the residue of the deuteron propagator, is by
construction exact at NLO:
\begin{equation}
  \label{eq:wavefuren}
  \calZ_t:=\left(\frac{\de}{\de p_0}\;\frac{1}{D_t(p_0,\vec{0})}
    \bigg|_{p_0=-\frac{\gamma_t^2}{M}}\right)^{-1}
  =\frac{2\gamma_t}{M}\;\left[\underbrace{\rule[-1ex]{0ex}{0ex}1
    }_{\mathrm{LO}}+
    \underbrace{\rule[-1ex]{0ex}{0ex}(Z_t-1)}_{\mathrm{NLO}}+
    \underbrace{\rule[-1ex]{0ex}{0ex}0
    }_{\text{\NXLO{n}}}\;\;,\;n\ge2
  \right]
\end{equation}

\absatz One could also replace directly the deuteron propagator in ERE by the
result in Z-parameter\-isation (\ref{eq:effrangeZ}), as e.g.~argued to be
computationally simpler by Beane and Savage~\cite{Beane:2000fi}. This would be
mandatory if $\gamma_t\rho_{0t}\approx 1$. Equivalently, if one re-sums all
orders in $c_{0t}$, one finds again the ERE-parameterisation
(\ref{eq:parametersERE}),
$c_{0t}=\frac{Z_t-1}{Z_t}\;\frac{M}{2\gamma_t}=\frac{\rho_{0t}}{2}\;M$.
However, there are a number of dis-advantages of this approach, as will be
discussed in Sect.~\ref{sec:allorders}.

%%%%%%%%%%%%%%%%%%%%%%%%%%%%%%%%%%
\subsection{Z-Parameterisation for the \oneS-channel?}
\label{sec:oneS}

The obvious question is: Why not also impose Z-parameterisation for the
\oneS-channel of $NN$ scattering? As its bound state is only virtual, one
usually performs the ERE around zero momentum,
\begin{equation}
  \label{eq:NNscatteringEREtransvestiteprelim}
  \calA_{NN}(\oneS,k)=
  -\frac{M}{4\pi}\;\frac{1}{-\frac{1}{a_s}-\frac{r_{0s}}{2}\; k^2- 
    \sum\limits_{n=1}^{\infty}r_{ns}k^{2n+2}+\ii\;k}\;\;,
\end{equation}
with $a_s=23.714\;\fm,\;r_{0s}=2.73\;\fm,\;\;r_{1s}=-0.48\;\fm^3$ the
scattering length, effective range and shape parameter. As
$r_{0s}/a_s\approx0.11$, the series-expansion of the residue of the virtual
bound-state converges much faster than in the deuteron channel. Still, in
order to simplify notation, consider a Lagrangean analogous to the
\threeS-channel:
\begin{eqnarray}
  \label{eq:oneSlagrangean}
  \calL_{2N,s}&=&
  -y\left( d_s^{A \dagger} (N^T P^A_s N) +\mathrm{H.c.}\right)\\ 
  &&
  +d_s^{A\dagger}\left[\Delta_s
  -c_{0s}\left(\ii\partial_0+\frac{\dev^2}{4M}+\frac{\gamma_s^2}{M}\right)
  -\sum\limits_{n=1}^\infty c_{ns}
  \left(\ii\partial_0+\frac{\dev^2}{4M}+\frac{\gamma_s^2}{M}\right)^{n+1} 
  \right]
  d_s^A\non
\end{eqnarray}
The auxiliary field $d_s$ represents the spin-singlet iso-spin-triplet state,
whose projector is $P^A_s=\frac{1}{\sqrt{8}} \sigma_2\tau_2\tau^A$. One now
first re-writes the ERE expansion (\ref{eq:NNscatteringEREtransvestiteprelim})
in the form analogous to (\ref{eq:NNscatteringERE}) in which the pole position
does not change from order to order,
\begin{eqnarray}
  \label{eq:NNscatteringEREtransvestite}
  \calA_{NN}(\oneS,k)=
  -\frac{M}{4\pi}\;\frac{1}{\gamma_s-\frac{\rho_{0s}}{2}\;(\gamma_s^2+k^2)- 
    \sum\limits_{n=1}^{\infty}\rho_{ns}(\gamma_s^2+k^2)^{n+1}+\ii\;k}\;\;\,
\end{eqnarray}
and determines the coefficients by matching as
\begin{eqnarray}
  \label{eq:matchingtransvestite}
  \gamma_s&=&\frac{1}{a_s}+\frac{r_{0s}}{2}\;
  \gamma_s^2-\sum\limits_{n=1}^\infty r_{ns}(-\gamma_s^2)^{n+1}\\
  \rho_{0s}&=&r_{0s}+2\sum\limits_{n=1}^\infty (n+1)\;r_{ns}(-\gamma_s^2)^{n}\\
  \rho_{ms}&=&r_{ms}+\sum\limits_{n=m}^\infty
  {n+1\choose m+1}\;r_{ns}(-\gamma_s^2)^{n-l}\;\;,\;m>0\;\;.
\end{eqnarray}  
Truncation at $\rho_{0s}$ ($\rho_{1s}$) leads to the numerical values
$\gamma_s=-7.8904\;\MeV \;(-7.8902\;\MeV)$, $\rho_{0s}=2.730\;\fm
\;(2.733\;\fm)$ (, $\rho_{1s}=r_{1s}$). The residue is indeed very close to
unity:
\begin{equation}
  Z_s=\frac{1}{1-\gamma_s\rho_{0s}}=0.9016 \;(0.9015)
\end{equation}  
At \NXLO{2}, the difference between the perturbatively built residue
$1+\gamma_s\rho_{0s}+(\gamma_s\rho_{0s})^2=1-0.1092+0.0119=0.9027$ and the
exact values is with $0.1\%$ considerably smaller than leaving out
relativistic and other effects. In the results presented in the following
Section, the difference between the ERE-parameterisation and
Z-parameterisation for the \oneS-channel cannot be discerned in the plots of
the phase-shifts.  Still, the re-formulation serves the purpose to compactify
and simplify formulae. The parameters of the Lagrangean are determined in
Z-parameterisation by the analogue to (\ref{eq:parametersZ}):
\begin{eqnarray}
  \label{eq:parametersZtarnsvestite}
  y^2&=&\frac{4\pi}{M}\sim Q^0\\
  \Delta_s+\mu&=&\gamma_s\sim Q\\
  c_{0s}^{(n)}&=&(-)^n\;(Z_s-1)^{n+1}\;\frac{M}{2\gamma_s}\sim Q^n\\
  c_{ns}&=&\rho_{ns}\;M^{n+1}\sim Q^0\;\;\forall n\ge 1\;\;.
\end{eqnarray}
Note that in contradistinction to previous work, $y$ is chosen to be identical
in the spin-triplet and spin-singlet channel and thus carries no sub-script.

%%%%%%%%%%%%%%%%%%%%%%%%%%%%%%%%%%%%%%%%%%%%%%%%%%%%%%%%%%%%%%%%%%%%
\section{The Three-Body System in Z-Parameterisation}
\setcounter{equation}{0}
\label{sec:results}

%%%%%%%%%%%%%%%%%%%%%%%%%%%%%%%%%%
\subsection{Formalism}
\label{sec:formalism}

With the parameters of the two-nucleon Lagrangean fixed by Z-parameterisation,
it is now straight-forward to state the equations governing neutron-deuteron
scattering. As they were derived repeatedly in the literature (see
e.g.~\cite{3stooges_doublet,chickenpaper,skorny}), the following presentation
focuses mainly on notation. Appendix~\ref{app:appendix} contains a brief
overview, defining also the pertinent projection operators on the various
partial waves and three-nucleon configurations.

Two cluster-configurations exist in the three-nucleon system: The
$Nd_t$-cluster with total spin $S=\frac{3}{2}$ or $S=\frac{1}{2}$, depending
on whether the deuteron and nucleon spins are parallel or anti-parallel; and
the $Nd_s$-cluster which has total spin $S=\frac{1}{2}$, as $d_s^A$ is a
scalar.  The leading-order three-particle amplitude is $\calO(Q^{-2})$ (before
wave-function renormalisation) and includes all diagrams built out of the
leading two-body interactions, i.e.~the ones proportional to $y$ and
$\Delta_t,\;\Delta_s$ in the two-nucleon Lagrangeans
(\ref{eq:threeSlagrangean}/\ref{eq:oneSlagrangean}). The resultant Faddeev
integral equation -- first derived by Skorniakov and
Ter-Martirosian~\cite{skorny} without three-body force -- is pictorially
represented in Fig.~\ref{fig:faddeeveq}.

As the Lagrangean up to \NXLO{2} does not mix partial waves or flip the spin
of the auxiliary fields, angular momentum is conserved in the quartet and
doublet channels. Strictly speaking, $\mathrm{SD}$-mixing in the deuteron
channel produces a splitting and mixing of the three-body amplitudes with the
same spin and angular momentum but different total angular momentum. However,
we limit ourselves to the averaged phase-shifts, as the spitting is at this
order not going to be realistic enough to describe spin-observables in the
three-nucleon system. This path was also pursued in~\cite{4stooges} and
\cite{chickenpaper} which used ERE-parameterisation, to which the findings in
Z-parameterisation will be compared. An analysis of these spitting is
postponed to a future presentation, which also deals with spin-observables in
the three-nucleon system.

\begin{figure}[!htb]
\begin{center}
  \ifx\feynVersion\AnswerYes
      %%% if Feynman graphs explicit
  
  \vspace*{3ex}
  
  \setlength{\unitlength}{0.7pt} \feyngraph{75}{64}{18}{0}{0}{0}{
    \fmfleft{i2,i1} \fmfright{o2,o1} \fmf{vanilla,width=2thick}{i1,v1,o1}
    \fmf{vanilla,width=thin}{i2,v5,o2}
    \fmfv{label=$\fs(E-\frac{\kv^2}{2M},,\kv)$,label.angle=90}{i1}
    \fmfv{label=$\fs(\frac{\kv^2}{2M},,-\kv)$,label.angle=-90}{i2}
    \fmfv{label=$\fs(E-\frac{\pv^2}{2M},,\pv)$,label.angle=90}{o1}
    \fmfv{label=$\fs(\frac{\pv^2}{2M},,-\pv)$,label.angle=-90}{o2} \fmffreeze
    \fmf{ellipse,foreground=red,rubout=1}{v1,v5} } \hq\hq\hq$=$\hq
  \feyngraph{75}{64}{0}{0}{0}{0}{ \fmfleft{i2,i1} \fmfright{o2,o1}
    \fmf{phantom}{i1,v1} \fmf{phantom}{v1,v2} \fmf{phantom}{v2,o1}
    \fmf{phantom}{i2,v5} \fmf{phantom}{v5,v6} \fmf{phantom}{v6,o2} \fmffreeze
    \fmf{vanilla,width=2thick}{i1,v1} \fmf{vanilla,width=2thick}{v6,o2}
    \fmf{vanilla,width=thin}{v1,v2,o1} \fmf{vanilla,width=thin}{i2,v5,v6}
    \fmf{vanilla,width=thin,label=$\fs\calK$,label.side=left}{v1,v6} }
  \hq$+$\hq \feyngraph{75}{64}{0}{0}{0}{0}{ \fmfleft{i2,i1} \fmfright{o2,o1}
    \fmf{vanilla,width=2thick}{i1,v,o1} \fmf{vanilla,width=thin}{i2,v,o2}
    \fmfv{decor.shape=circle,decor.filled=full,decor.size=3thick,
      label=$\fs\blue{\calH}$,label.angle=90,label.dist=0.3h }{v} } \hq $+$\hq
  \feyngraph{120}{64}{0}{0}{0}{0}{ \fmfleft{i2,i1} \fmfright{o2,o1}
    \fmf{phantom}{i1,v1} \fmf{phantom}{i2,v5} \fmf{phantom,tension=0.5}{v1,v2}
    \fmf{phantom,tension=0.5}{v5,v6} \fmf{phantom,tension=2}{v3,o1}
    \fmf{phantom,tension=2}{v7,o2} \fmf{phantom}{v2,v3} \fmf{phantom}{v6,v7}
    \fmffreeze \fmf{vanilla,width=2thick}{i1,v1} \fmf{vanilla,width=2thick,
      label=$\fs\calD$,label.side=left}{v1,v2}
    \fmf{vanilla,width=2thick}{v7,o2} \fmf{vanilla,width=thin}{v2,v3,o1}
    \fmf{vanilla,width=thin}{i2,v5,v6,v7}
    \fmf{vanilla,width=thin,label=$\fs\calK$,label.side=left}{v2,v7}
    \fmf{ellipse,foreground=red,rubout=1}{v1,v5} \fmffreeze } \hq$+$\hq
  \feyngraph{120}{64}{0}{0}{0}{0}{ \fmfleft{i2,i1} \fmfright{o2,o1}
    \fmf{phantom,tension=1.5}{i1,v1} \fmf{phantom,tension=1.5}{i2,v5}
    \fmf{phantom}{v1,v2} \fmf{phantom}{v5,v6} \fmf{phantom}{v2,v3}
    \fmf{phantom}{v6,v7} \fmf{phantom,tension=2}{v3,o1}
    \fmf{phantom,tension=2}{v7,o2} \fmffreeze
    \fmf{vanilla,width=2thick}{i1,v1} \fmf{vanilla,width=2thick,
      label=$\fs\calD$,label.side=left}{v1,v2}
    \fmf{vanilla,width=2thick}{v3,o1} \fmf{vanilla,width=thin}{v7,o2}
    \fmf{vanilla,width=thin}{i2,v5,v6} \fmf{vanilla,width=2thick}{v2,v,v3}
    \fmf{vanilla,width=thin}{v6,v,v7}
    \fmfv{decor.shape=circle,decor.filled=full,decor.size=3thick,
      label=$\fs\blue{\calH}$,label.angle=90,label.dist=0.4h }{v}
    \fmf{ellipse,foreground=red,rubout=1}{v1,v5} }
  \setlength{\unitlength}{1pt}
  
  \vspace*{4ex}
           
  \else
      %%% if Feynman graphs as .eps file
  \includegraphics*[width=1.00\textwidth]{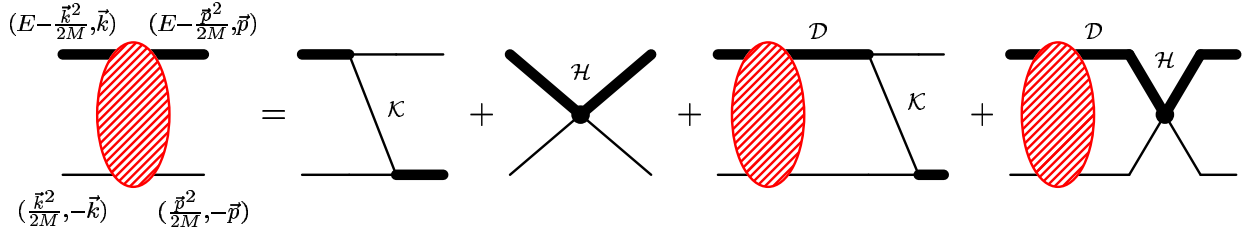} \fi
\caption{The Faddeev equation for $Nd$-scattering to \NXLO{2}. Thick solid
  line: propagator of the two intermediate auxiliary fields $d_s$ and $d_t$,
  denoted by $\calD$, see (\ref{eq:totaltwoparticlepropagator}); $\calK$:
  propagator of the exchanged nucleon, see (\ref{eq:projectedNpropagator});
  $\calH$: three-body force, see (\ref{eq:calH}).}
\label{fig:faddeeveq}
\end{center}
\end{figure}

\absatz The spin-quartet channel of the $Nd$-system is particularly simple.
The $l$th partial wave of the amplitude $t^{(l)}_q$ in the centre-of-mass (cm)
frame is given by~\footnote{The sub-script $q$ ($d$) denotes quantities in the
  spin-quartet (doublet) channel of the three-nucleon system.}
\begin{eqnarray}
  \label{eq:quartetpw}
  t^{(l)}_q(E;k,p)=-4\pi\; \calK^{(l)}(E;k,p)
  +\frac{2}{\pi}\int\limits_0^\infty\deint{}{q} q^2\;\calK^{(l)}(E;q,p)\;
  D_t(E-\frac{q^2}{2M},q)\;t^{(l)}_q(E;k,q)\;\;.
\end{eqnarray}
The total non-relativistic energy is
$E:=\frac{3\kv^2}{4M}-\frac{\gamma_t^2}{M}$; the incoming (outgoing)
deuteron-momentum $\kv$ ($\pv$); and the projected propagator of the exchanged
nucleon on angular momentum $l$~\footnote{This corrects a typographical error
  in a previous paper on the higher partial waves in the
  $Nd$-system~\cite{chickenpaper}, affecting the odd partial waves.}
\begin{equation}
  \label{eq:projectedNpropagator}
  \calK^{(l)}(E;q,p):=\frac{1}{2}\;\int\limits_{-1}^1\deint{}{x}
  \frac{P_l(x)}{p^2+q^2-ME-\ii\epsilon+ pqx}
  =\frac{(-1)^l}{pq}\;Q_l\left(\frac{p^2+q^2-ME-\ii\epsilon}{pq}\right)\;\;,
\end{equation}
where the $l$th Legendre polynomial of the second kind with complex argument
is as in~\cite{Gradstein}
\begin{equation}
  \label{eq:legendreQ}
   Q_l(z)=\half\int\limits_{-1}^1\deint{}{t}\frac{P_l(t)}{z-t}\;\;.
\end{equation}

In the doublet channel, the Faddeev equation is two-dimensional in
cluster-configuration space as both $Nd_t$- and $Nd_s$-configurations
contribute:
\begin{eqnarray}
  \label{eq:doubletpw}
  \vec{t}^{(l)}_d(E;k,p)&=&2\pi\;
  \left[\calK^{(l)}(E;k,p)\;{1\choose -3}
    +\delta^{l0}\;\calH(E;\Lambda)\;{1\choose -1}\right]\\
  &&-\;\frac{1}{\pi}\int\limits_0^\infty\deint{}{q} q^2\;\left[
    \calK^{(l)}(E;q,p)\;\begin{pmatrix}1&-3\\-3&1\end{pmatrix}
    +\delta^{l0}\;\calH(E;\Lambda)\;\begin{pmatrix}1&-1\\-1&1\end{pmatrix}
  \right]\;
  \non\\&&
  \hq\hq\hq\hq\hq\hq\hq\hq\hq\hq
  \times\;\calD(E-\frac{q^2}{2M},q)\;\vec{t}^{(l)}_d(E;k,q)
  \non
\end{eqnarray}
The three-body force $\calH$ will be discussed in the next Sub-section. The
vector
\begin{equation}
  \label{eq:tvector}
  \vec{t}^{(l)}_d := {t^{(l)}_{d,tt}\choose t^{(l)}_{d,ts}}
\end{equation}
is built out of the two amplitudes which get mixed: $t^{(l)}_{d,tt}$ for the
$Nd_t\to Nd_t$-process, and $t^{(l)}_{d,ts}$ for the $Nd_t\to Nd_s$-process.
Furthermore,
\begin{equation}
  \label{eq:totaltwoparticlepropagator}
  \calD(p_0,p):=\begin{pmatrix}D_t(p_0,p)&0\\0&D_s(p_0,p)\end{pmatrix}
\end{equation}
is the propagator of the two intermediate auxiliary fields.

\absatz How to calculate higher-order corrections? The re-summation of all
interactions proportional to $y$ and $\Delta_{s/t}$ into a Faddeev equation is
mandatory since the power-counting classifies all diagrams which are built
only out of them as contributing equally strong to the final amplitude.
Ref.~\cite{4stooges} proposed to calculate the higher-order corrections by
expanding the kernel and inhomogeneous part of the integral equation in powers
of $Q$ to the desired order of accuracy, and iterate then by inserting it into
a Faddeev equation (\emph{partially re-summed \NXLO{n}-calculation}). This
re-summation of some higher-order effects does not increase the accuracy of
the calculation, which is still set by the accuracy to which the kernel is
expanded. It does however simplify the numerical treatment, as no divergences
in the amplitude or in three-body forces $\calH$ are encountered as the
numeric cut-off is removed. It also leads to a simple, analytical argument at
which orders three-body forces with derivatives enter, as discussed below. In
contradistinction, a \emph{strict perturbation} around the LO solution soon
becomes cumbersome numerically, as full off-shell amplitudes need to be
computed and the numerical integrals soon start to diverge, making a numerical
renormalisation necessary~\cite{doubletNLO}.

The computational effort to solve the integral equations
(\ref{eq:quartetpw}/\ref{eq:doubletpw}) numerically is trivial, as all
potentials are separable. A simple Mathematica-code can be down-loaded from
\texttt{http://www.physik.tu-muenchen.de/\~{}hgrie}. Here, a step-function
cut-off $\Lambda$ in momentum space was chosen. The cut-off dependence of the
results is discussed below.  To reduce the numerical instabilities from the
poles of the two-nucleon amplitudes and logarithmic singularities of the
projected nucleon propagator in the kernel, the integral equations are --
following Hetherington and Schick~\cite{HetheringtonSchick} -- first solved on
a contour in the complex plane. The amplitudes on the real axis are then
re-constructed by another use of the equations. A grid of $70$ points does
more than suffice for numerical stability. For example, the imaginary parts of
$k\cot\delta$ vanish below the deuteron-dis-integration threshold easily to 1
part in $10^{8}$.

Finally, the scattering phase-shift of the $l$th partial wave in the quartet
and doublet channel is related to the renormalised on-shell amplitudes by
\begin{equation}
  \label{eq:phaseshifts}
  T_q^{(l)}=\calZ_t\;t_q^{(l)}=
  \frac{3\pi}{M}\;\frac{1}{k \cot\delta_q^{(l)}-\ii k}\;\;,\;\;
  T_{d,xy}^{(l)}= 
  \frac{3\pi}{M}\;\frac{1}{k \cot\delta_{d,xy}^{(l)}-\ii k}\;\;.
\end{equation}
where $x,y=s,t$ label the matrix entries in cluster-configuration space, and
\begin{equation}
  \vec{T}_{d}^{(l)}=\calZ \vec{t}_{d}^{(l)} \;\;\mbox{ with }
  \;\;\calZ:=\begin{pmatrix}\calZ_t&0\\0&\sqrt{\calZ_t\calZ_s}\end{pmatrix}
\end{equation}
is the renormalised doublet-amplitude and its wave-function renormalisation.
In the doublet channel, the only observable process is nucleon-deuteron
scattering, $Nd_t\to Nd_t$, i.e.~$x=y=t$.  Results are not discussed for the
un-physical processes $Nd_t\leftrightarrow Nd_s$ and $Nd_s\to Nd_s$.

The elastic differential cross-section from $Nd$-scattering in the cm-frame is
finally~\cite{GoldbergerWatson}
\begin{equation}
  \label{eq:diffcrosssection}
  \frac{\dd\sigma}{\dd\Omega}=\frac{1}{3}\left[
    \left|\sum\limits_{l=0}^{\infty} (2l+1)
      \;\frac{P_l(\cos\theta)}{k\cot\delta_{d,tt}^{(l)} -\ii k}\right|^2+
    2 \left|\sum\limits_{l=0}^{\infty} (2l+1)
      \;\frac{P_l(\cos\theta)}{k\cot\delta_{q}^{(l)} -\ii k}\right|^2
    \right]\;\;.
\end{equation}  
A maximum angular momentum $l_\mathrm{max}=4$ serves to converge the
differential cross-section.

\absatz The recipe to compute $Nd$-scattering up to \NXLO{2} is hence as
follows: Expand the auxiliary-field propagators and their wave-function
renormalisations in the integral equations
(\ref{eq:quartetpw}/\ref{eq:doubletpw}) to the desired level of accuracy in
Z-parameterisation as in (\ref{eq:dpropexpanded}/\ref{eq:wavefuren}), and then
solve the Faddeev equation. To \NXLO{2}, the projected nucleon propagator
$\calK^{(l)}$ is unchanged.

Nucleon-deuteron scattering is to \NXLO{2} thus completely determined by four
simple observables of $NN$-scattering: the deuteron binding energy and residue
(or effective range), and the scattering length and effective range of the
\oneS-channel. Only the \twoS-channel has further unknowns, namely the
strength of the three-body interaction $H_0$ at LO and NLO, and in addition of
$H_2$ at \NXLO{2}, as discussed momentarily. They are determined by its
measured scattering length $a_d$~\cite{doublet_sca} and the triton binding
energy $B_d$, respectively:
\begin{equation}
  \label{eq:threebodyexpvalues}
  a_d=(0.65\pm0.04)\;\fm\;\;,\;\;B_d=8.48\;\MeV
\end{equation}
But why are two and only two three-body force needed at \NXLO{2}?

%%%%%%%%%%%%%%%%%%%%%%%%%%%%%%
\subsection{Three-body Force and Lagrangean}
\label{sec:threebodyforce}

We now turn to the three-body force $\calH$ already included in the
doublet-channel Faddeev equation (\ref{eq:doubletpw}). The Pauli principle in
the quartet channel and the centrifugal barrier in all partial waves $l>0$
forbids three-body forces in nearly all partial waves at \NXLO{2}; see also
\cite{suppressed3bfs}. However, neither rules out a three-body force without
derivatives in the \twoS- wave, the physically most interesting channel, which
contains the triton and ${}^3$He as real bound states. While one would
na\"ively have guessed that such an energy-independent three-body force scales
as $Q^0$ and hence enters only at \NXLO{2}, an unusual renormalisation of the
three-nucleon system in the triton channel mandates its inclusion into the
Faddeev equation as a LO term~\cite{3stooges_boson,3stooges_doublet}: As was
first pointed out by Minlos and Faddeev~\cite{faddeev}, this channel suffers
from a peculiar cut-off sensitivity of the on-shell amplitudes if three-body
forces are absent. The kernel of (\ref{eq:doubletpw}) is for $l=0$ at short
distances identical to the one of an attractive $1/r^2$-potential between the
auxiliary fields and the nucleon~\cite{seattle_review}. Without a three-body
force, the wave function would hence collapse, and all observables would
become sensitive to Physics at very short distances, a phenomenon well-known
as Thomas effect~\cite{thomas}. Its mathematical origin lies in the fact that
in the absence of a three-body force, the solution to the integral equation is
not unique because it allows for a zero mode, as the kernel of the
doublet-$\mathrm{S}$ wave Faddeev equation is not compact~\cite{danilov}.

\emph{In praxi}, the integral equation is solved numerically by imposing a
cut-off $\Lambda$, which should not be confused with the breakdown-scale
$\LambdaNoPion$ of \EFTNoPion. In that case, a unique solution exists in the
\twoS-channel for each $\Lambda$ and $\calH=0$, but no unique limit as
$\Lambda\to\infty$.  The $Nd$-scattering length, for example, can be tuned as
a function of $\Lambda$ to have any value between minus infinity and plus
infinity.  As long-distance phenomena must however be insensitive to details
of the short-distance Physics (and in particular of the regulator chosen),
Bedaque et al.~\cite{3stooges_boson,3stooges_doublet,4stooges} showed that the
system must be stabilised by a three-body force
\begin{equation}
  \label{eq:calH}
   \calH(E;\Lambda)=
   \frac{2}{\Lambda^2}\sum\limits_{n=0}^\infty\;H_{2n}(\Lambda)\;
   \left(\frac{ME+\gamma_t^2}{\Lambda^2}\right)^n
   =\frac{2H_0(\Lambda)}{\Lambda^2}+
   \frac{2H_2(\Lambda)}{\Lambda^4}\;(ME+\gamma_t^2)+\dots
\end{equation}  
which absorbs all dependence on the cut-off as $\Lambda\to\infty$. It is
analytical in $E$ and can be obtained from a three-body Lagrangean, employing
a three-nucleon auxiliary field analogous to the treatment of the two-nucleon
channels, see~\cite{4stooges} and the end of this Sub-Section.

$H_{2n}$ is dimension-less but depends on the cut-off $\Lambda$ in a
non-trivial way, as a renormali\-sation-group analysis reveals: Instead of
approaching a fixed-point as $\Lambda\to\infty$, it shows an oscillatory
behaviour known as ``limit
cycle''~\cite{3stooges_boson,wilson,Braaten:2003eu}.

As one needs a three-body force at LO, $H_0\sim Q^{-2}$, to prevent the system
from collapse, all three-body forces obtained by expanding $\calH$ in powers
of $E$ are also enhanced, with the interactions proportional to $H_{2n}$
entering at \NXLO{2n}~\cite{4stooges}. Since a numerical verification of this
analytical observation is one of the prime advantages Z-parameterisation has,
the argument is recalled here: Bedaque et al.~\cite{3stooges_doublet} noted
that a re-parameterisation of the amplitudes relates the problem for the LO
three-body force to a particularly interesting symmetry, and this was extended
by Ref.~\cite{4stooges}. Building the linear combination
\begin{equation}
  \label{eq:lincombforWigner}
  \vec{t}^{(0)}_\mathrm{Wigner}:={t_{\mathrm{Wigner},-}^{(0)}
    \choose t_{\mathrm{Wigner},+}^{(0)}}
  =\half\begin{pmatrix}1&-1\\1&1\end{pmatrix}
  \;{t^{(0)}_{d,tt}\choose t^{(0)}_{d,ts}}\;\;,
\end{equation}
the Faddeev equation becomes in the \twoS-channel
\begin{eqnarray}
  \label{eq:faddeevWigner}
  \vec{t}^{(0)}_\mathrm{Wigner}(E;k,p)&=&4\pi\;
  \left[\calK^{(0)}(E;k,p)\;{1\choose -\half}
    +\calH(E;\Lambda)\;{\half\choose 0}\right]\\
  &&-\frac{2}{\pi}\int\limits_0^\infty\deint{}{q} q^2\;
    \begin{pmatrix}2\calK^{(0)}(E;q,p)+\calH(E;\Lambda)&0\\
      0&-\calK^{(0)}(E;q,p)\end{pmatrix} \non\\&&
    \hq\hq\hq\hq\hq\hq\hq\hq\hq\hq \times\;
    \begin{pmatrix}\Sigma(E-\frac{q^2}{2M},q)&\Delta(E-\frac{q^2}{2M},q)\\
      \Delta(E-\frac{q^2}{2M},q)&\Sigma(E-\frac{q^2}{2M},q)\end{pmatrix}
    \;\vec{t}^{(0)}_\mathrm{Wigner}(E;k,q)\;\;.  \non
\end{eqnarray}
While $\Sigma:=\half(D_t+D_s)=\half\tr\calD$ is the ``average''
$NN$-$\mathrm{S}$-wave scattering amplitude,
$\Delta:=\half(D_t-D_s)=\half\mathrm{str}\calD$ parameterises the degree to
which the \threeS- and \oneS-channel differ. In the Wigner-$SU(4)$-limit of
arbitrary combined spin- and iso-spin rotation~\cite{su4}, the two amplitudes
are identical, $\gamma_t=\gamma_s,\;\rho_{nt}=\rho_{ns}\;\forall n$. In that
case, $\Delta=0$, and the two equations for $t^{(0)}_{\mathrm{Wigner},\mp}$
decouple, as first observed in~\cite{3stooges_doublet}.  The essential
observation is now that for the UV-behaviour of the amplitude, in which all
scales are discarded except for the off-shell momentum $q\sim\Lambda$ in the
loop, the Wigner-$SU(4)$-limit is recovered automatically.
$t_{\mathrm{Wigner},+}^{(0)}$ obeys the same integral equation as the
quartet-$\mathrm{S}$-wave in (\ref{eq:quartetpw}), where the Pauli principle
forbids three-body forces without derivatives. Only
$t_{\mathrm{Wigner},-}^{(0)}$ is subject to the Wigner-$SU(4)$ symmetric
three-body force $\calH$. Its integral equation is the same as for three
spin-less bosons, whose wave-function is well-known to collapse in the absence
of three-body forces~\cite{3stooges_boson,faddeev,danilov}. $H_0$ therefore
must be LO to ensure that the physical on-shell amplitude is cut-off
independent.

To determine the running of $H_2$, one expands the quantities of
(\ref{eq:faddeevWigner}) for $q\gg k,p,\gamma,\dots$:
\begin{equation}
  \label{eq:expandWigner}
  \begin{array}{rclll}
    &&\mathrm{LO}&\mathrm{NLO}&\text{\NXLO{2}}\\[1ex]
  \Sigma(E-\frac{q^2}{2M},q)
  &\to&\dis-\sqrt{\frac{4}{3}}\;\frac{1}{q}
  &\dis-\frac{4\gamma}{3q^2}-\frac{Z-1}{2\gamma}
  &\dis-\frac{4ME+8\gamma^2}{3\sqrt{3}\,q^3}-
                \sqrt{\frac{4}{3}}\frac{Z-1}{q}-
                \frac{\sqrt{3}(Z-1)^2}{8\gamma^2}\,q+\dots\\[3ex]
  \calK^{(0)}(E;q,p)&\to&\dis\frac{1}{q^2}&
  &\dis+\frac{ME-\frac{2q^2}{3}}{q^4}+
  \dots \\[3ex]
  \Folgt \calH(E;\Lambda)&\to&\dis\frac{2H_0^\mathrm{LO}(\Lambda)}{\Lambda^2}
  &\dis+\frac{2H_0^\mathrm{NLO}(\Lambda)}{\Lambda^2}
  &\dis+\frac{2H_0^\text{\NXLO{2}}(\Lambda)}{\Lambda^2}+
  \frac{ME+\gamma^2}{\Lambda^2}\;\frac{2H_2(\Lambda)}{\Lambda^2}+
  \dots
  \end{array}
\end{equation}
At LO, $H_0$ is independent of two-body observables and is fixed by a
three-body datum. At NLO, only $\Sigma$ has non-vanishing contributions.  They
depend only on the low-energy two-body observables $\gamma\sim Q$ and
$\rho_0\sim Q^0$, but not on the total cm-energy $E\sim Q^2$ -- or
equivalently, not on the on-shell momentum $k\sim Q$.  The cut-off dependence
they induce can hence be absorbed by $H_0^\mathrm{NLO}(\Lambda)$, a
momentum-independent correction to the LO three-body force whose value now
depends also on the two-body observables $\gamma$ and $\rho_0$.

At \NXLO{2}, more corrections arise which depend on $\gamma$ and $\rho_0$. In
addition, both the projected nucleon propagator $\calK$ and the
$NN$-scattering amplitude $\Sigma$ show corrections proportional to $ME$. To
match the behaviour of the off-shell (UV-)amplitude on the physical on-shell
momentum $E$, the inclusion of an \emph{energy-dependent} three-body force
$H_2$ is thus mandatory. Its value is not determined by two-body observables.
The expansion in powers of $E\sim Q^2$ proceeds straightforwardly: Another
three-body force $H_{2n}(\Lambda)$ proportional to $E^n$ is needed every even
order in $Q$ from expanding both $\Sigma$ and $\calK^{(0)}$. $H_{2n}(\Lambda)$
first enters at \NXLO{2n}, independent of two-body observables. To determine
its strength, one additional three-body datum is needed. The power-counting
for the three-body forces is hence
\begin{equation}
  \label{eq:calHpowercounting}
  H_0\sim Q^{-2}\;\;,\;\;H_2\sim Q^{-2}\;\;,\;\;H_{2n}\sim Q^{-2}\;\;.
\end{equation}

Corrections to the limit $q\to\infty$ from the breaking of
Wigner-$SU(4)$-symmetry can only induce a momentum-dependent three-body force,
because the only three-body force without derivatives is necessarily
Wigner-$SU(4)$-symmetric~\cite{3stooges_doublet}. They can be shown to be of
higher order, see~\cite{4stooges} for a more detailed discussion.

%%%%%%%%%%%%%%%%%%%%%%%%%%%%%%%%%%
\absatz The chain of isotropic ($\mathrm{S}$-wave) Wigner-$SU(4)$-symmetric
three-body forces is also found from a three-nucleon auxiliary-field
Lagrangean, in complete analogy to the treatment of the two-nucleon auxiliary
fields in
Sects.~\ref{sec:auxiliaryfield}/\ref{sec:oneS}~\cite{3stooges_doublet,4stooges}:
\begin{eqnarray}
  \label{eq:threeNlagrangean}
  \calL_{3N}&=&
  -\frac{y_3(\Lambda)}{\sqrt{3}}\;\left[t^\dagger\left((\sigma^iN)d_t^i-
      (\tau^AN)d_s^A\right)+\mathrm{H.c.}\right]\\
  &&
  +t^\dagger\left[\Omega-\sum\limits_{n=1}^\infty h_{2n}(\Lambda)
  \left(\ii\de_0+\frac{\dev^2}{6M}+\frac{\gamma_t^2}{M}\right)^{n}\right]t
  \;\;,\non
\end{eqnarray}
where the auxiliary field $t$ has the quantum numbers of the triton/${}^3$He,
i.e.~spin and iso-spin $\half$. The relative coupling strength between the
processes $t\to Nd_t$ and $t\to Nd_s$ is fixed because there is only one
three-body force without derivatives, which also happens to be
Wigner-$SU(4)$-symmetric~\cite{3stooges_doublet,4stooges}. This field is
treated analogously to the two-nucleon auxiliary fields: At LO, the propagator
of a ``triton'' with kinetic energy $E$ and momentum $\pv$ is proportional to
$\frac{1}{\Omega}$, with insertions proportional to $h_{2n}(\Lambda)$
suppressed by $n$ powers of the triton kinetic energy:
\begin{equation}
  \label{eq:tritonpropagator}
%   \feyngraph{40}{10}{0}{0}{0}{0}{ \fmfleft{i}
%    \fmfright{o}
%    \fmf{triple}{i,o}}\hq\hq=
  \frac{\ii}{\Omega}\left[1+\frac{h_2(\Lambda)}{M\Omega}
    \left(ME-\frac{\pv^2}{6}+\gamma_t^2\right)+\dots\right]
%  \to\frac{\ii M}{M\Omega-\sum\limits_{n=1}^\infty
%    \dis\frac{h_{2n}(\Lambda)}{M^{n-1}}(ME-\frac{\pv^2}{6}+\gamma_t^2)^n}
\end{equation}
Comparing the final result of the resultant three-body force in the \twoS-wave
(\ref{eq:calHinteraction}) with (\ref{eq:doubletpw}), one finds that the
couplings $y_3(\Lambda)$ and $h_{2n}(\Lambda)$ are in the cm-frame related to
$\calH$ by
\begin{equation}
  \label{eq:calHfrom3Blagr}
   \calH(E;\Lambda)= -\frac{2y_3^2(\Lambda)}{y^2\;M\Omega}\left[
     1+\;\sum\limits_{m,n=1}^\infty\left[
     \frac{h_{2n}(\Lambda)}{\Omega\;M^{n}}\;(ME+\gamma_t^2)^{n}\right]^m\right]
\end{equation}  
or matching order by order the dimension-less three-body strengthes
$H_{2n}(\Lambda)$ of (\ref{eq:calH}),
\begin{equation}
  H_0(\Lambda)=-\frac{y_3^2(\Lambda)}{y^2}\;\frac{\Lambda^2}{M\Omega}\;\;,\;\;
  H_2(\Lambda)=-\frac{y_3^2(\Lambda)}{y^2}\;\frac{\Lambda^4}{(M\Omega)^2}
  \;h_2(\Lambda)\;\;,\;\dots,
\end{equation}
so that the couplings are particularly simple when one chooses
\begin{equation}
  \label{eq:threebodyparameters}
  \Omega=\frac{\Lambda^2}{M}\;\;,\;\;y_3^2(\Lambda)=-y^2 \;H_0(\Lambda)
  \;\;,\;\;h_2(\Lambda)=\frac{H_2(\Lambda)}{H_0(\Lambda)}\;\;,
\end{equation}
as they follow then from (\ref{eq:calHpowercounting}) the natural scaling laws
-- with the exception of $y_3(\Lambda)$:
\begin{equation}
  y_3^2(\Lambda)\sim Q^{-2}\;\;,\;\;\Omega\sim Q^0\;\;,
  \;\;h_{2n}(\Lambda)\sim Q^0
\end{equation}
Notice that in contradistinction to $D_{s/t}$ in
Sect.~\ref{sec:auxiliaryfield}, the power counting does not require even a
partial re-summation of the triton propagator.

%%%%%%%%%%%%%%%%%%%%%%%%%%%%%%%%%%
\subsection{The \twoS- (Triton-) Channel}
\label{sec:tritonresults}

That Z-parameterisation improves convergence is most importantly seen in the
only channel in which a bound state is found, and in which the three-body
force enters already at LO to stabilise the system from collapse. In
particular, it can be instrumentalised to confirm that the first
momentum-dependent three-body force enters at \NXLO{2}, as the analytical
argument of the preceding section demands.

Previously, Hammer and Mehen calculated the \twoS-wave phase shifts below the
deuteron break-up to NLO by sandwiching the ERE-corrections to the
two-particle propagators between the LO half off-shell
amplitudes~\cite{doubletNLO}.  ERE-parameterisation was also used in the
recent article which proposed partially re-summed \NXLO{n}-calculations to
simplify numerics and showed that the momentum-dependent three-body force
$H_2$ is conceptually necessary at \NXLO{2}~\cite{4stooges}. An estimate of
NLO-effects in Z-parameterisation by Afnan and Phillips~\cite{Afnan:2003bs}
included the part of the two-particle propagator
(\ref{eq:dpropexpanded}/\ref{eq:wavefuren}) which leads to the correct
residues $Z_{s/t}$, but not the part which changes the off-shell propagation
of the deuteron field, iterating again the perturbatively expanded
kernel. %Here, the complete NLO and \NXLO{2}-result in Z-parameterisation is
%given.

\begin{figure}[!htb]
  \hspace*{-4ex}
  \begin{minipage}{0.52\linewidth}
    \begin{flushright}
      \includegraphics*[height=0.59\textwidth]
      {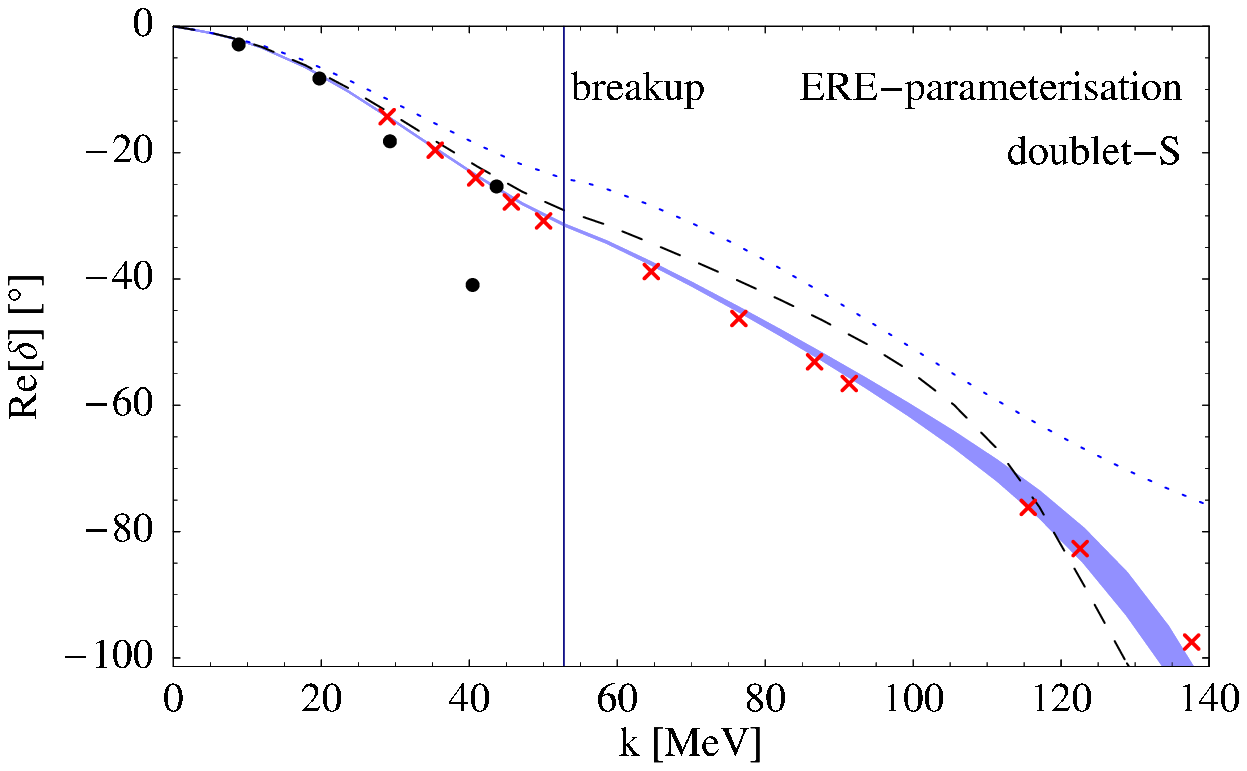}\\[2ex]
      \includegraphics*[height=0.58\textwidth]
      {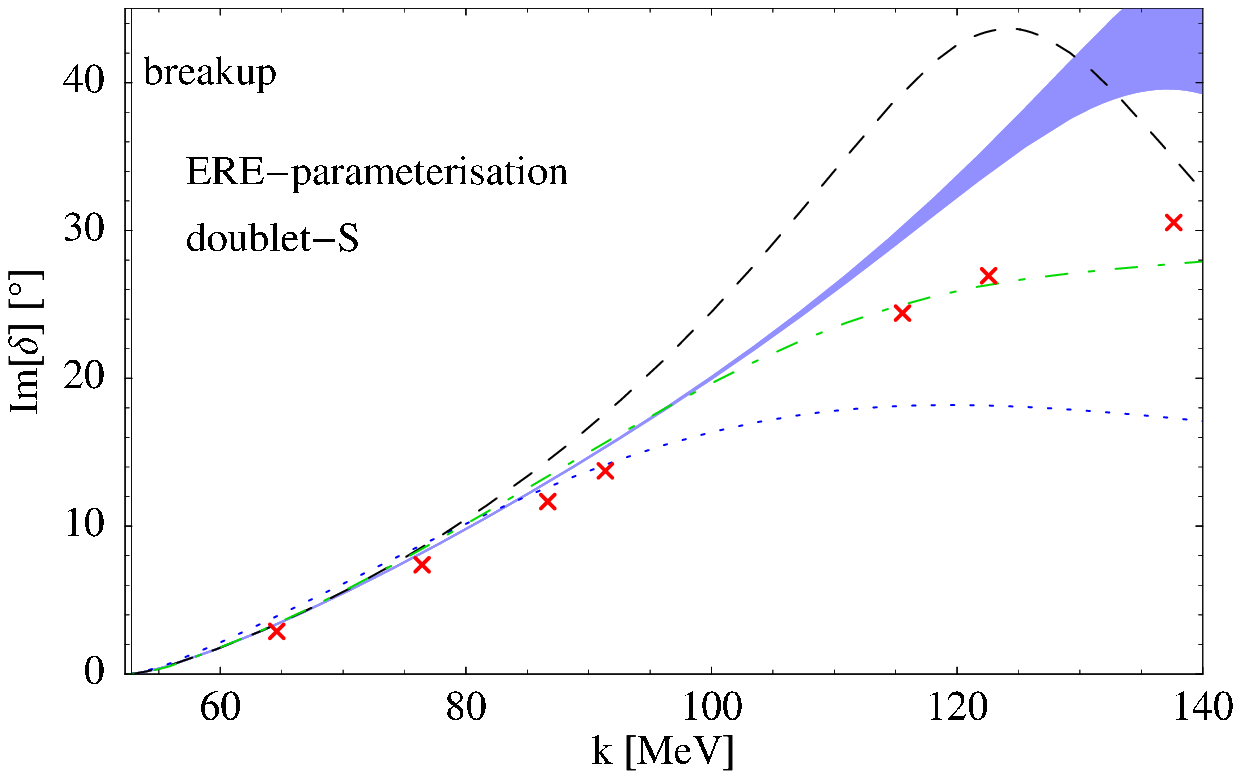}
    \end{flushright}
  \end{minipage}
  \begin{minipage}{0.52\linewidth}
    \begin{flushright}
      \includegraphics*[height=0.59\textwidth]
      {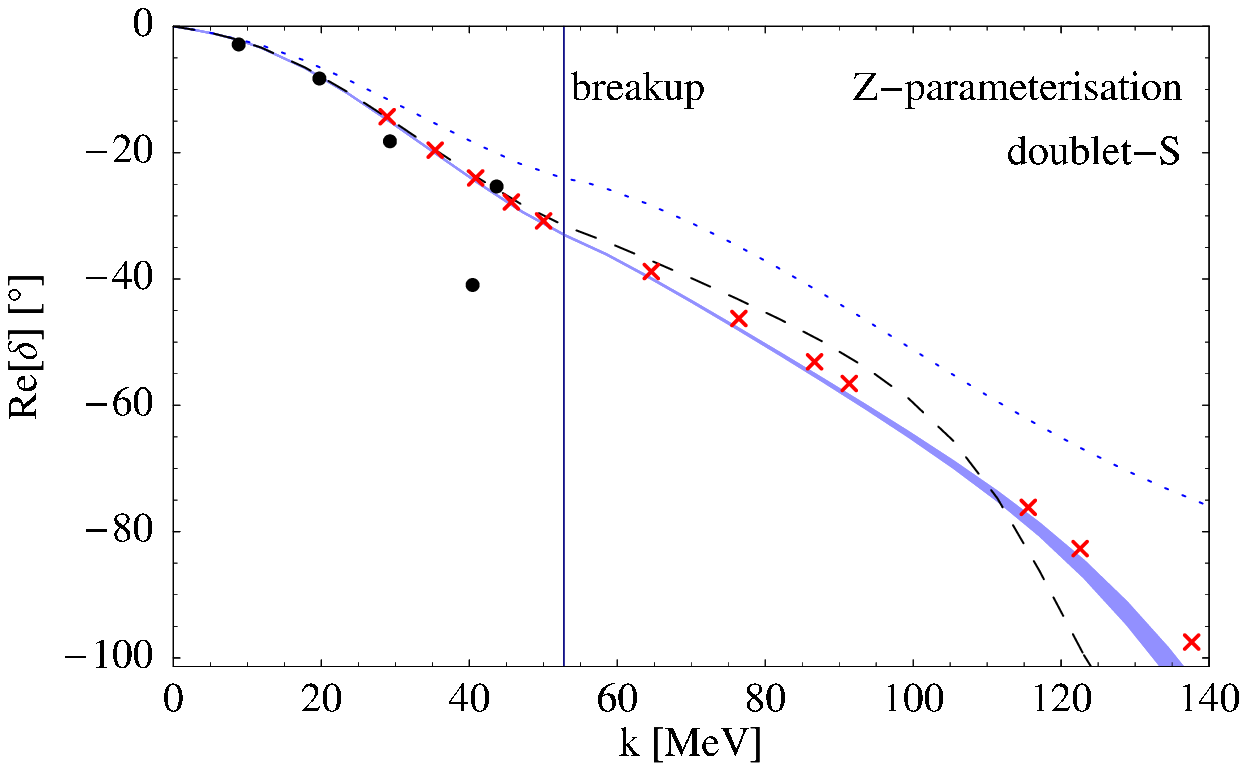}\\[2ex]
      \includegraphics*[height=0.58\textwidth]
      {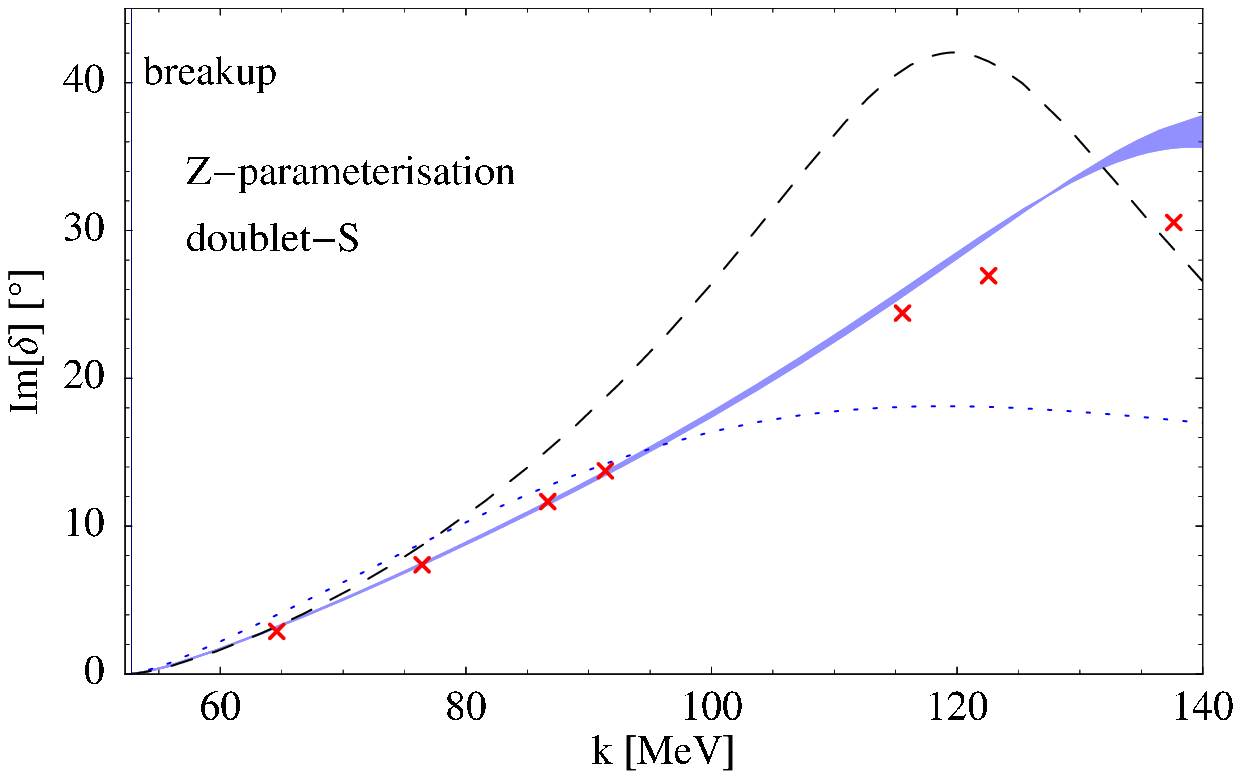}
    \end{flushright}
  \end{minipage}
\begin{center}
\caption{Comparison of the real (top) and imaginary (bottom) parts of the
  neutron-deuteron \twoS-phase-shift at LO (dotted), NLO (dashed) and \NXLO{2}
  (dark band) in ERE-parameterisation (left) and Z-parameterisation (right) as
  function of the cm-momentum $k$. The band in the \NXLO{2}-curve shows the
  variation induced by shifting the cut-off from $\Lambda=200\;\MeV$ to
  $\infty$. Dots: phase-shift analysis from 1967~\cite{doublet_PSA}; crosses:
  Argonne V18 results including the Urbana IX three-body force
  from~\cite{kievsky1996} below, and from~\cite{kievskyprivcommun} above
  break-up. At LO and NLO, the results for $\Lambda=900\;\MeV$ are shown. The
  dot-dashed curve in the imaginary parts of the ERE-result is a partial
  \NXLO{3}-calculation as described in the text.}
\label{fig:doublets}
\end{center}
\end{figure}

In the following, one has to differentiate between two kinds of convergence:
First, the results should of course agree with available measurements, at the
level of accuracy predicted by EFT.  Second, it is vital for a reliable
error-estimate and hence for predicting the accuracy of the calculation that
EFT can demonstrate that contributions which are classified as higher-order
are indeed suppressed, i.e.~that corrections from order to order become
smaller in the range of validity. In particular whenever no or scarce data are
available, this \emph{a-priori} error-estimate is essential to minimise the
theoretical bias in the description of few-body properties.
Z-parameterisation improves both variants over ERE-parameterisation.

The ERE- and Z-parameterisation results for the phase-shifts of the
\twoS-channel at LO, NLO and \NXLO{2} are shown in Fig.~\ref{fig:doublets},
together with the only available phase-shift analysis~\cite{doublet_PSA},
dating from 1967, and results using the Argonne V18 potential supplemented
with the Urbana IX three-body force~\cite{kievsky1996,kievskyprivcommun}. The
bands in the \NXLO{2}-results come from varying the cut-off in the Faddeev
equation between $200$ and $900\;\MeV$, when $\Lambda$-independence is
achieved within drawing-accuracy, except at those points at which the
strengthes of the three-body forces diverge, see later
Fig.~\ref{fig:H0H2variation}.

At this order, considerably more sophisticated
and involved potential-model calculations must agree with the EFT-predictions,
if all are fitted to the same low-energy observables. For lack of a direct
phase-shift analysis of the three-nucleon system, convergence to experiment is
discussed by comparison to such a potential-model calculation which reproduces
the same two-nucleon observables, triton binding energy and scattering length
in the \twoS-wave. At higher orders, EFT-calculations may be more accurate
than those of potential models, because the EFT-result will depend on new
three-body low-energy coefficients which do \emph{not} enter in potential
models, while a systematic treatment of three-body forces is built into
\EFTNoPion.

At LO, both parameterisations are identical. As expected from the discussion
in Sect.~\ref{sec:zparameterisation}, the NLO corrections are larger for
Z-parameterisation than for the ERE-version due to the inclusion of the full
strength of the deuteron residue. Higher-order corrections in
Z-parameterisation should be considerably smaller than in
ERE-parameterisation, because their typical scale from $NN$-scattering is in
the former set by $(Z_t-1)/2\approx 0.3$, while in the latter, it is
$\gamma_t\rho_{0t}\approx0.4$. This amounts to a difference in the predicted
accuracy of the \NXLO{2}-calculation of $(0.3)^3\approx 3\%$ in
Z-parameterisation versus $(0.4)^3\approx 6.5\%$ in ERE-parameterisation for
typical momenta in the three-body problem. Indeed, the correction from NLO to
\NXLO{2} is in general larger for the ERE-parameterisation than for the
Z-parameterisation, and the result closer to the potential-model values. The
plot of the logarithmic deviation of the on-shell point of the inverse
$K$-matrix, $k\cot\delta$, from the potential-model numbers at each order in
both parameterisations verifies these findings on the quantitative level, see
left panel in Fig.~\ref{fig:lepage}.

In particular, the ERE-result for the imaginary part of the phase-shift has at
large momenta not converged to the potential-model results at \NXLO{2}, while
Z-parameterisation agrees with these computations. As the effective-range
effects are expected to be the dominant corrections in ERE-parameterisation,
the imaginary part of the ERE-result in Fig.~\ref{fig:doublets} contains also
a computation in which these are included in the deuteron propagator
(\ref{eq:dprop}) to order $(\gamma_t\rho_{0t})^3$. This is only a partial
\NXLO{3}-calculation, neglecting shape-parameter, relativistic,
$\mathrm{SD}$-mixing and other corrections, but improves the agreement to the
level achieved already at \NXLO{2} in Z-parameterisation.  Close to
$k=\mpi\approx\LambdaNoPion$, both parameterisations become un-reliable as the
NLO and \NXLO{2}-corrections are comparable in size.

\begin{figure}[!htbp]
\begin{center}
  \includegraphics*[height=0.28\textwidth]
  {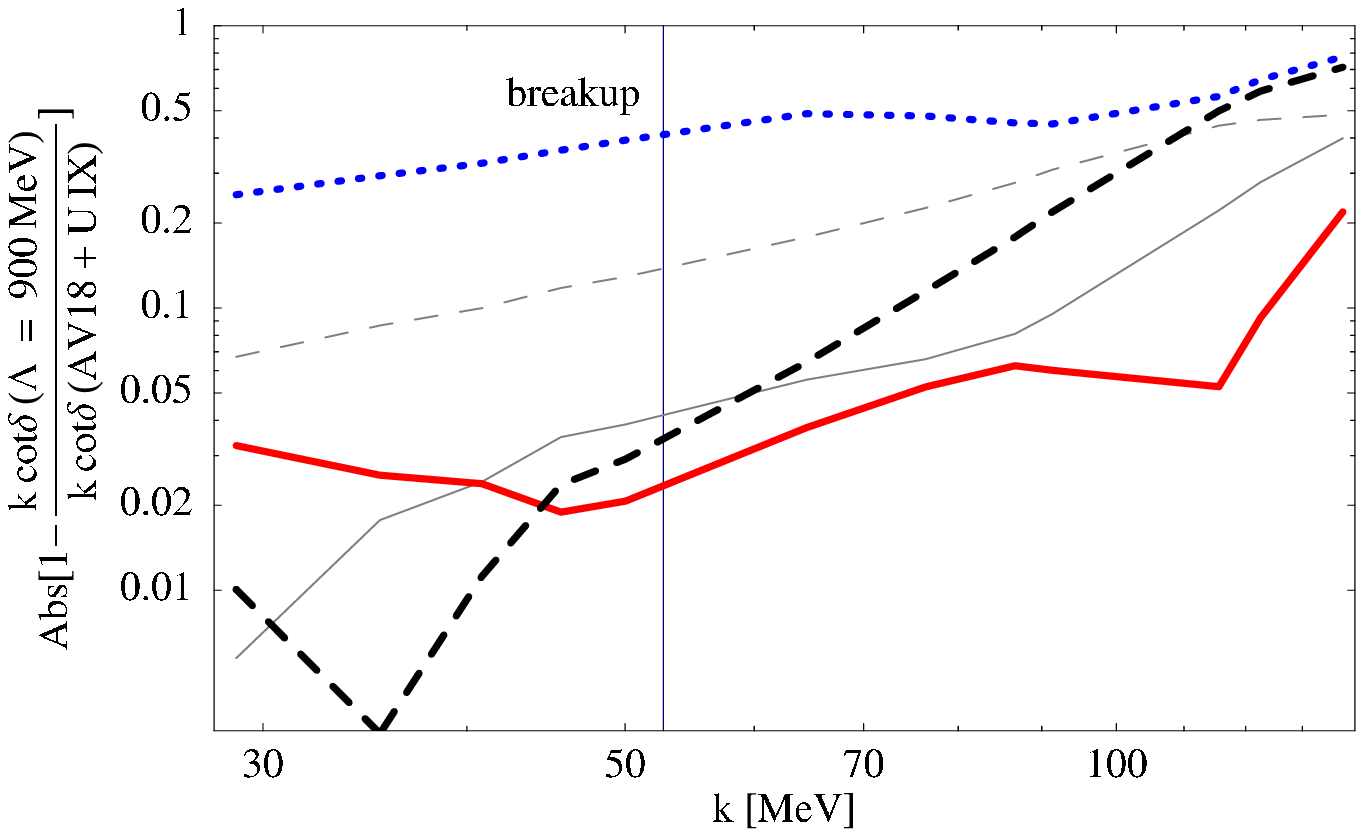}\hq\hq\hq\hq
  \includegraphics*[height=0.28\textwidth] {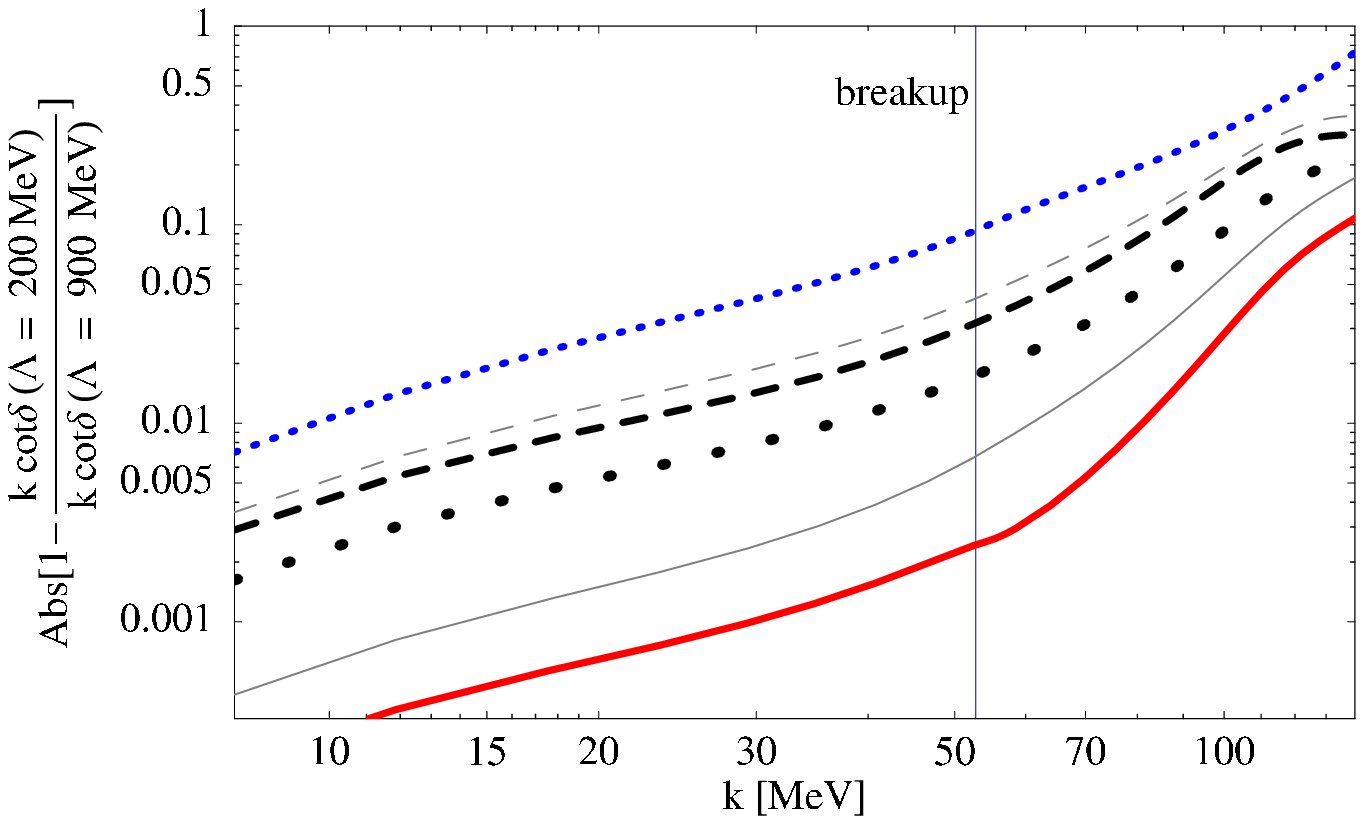}
\caption{Convergence of $k\cot\delta$ as function of the cm
  momentum. Left: deviation from the result of the potential-model
  calculation~\cite{kievsky1996,kievskyprivcommun}. Right: dependence on
  cut-off variations between $\Lambda=200$ and $900\;\MeV$. Dotted: LO;
  dashed: NLO; solid: \NXLO{2}; thin lines: ERE-parameterisation; thick lines:
  Z-parameterisation. Thick dots on the right: ``\NXLO{2}''-calculation in
  Z-parameterisation with $H_2$ set to zero.}
\label{fig:lepage}
\end{center}
\end{figure}

\absatz At least as important is that the internal convergence is drastically
improved in Z-parameter\-isation. This is evident from the right side of
Fig.~\ref{fig:lepage} which compares the uncertainties induced into the
\NXLO{2} phase-shifts by cut-off variations. As outlined in the Introduction,
the main strength of the EFT-approach is that it allows for an \emph{a-priori}
estimate of the theoretical uncertainties of a calculation because
contributions to the amplitudes are ordered by a small expansion parameter
$Q$.  \NXLO{n} corrections to $k\cot\delta$ should typically be of the order
\begin{equation}
  \label{eq:powercounting}
  \Delta(k\cot\delta)\sim Q^{n}=
  \left(\frac{p_\mathrm{typ}}{\LambdaNoPion}\right)^{n}
\end{equation}
compared to the LO result.  Typical low-momentum scales $p_\mathrm{typ}$ in
the three-body system are the binding momenta of the two-nucleon real and
virtual bound states, $\gamma_s\approx -8.0\;\MeV,\;\gamma_t\approx 45\;\MeV$
and the scattering momentum $k$. In addition, the three-body forces are
determined in part by the typical three-nucleon bound-state momentum
$\gamma_d\sim\sqrt{MB_d}\approx 90\;\MeV$, $B_d$ the triton binding energy.
The breakdown scale $\LambdaNoPion\approx\mpi$ of the theory is the scale at
which higher-order corrections become comparable in size. Like the actual size
of the expansion parameter $Q$, its value must be verified in actual
calculations.

As observables at low energies must be independent of details of
short-distance physics, they must be independent of the arbitrary regulator
$\Lambda$ up to the order of the expansion. In other words, the physical
scattering amplitude (or more accurately its physically relevant part,
$k\cot\delta$) must be dominated by integrations over off-shell momenta $q$ in
the integral equations (\ref{eq:quartetpw}/\ref{eq:doubletpw}) in the region
in which the EFT is applicable, $q\lesssim\LambdaNoPion$. As argued e.g.~by
Lepage~\cite{Lepage:1997cs}, one can therefore estimate sensitivity to
short-distance Physics, and hence provide a reasonable error analysis, by
varying the cut-off $\Lambda$ between the break-down scale $\LambdaNoPion$ and
$\infty$. \emph{In praxi}, an upper limit $\Lambda_\mathrm{max}=900\;\MeV$
suffices since the phase-shifts are essentially cut-off independent beyond
$\Lambda\approx 600\;\MeV$.

On the right in Fig.~\ref{fig:lepage}, the logarithmic cut-off variation is
displayed as function of the cm momentum, both for ERE- and
Z-parameterisation. Several points are worth noticing.

First, the cut-off variation is substantially smaller in Z-parameterisation
over the whole range.  For example, at the typical three-body scale
$k\sim\sqrt{M B_d}\approx 90\;\MeV$, the cut-off variation is decreased from
$15\%$ and $4\%$ at NLO and \NXLO{2} in ERE-parameterisation to $12\%$ and
$1.3\%$ in Z-parameterisation~\footnote{The deviation at \NXLO{2} to the
  potential-model results is at the same scale improved from $9.1\%$ to
  $5.7\%$.}.

With $k$ the dominant low-energy scale for momenta above the breakup point
$(4\gamma_t^2/3)^{1/2}\approx 52\;\MeV$, one can secondly verify the estimate
(\ref{eq:powercounting}) for the corrections by fitting $n$ to the nearly
straight lines in the right panel of Fig.~\ref{fig:lepage} in the momentum
range between $70$ and $100$ to $130\;\MeV$. At lower momenta, the slope does
not change significantly from order to order because the dominant corrections
are $\sim(\gamma_{s,t}/\Lambda)^n$ and hence nearly independent of $k$.  The
power-law obtained for each order and parameterisation is listed in
Table~\ref{tab:powerlaw}. The slope increases in Z-parameterisation by one
unit from LO to NLO, and by two units from NLO to \NXLO{2}.  The latter may
stem from the partial inclusion of higher-order graphs in the partially
re-summed integral equation.  The ERE-parameterisation follows a weaker
power-law than Z-parameterisation, indicating again greater cut-off
sensitivity and less accuracy.

Thirdly, extrapolating the relative errors from the fit region, the cut-off
variations at NLO and \NXLO{2} become comparable at $k\sim150-200\;\MeV$,
matching the estimate $\LambdaNoPion\sim\mpi$.

\begin{table}[!htbp]
  \centering
  \begin{tabular}{|l||c|c|}
    \hline
   order \rule[-1.5ex]{0ex}{4ex}& $n$ (Z-param.) & $n$ (ERE-param.)\\
   \hline
   \hline
   LO\rule[-1.5ex]{0ex}{4ex}&\multicolumn{2}{c|}{$1.8-2.1$}\\
   \hline
   NLO\rule[-1.5ex]{0ex}{4ex}&$2.9$&$2.6$\\
   \hline
   \NXLO{2}\rule[-1.5ex]{0ex}{4ex}&$4.8$&$3.7$\\
   \hline
   \hline
   ``\NXLO{2}'', $H_2=0$\rule[-1.5ex]{0ex}{4ex}& $3.1$ &$2.8$\\
   \hline
  \end{tabular}
  \caption{Fit of the logarithmic cut-off variation 
    $\left|1-\frac{k\cot\delta(\Lambda=200\;\MeV)}
      {k\cot\delta(\Lambda=900\;\MeV)}\right|$ of $k\cot\delta$,
    shown in Fig.~\ref{fig:lepage}, for momenta higher than  $70$ and lower
    than $100$ to $130\;\MeV$ to a power-law $(k/\LambdaNoPion)^n$
    for ERE- and Z-parameterisation at LO, NLO and \NXLO{2}, and at \NXLO{2}
    without momentum-dependent three-body force $H_2$. Varying the upper limit
    changes only the numbers for LO.}
  \label{tab:powerlaw}
\end{table}

Finally, one confirms numerically the analytical finding in \cite{4stooges},
recalled in Sect.~\ref{sec:formalism}, that a momentum-dependent three-body
force $H_2$ enters at \NXLO{2} to increase accuracy. Setting $H_2=0$ does not
only substantially increase the cut-off dependence of the
\NXLO{2}-calculation, putting it close to the NLO dependence. As the slope at
large momenta is also practically unchanged compared to the NLO value, $H_2$
is necessary to improve cut-off independence.

\begin{figure}[!htb]
\begin{center}
  \includegraphics*[height=0.35\textwidth] {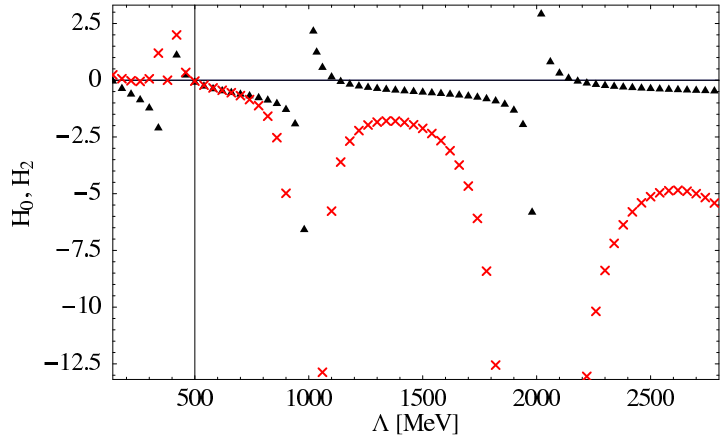}
\caption{Dependence of the dimension-less three-body couplings at \NXLO{2}
  on the sharp momentum cut-off $\Lambda\in[140;2800]\;\MeV$ in
  Z-parameterisation. Triangles: $H_0(\Lambda)$; crosses: $H_2(\Lambda)$.}
\label{fig:H0H2variation}
\end{center}
\end{figure}

Notice that observables are at \NXLO{2} converged to be practically cut-off
independent as low as $600\;\MeV$, while the dimension-less three-body forces
$H_0$ and $H_2$ vary dramatically even in a small window of cut-off
variations, see Fig~\ref{fig:H0H2variation}. Only when the absolute value of
$H_0$ or $H_2$ is unnaturally large ($\gtrsim 10$) is cut-off sensitivity felt.
In that case, more of the string of three-nucleon forces would be needed. Such
singular points (e.g.~at$\Lambda\sim1000\MeV$) are not considered here. The
actual numerical values of $H_0$ and $H_2$ are quite sensitive to the
regulator chosen, and to the precise values for $\gamma_{s/t}$ and $Z_{s/t}$
used. It is thus not the three-body forces which are large but the effect of
variations in the cut-off $\Lambda$ on observables. No simple connection
between $H_0$ and $H_2$ exists at \NXLO{2}, except that they both show a
limit-cycle with the same period, which depends on $\gamma_{s/t}$ and
$Z_{s/t}$.

\begin{figure}[!htb]
\begin{center}
  \includegraphics*[height=0.3\textwidth]
  {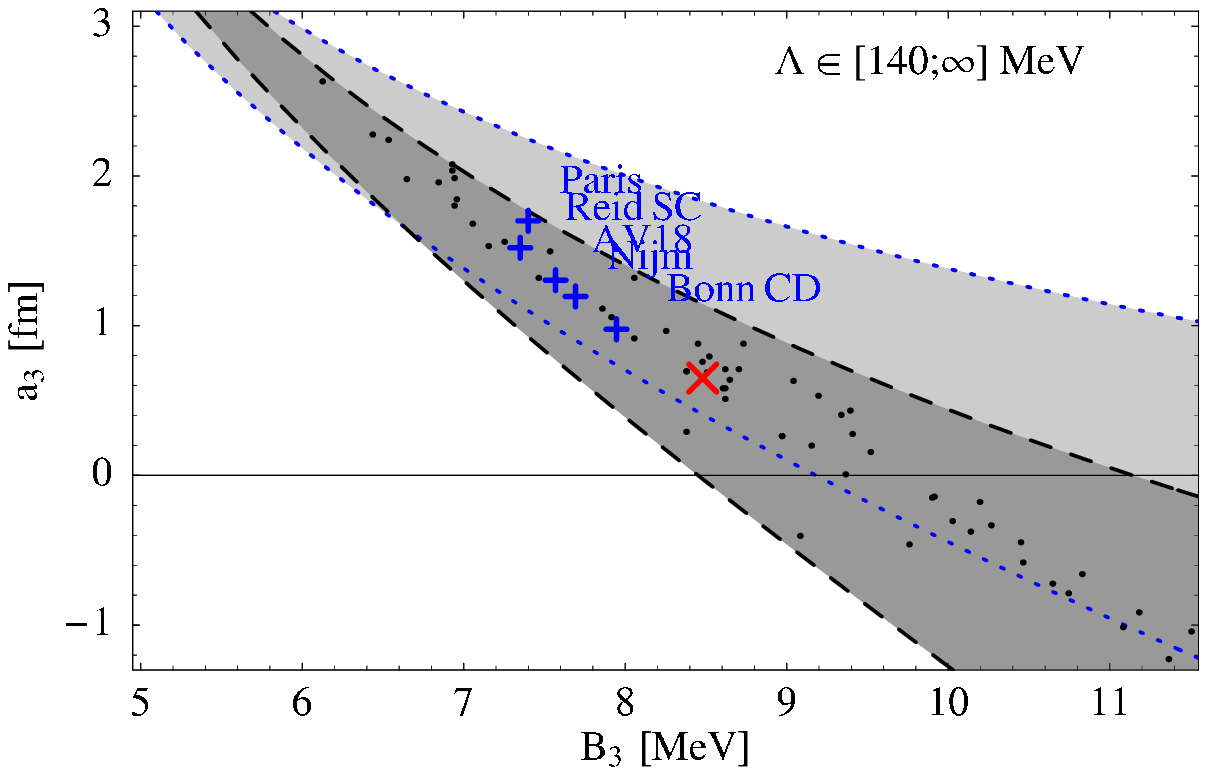}\hq\hq\hq\hq
  \includegraphics*[height=0.3\textwidth]
  {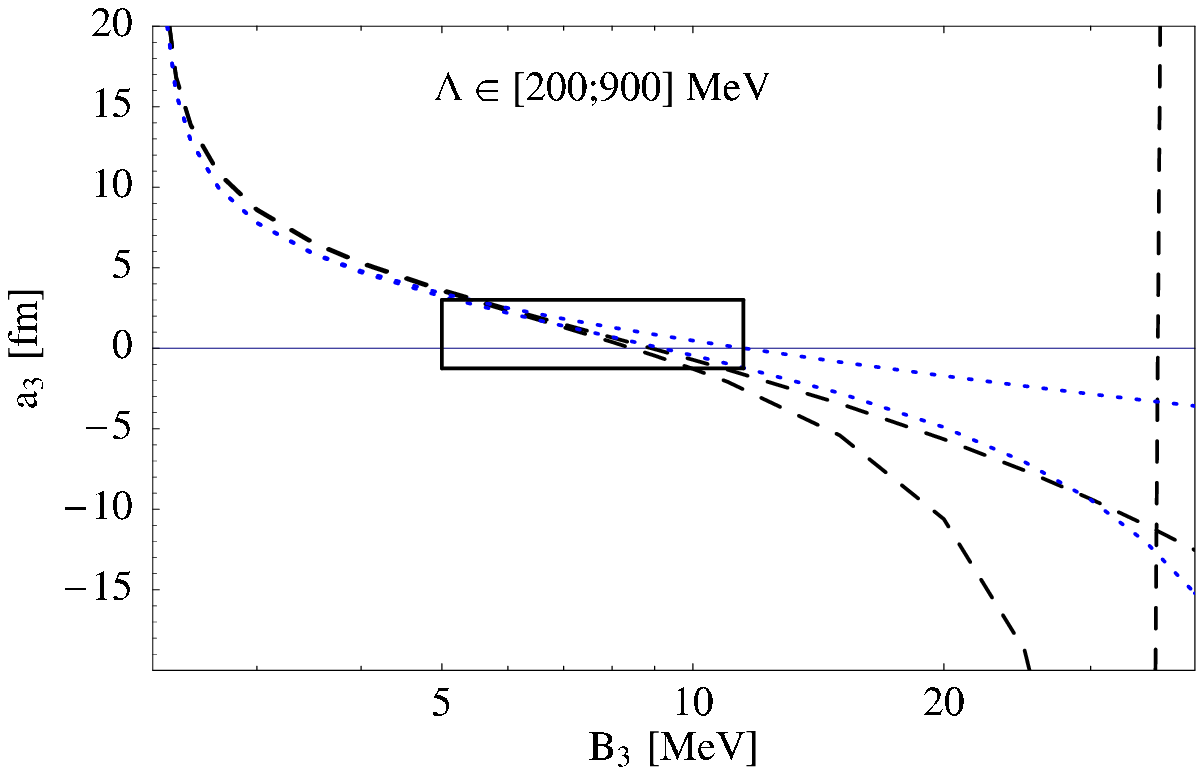}
\caption{Prediction for the Phillips line at LO (dotted; lightly shaded
  region) and NLO (dashed; dark shaded region) in Z-parameterisation when the
  cut-off is varied in the region $\Lambda\in[140;900]\;\MeV$. Left: The dots
  correspond to the predictions for the triton binding energy and doublet
  scattering length in different models with the same two-body scattering
  lengths and effective ranges as inputs~\cite{tkachenko}. $\times$:
  experimental result.  The outcome in various modern high-precision
  $NN$-potentials is indicated by crosses~\cite{Witala:2003mr}. Right: Larger
  range, the rectangle showing the region covered on the left.}
\label{fig:phillips}
\end{center}
\end{figure}

To complete the discussion, Fig.~\ref{fig:phillips} displays the
NLO-prediction for the Phillips line in Z-parameterisation, compared to
various old and modern potential model calculations which share the same
two-body on-shell Physics. In the three-body system, their widely different
off-shell behaviour predicts however different three-body bound state energies
and $Nd$ scattering lengthes. The correlation between scattering length and
bound state energy was first discussed by Phillips~\cite{phillips} and
explained by Efimov~\cite{tkachenko}. The spread induced by varying the
cut-off between $140$ and $900\;\MeV$ covers the off-shell dependence of all
potential models. Note the pole in the scattering length at
$B_d^\prime\approx36\;\MeV$ at NLO for a cut-off $\Lambda=900\;\MeV$. This
intrusion of another bound triton state is however outside the range of
applicability of \EFTNoPion. As
$k_\mathrm{typ}\approx180\;\MeV\gg\LambdaNoPion$ and the widthes of the LO and
NLO band are comparable, the EFT-calculation has not converged there.

%%%%%%%%%%%%%%%%
\subsection{Including the Effective Range to All Orders?}
\label{sec:allorders}

Why not re-sum all effective-range parameters up to a given order and replace
the expanded version of the two-body scattering amplitudes
(\ref{eq:effrangeZ}) with the version
(\ref{eq:NNscatteringERE}/\ref{eq:NNscatteringEREtransvestiteprelim}),
\begin{equation}
  \label{eq:EREpropagain}
  D_{s/t}(p_0,\pv)\stackrel{?}{\to}
  \frac{1}{\gamma_{s/t}-
    \frac{\rho_{0s/t}}{2}(\gamma_{s/t}^2+\frac{\pv^2}{4}-Mp_0)-
 \sum\limits_{n=1}^{\infty}\rho_{ns/t}(\gamma_{s/t}+\frac{\pv^2}{4}-Mp_0)^{n+1}
  -\sqrt{\frac{\pv^2}{4}-Mp_0}}\;\;,
\end{equation}
possibly with all coefficients after $\rho_{ns/t}$ set to zero at \NXLO{2n+1}?
This would immediately restore all effective-range corrections, give the
correct pole residues, and is mandatory if $\rho_0\gamma\approx 1$. With this
re-summation, the scattering length in the \fourS-channel and phase-shifts of
the higher partial
waves~\cite{2stooges_quartet,3stooges_quartet,pbhg,chickenpaper,Rupak:2001ci}
were found with an accuracy of $\lesssim 4\%$.  It was later also shown to
simplify analytical calculations in the two-nucleon
system~\cite{Beane:2000fi}.

However, the improvement in accuracy is only superficial: The higher
ERE-parameters $\rho_{ns/t}$ are known with less and less accuracy, and other
higher-order terms like $\mathrm{P}$-wave interactions and relativistic
corrections are still not included.

More severely, the propagator (\ref{eq:EREpropagain}) has additional spurious
poles, see Table~\ref{tab:poles}: one for each $\mathrm{S}$-wave at NLO
($\rho_{0s/t}\not=0$), and two more at \NXLO{3} ($\rho_{1s/t}\not=0$).  Their
positions change as more parameters $\rho_{ns/t}$ are included, and their
residues add with the residue of the low-lying real and virtual bound state to
zero. Albeit they all lie strictly speaking outside the range of validity of
\EFTNoPion, $\gamma_{\mathrm{spur},s/t}>\LambdaNoPion$, they are in particular
in the \oneS-channel close enough to influence observables at momenta $k$
above the break-up threshold, in particular as the closest of the spurious
bound states have rather large residues. Recall that the half-off-shell
amplitude goes to zero in the \twoS-channel only slowly, $\calA(k,p)\propto
1/p$~\cite{3stooges_boson,faddeev}.

\begin{table}[!htbp]
  \centering
  \begin{tabular}{|l||c|c||c|c|}
    \hline
   \rule[-1.5ex]{0ex}{4ex}& \multicolumn{2}{c||}{\threeS-channel}
    & \multicolumn{2}{c|}{\oneS-channel} \\
   \hline
    \rule[-1.5ex]{0ex}{4ex} order
     &$\gamma_{\mathrm{spur},t}\;[\MeV]$& $Z_{\mathrm{spur},t}$
     &$\gamma_{\mathrm{spur},s}\;[\MeV]$& $Z_{\mathrm{spur},s}$\\
   \hline
   \hline
   NLO\rule[-1.5ex]{0ex}{4ex}&$178$&$-1.69$&$152$&$-0.90$\\
   \hline
   \NXLO{3}\rule[-1.5ex]{0ex}{4ex}&$-385$&$-0.14$&$133$&$-0.71$\\
   \rule[-1.5ex]{0ex}{4ex}&$170\pm108\;\ii$&$-0.77\mp0.60\;\ii$&
   $-62\pm351\;\ii$&$-0.10\mp0.12\;\ii$\\
   \hline
  \end{tabular}
  \caption{Binding momentum $\gamma_\mathrm{spur}=\sqrt{MB_\mathrm{spur}}$ and
   dimension-less residue analogous to (\ref{eq:residue}/\ref{eq:wavefuren})
   of the additional spurious bound states in the \threeS- and
   \oneS-scattering amplitudes when the effective-range  parameters are
   re-summed as in (\ref{eq:EREpropagain}). }
  \label{tab:poles}
\end{table}

In the $Nd$-system, these additional poles generate new cuts in the solution
to the integral equations: In the doublet channels, they conspire to pose an
additional technical problem when one performs a rotation of the integration
contour into the complex plane in the Hetherington-Schick procedure. On the
other hand, expanding the two-particle propagator to a given order as in
(\ref{eq:dpropexpanded}), only the physical pole with
$\gamma_{s/t}\ll\LambdaNoPion$ is found in the two-particle scattering
amplitude, and all problems with spurious bound-states are easily avoided.

Most severely, the asymptotic behaviour of the two-nucleon propagator is
changed from $1/p$ to $1/p^2$ at large $p\gg 1/\rho_{0s/t}$. This leads to a
non-trivial modification of the argument for the inclusion of a three-body
force in the \twoS-channel, as the resulting integral equation is then not
analytically soluble in the UV-limit. Therefore, it is not straight-forward to
construct analytically the cut-off dependence of the three-body force as in
\cite{3stooges_doublet,4stooges} and demonstrate that it solves the Thomas
problem of an attractive $1/r^2$-potential. Gabbiani argued that with the
kernel becoming less attractive, behaving only like $1/p^2$ (i.e.~in position
space like the Coulomb potential $1/r$), the cut-off sensitivity of the
Wigner-$SU(4)$ symmetric component (\ref{eq:faddeevWigner}) of the integral
equation in this channel disappears~\cite{gabbiani}. Thus, he claims that no
limit cycle is found, and no three-body force is needed to achieve cut-off
independence.  According to him, the resulting phase-shifts without three-body
forces in the \twoS-channel fail however miserably. I cannot confirm these
findings but see that the phase-shifts in the \twoS-channel are markedly
cut-off sensitive even at high $\Lambda$, and that three-body forces are
necessary when the effective range is included to all orders. These findings
will be discussed in a sub-sequent article~\cite{hgrieonchicken}.

Re-summing the effective-range corrections poses therefore numerically as well
as technically non-trivial problems most importantly in the \twoS-wave (but
also in the higher doublet-channels) while no increase in accuracy is
achieved. To insert on the other hand the expanded two-nucleon amplitudes
(\ref{eq:dpropexpanded}) in the kernel of the integral equation, as proposed
in~\cite{4stooges} and done here is a straightforward way to calculate
phase-shifts to high accuracy, free of the problems mentioned above.  In the
end, all regularisation and re-summation schemes -- including those which
generate spurious bound states outside the range of validity of the EFT --
must agree to the order of accuracy of the calculation.  They must also have
the same number of independently fixed three-body forces.  To set up a
consistent power-counting scheme and computations with good numerical
convergence in a particular version can however be cumbersome and
time-consuming.

%%%%%%%%%%%%%%%%
\subsection{More Partial Waves}
\label{sec:higherpws}

After discussing the merits of Z-parameterisation in the only channel in which
a three-body force enters already at LO and which exhibits a bound state of
three nucleons, we now turn to the other partial waves. The solutions of the
integral equations (\ref{eq:quartetpw}/\ref{eq:doubletpw}) for the real and
imaginary parts of the phase-shifts are displayed in Fig.~\ref{fig:doubletPWs}
for the doublet channel, and in Fig.~\ref{fig:quartetPWs} for the quartet
channel. The quartet-$\mathrm{S}$ wave is shown separately in
Fig.~\ref{fig:quartetSPW}. In the following, they are compared to
potential-model calculations, and their convergence from order to order is
discussed and juxtaposed with the phase-shifts obtained when the effective
range is included to all orders. Table~\ref{tab:ERparameters3N} lists
low-energy parameters of lower partial waves.

Previous work on the higher phase-shifts in the $Nd$-system in \EFTNoPion
focused mainly on the \fourS-channel, where
$pd$-scattering~\cite{Rupak:2001ci} and the scattering
length~\cite{2stooges_quartet,3stooges_quartet,pbhg} and phase
shifts~\cite{3stooges_quartet,pbhg,chickenpaper} of $nd$-scattering were
computed up to \NXLO{2}. The phase-shifts of the other partial waves were
presented in Ref.~\cite{chickenpaper} to \NXLO{2}. The NLO-results
of~\cite{pbhg} and \cite{chickenpaper} were obtained in ``strict
perturbation'', i.e.~sandwiching the NLO-correction to the deuteron propagator
proportional to $c_{0s/t}$ once between the LO-solution to the half-off-shell
amplitude.  Ref.~\cite{chickenpaper} used already the Z-parameterisation to
fix $c_{0t}$ at NLO. In each case, the \NXLO{2}-calculations were carried out
with the re-summed deuteron propagator (\ref{eq:EREpropagain}).

Here, we use as in the \twoS-channel Z-parameterisation at each order and
include perturbatively all corrections up to NLO and \NXLO{2} respectively
into the kernel of the integral equation. As already discussed in
Sect.~\ref{sec:formalism}, this contains some higher-order effects and leads
to technical simplifications but -- like a re-summed deuteron propagator --
not to increased accuracy. The goal is again to show that convergence and
agreement with results from potential-model calculations is improved.

\begin{figure}[!htbp]
  \begin{minipage}{0.495\linewidth}
    \begin{flushright}
      \includegraphics*[height=0.58\textwidth]
      {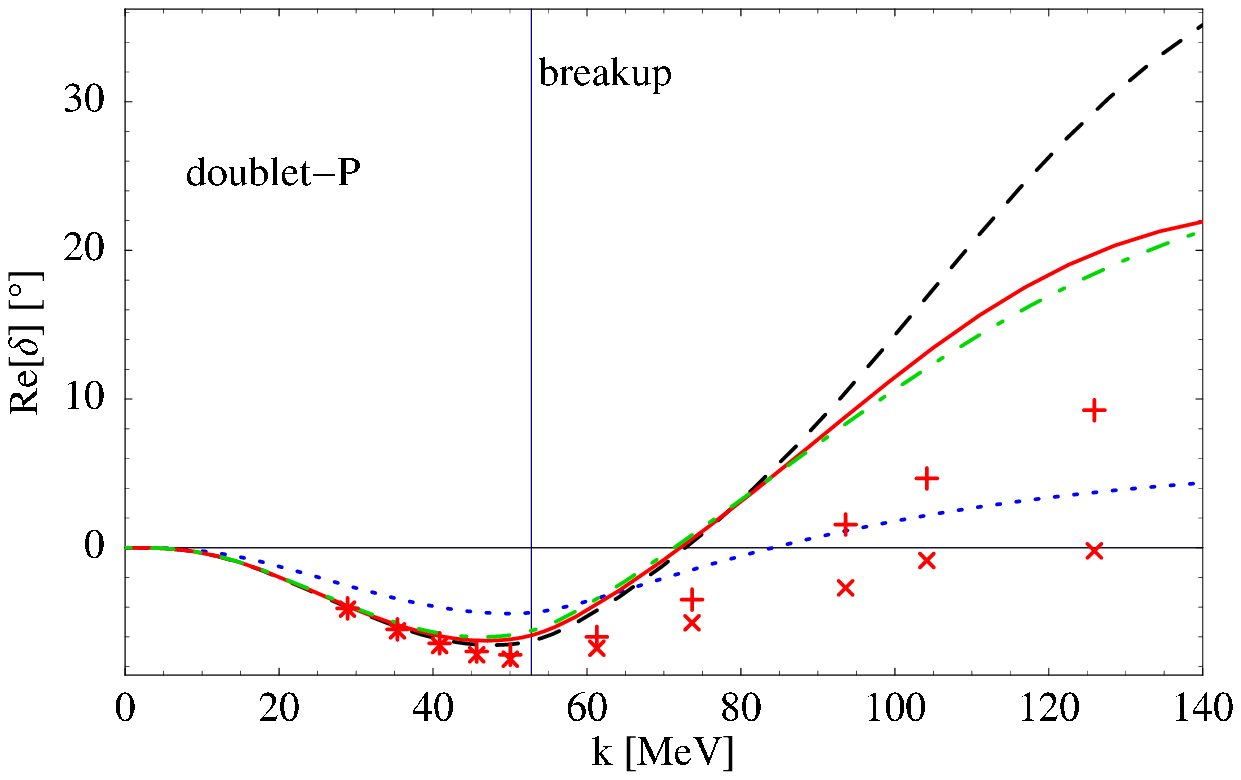}\\[1ex]
      \includegraphics*[height=0.58\textwidth]
      {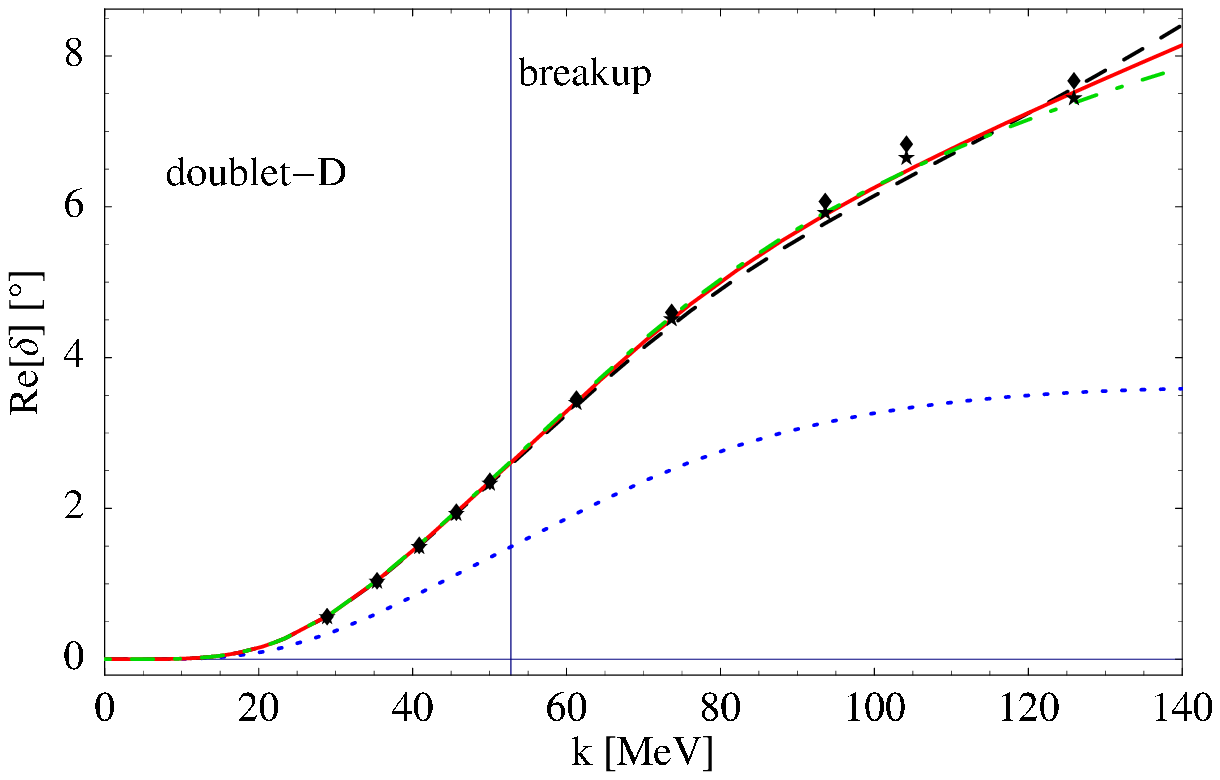}\\[1ex]
      \includegraphics*[height=0.58\textwidth]
      {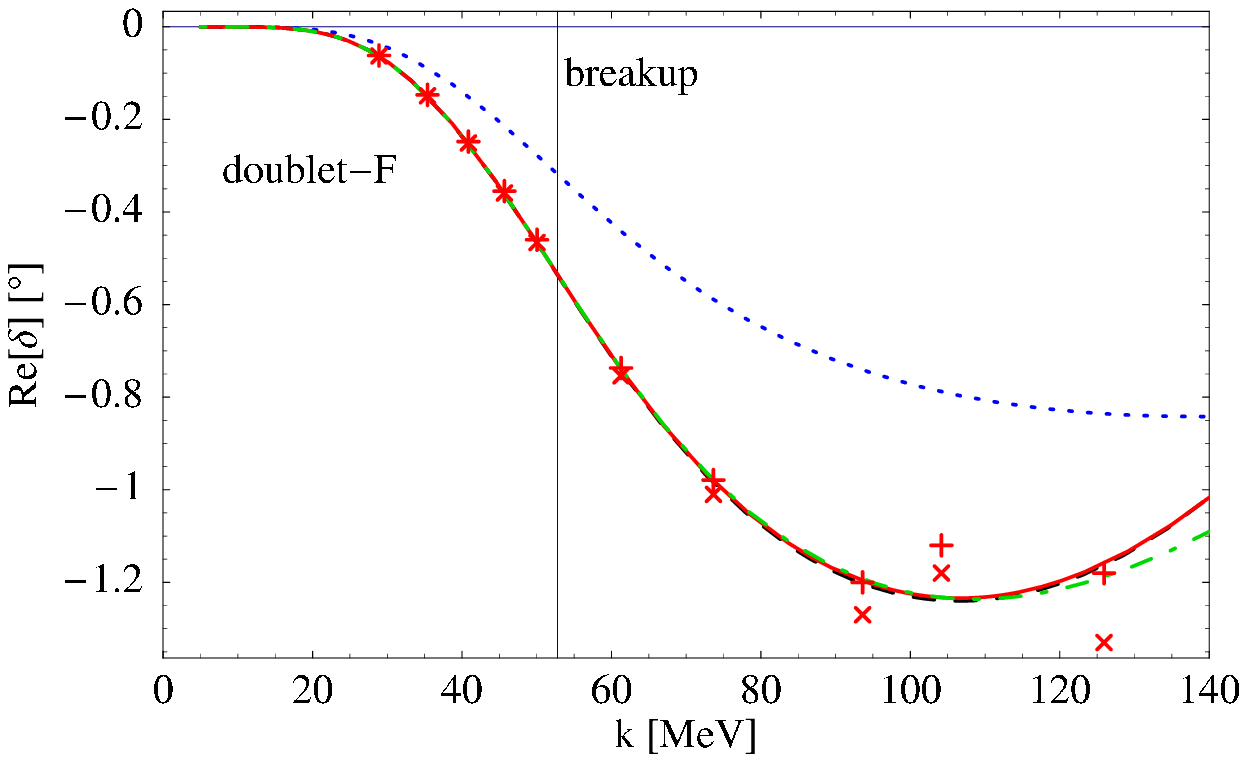}\\[1ex]
      \includegraphics*[height=0.58\textwidth]
      {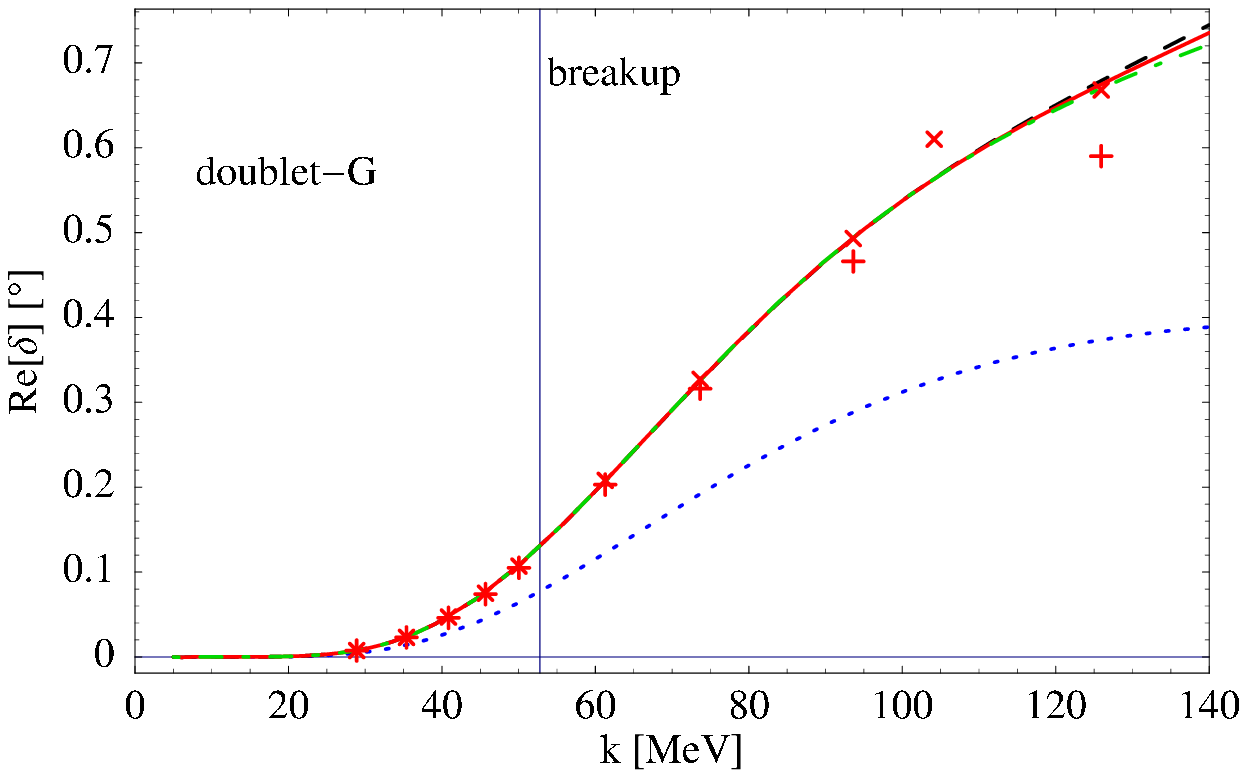}
    \end{flushright}
  \end{minipage}
  \begin{minipage}{0.495\linewidth}
    \begin{flushright}
      \includegraphics*[height=0.58\textwidth]
      {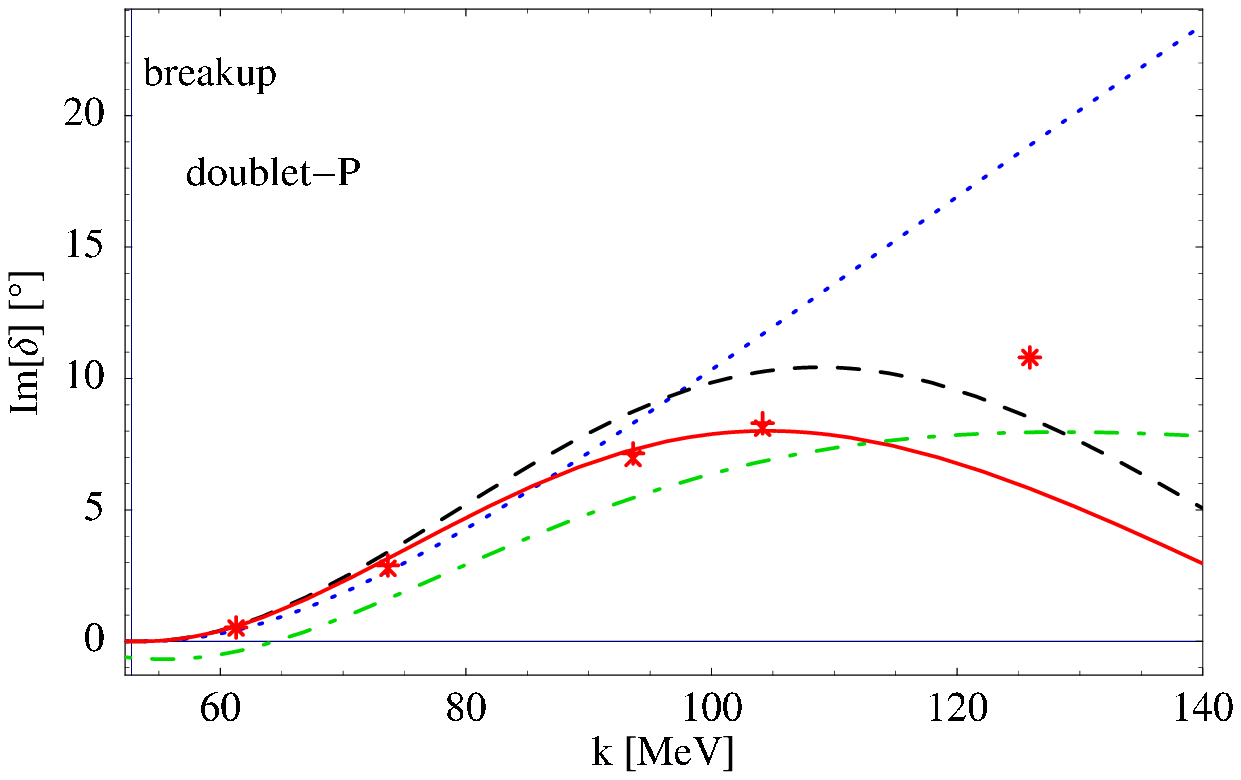}\\[1ex]
      \includegraphics*[height=0.58\textwidth]
      {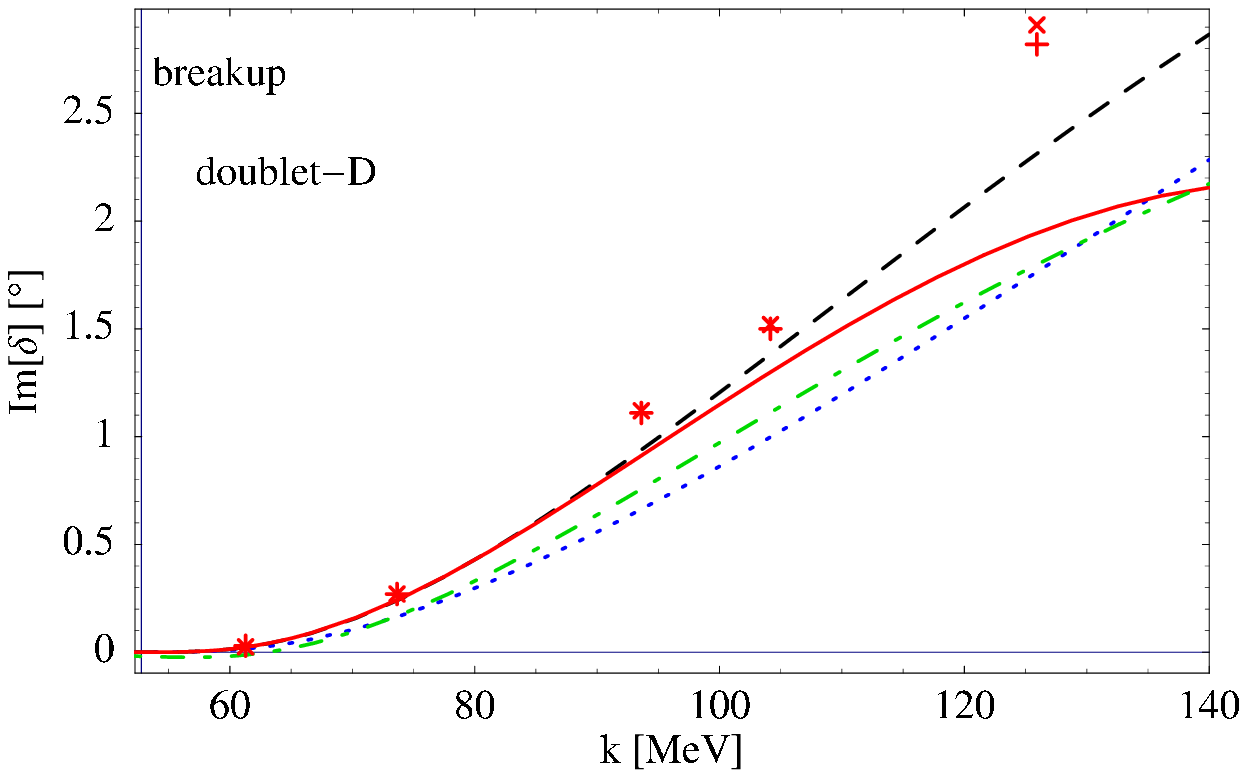}\\[1ex]
      \includegraphics*[height=0.58\textwidth]
      {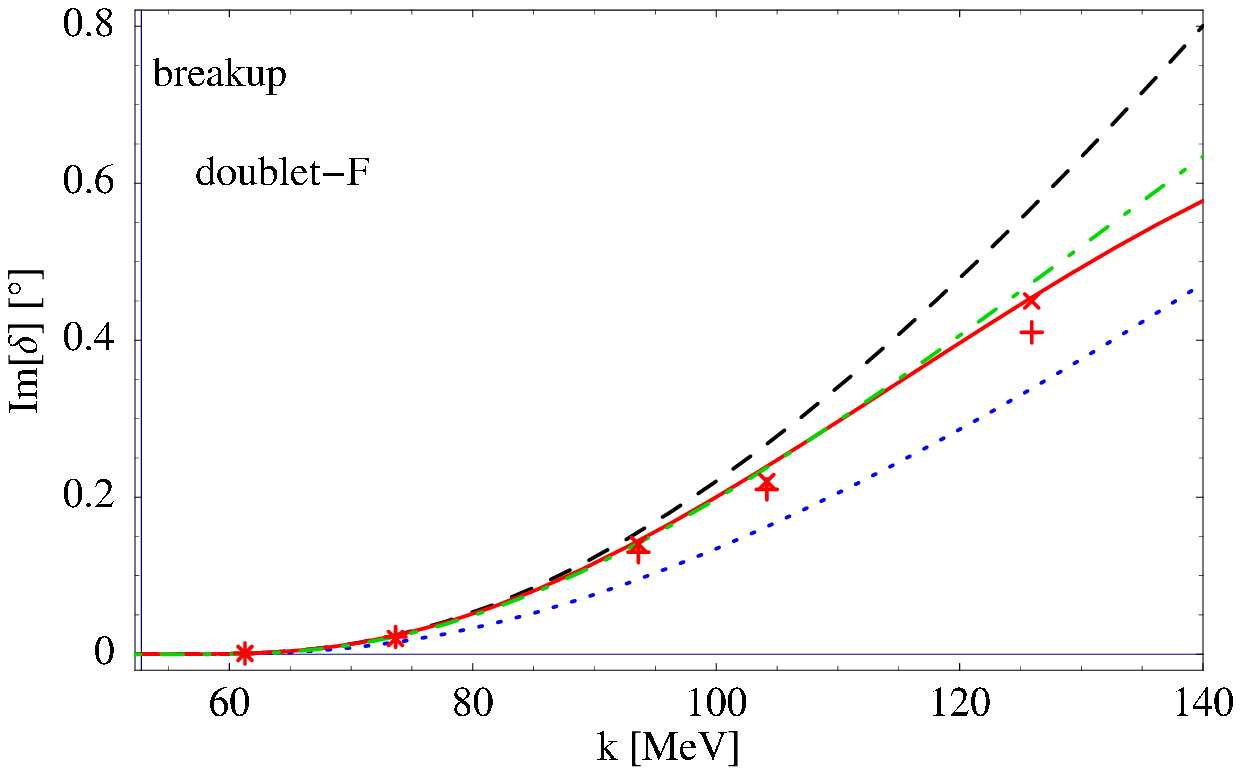}\\[1ex]
      \includegraphics*[height=0.58\textwidth]
      {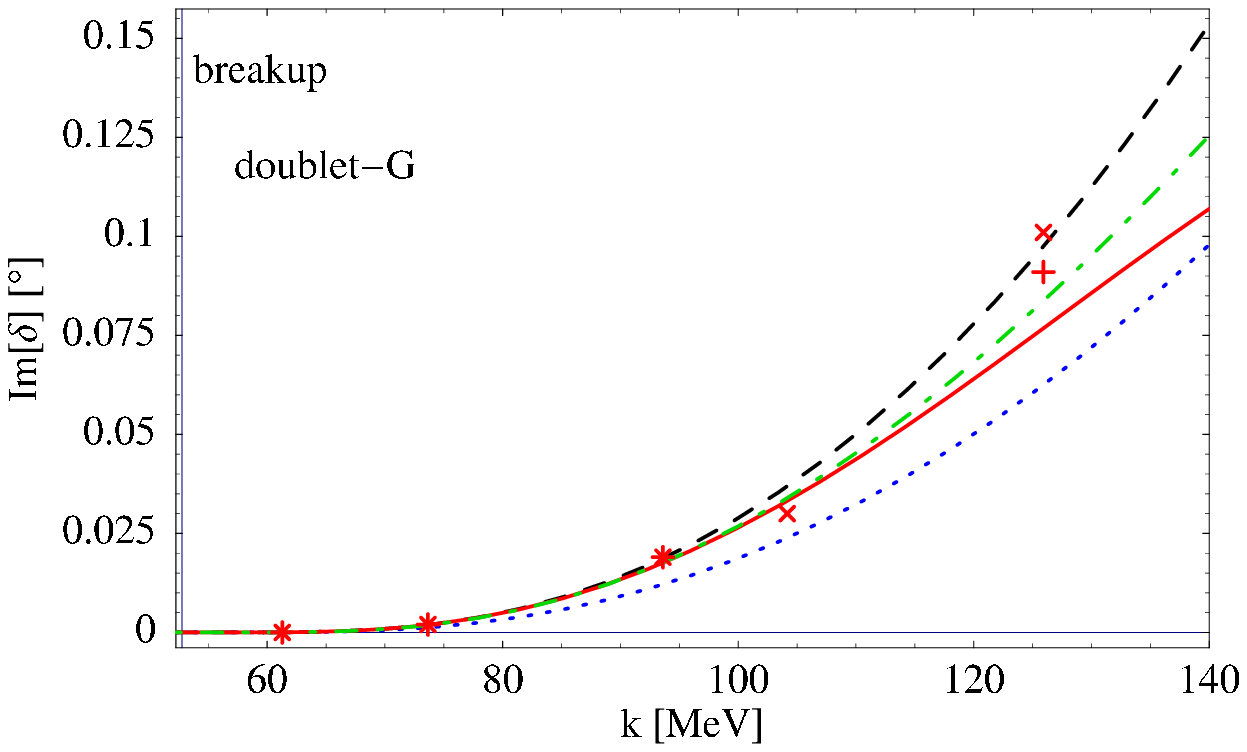}
    \end{flushright}
  \end{minipage}
\begin{center}
\caption{Real (left) and imaginary (right) part of the phase-shifts of the
  higher partial waves in the doublet channel in Z-parameterisation as
  function of the cm momentum $k$, $\Lambda=1000\;\MeV$. Dotted: LO; dashed:
  NLO; solid: \NXLO{2}; dot-dashed: effective-range corrections summed to all
  orders. Result of AV18 $+$ U IX from~\cite{kievsky1996} below break-up,
  from~\cite{Hueberetal} above, and at $k=104\;\MeV$
  from~\cite{Gloeckle:1995jg}. ${}^2l_j$ wave denoted by $\times$:
  $j=l-\half$; $+$: $j=l+\half$.}
\label{fig:doubletPWs}
\end{center}
\end{figure}

\begin{figure}[!htbp]
\begin{center}
  \begin{minipage}{0.495\linewidth}
    \begin{flushright}
      \includegraphics*[height=0.59\textwidth]
      {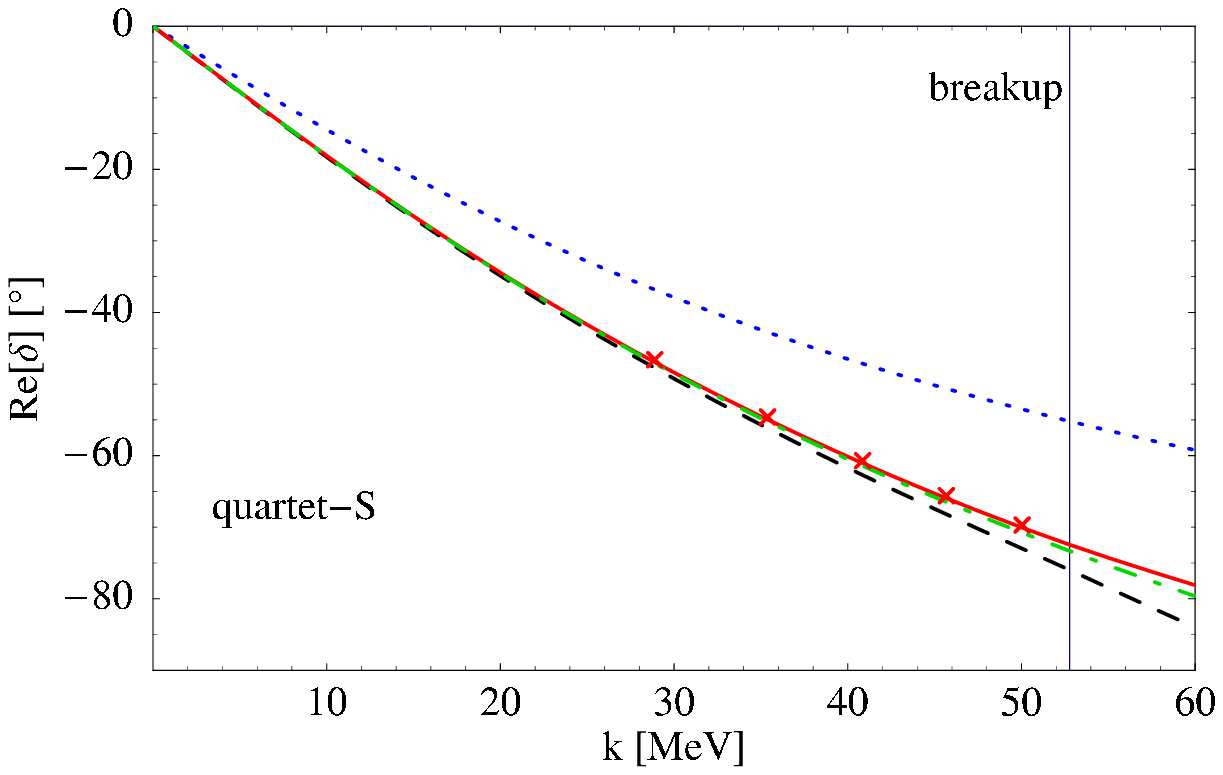}
    \end{flushright}
  \end{minipage}
  \begin{minipage}{0.495\linewidth}
    \begin{flushright}
      \includegraphics*[height=0.59\textwidth]
      {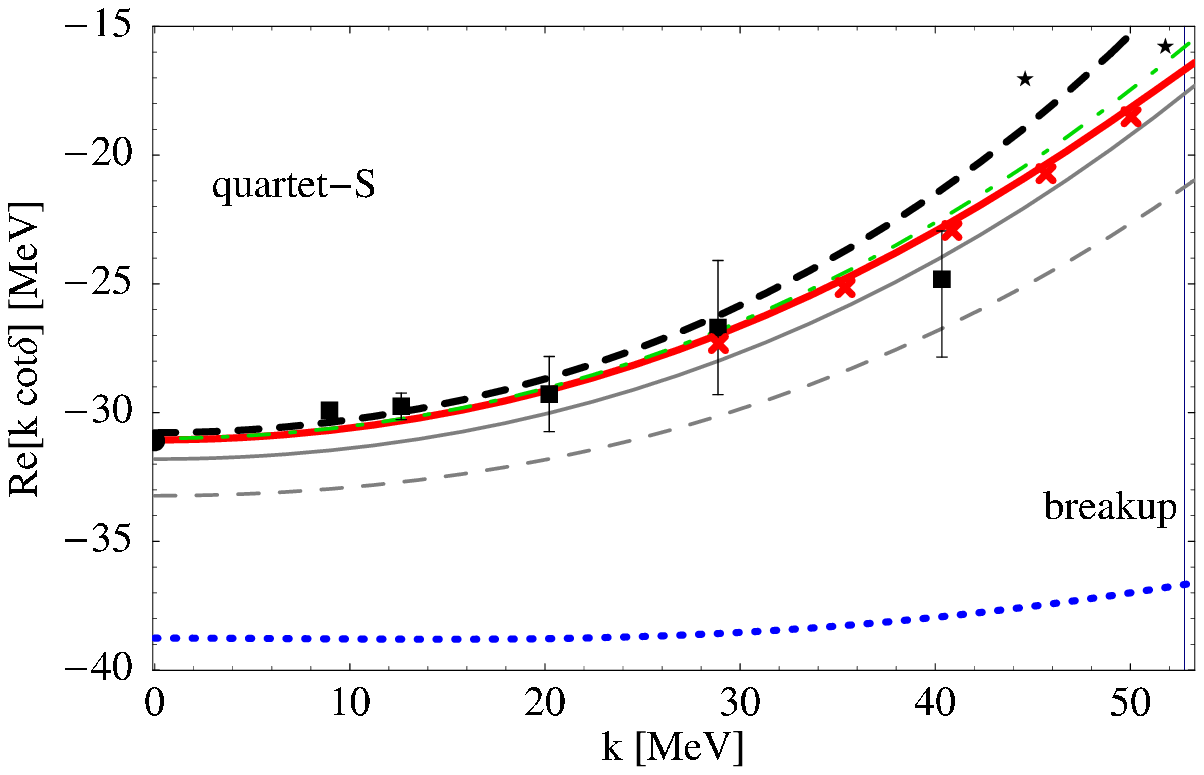}
    \end{flushright}
  \end{minipage}
\caption{Phase-shift in Z-parameterisation (left) and comparison of ERE- and
  Z-parameterisation for $k\cot\delta$ (right) of the \fourS-wave as function
  of the cm momentum $k$; $\Lambda=1000\;\MeV$. Notation as in
  Figs.~\ref{fig:doubletPWs} and~\ref{fig:lepage}, and in addition on the
  right: boxes/stars: partial wave analysis~\cite{doublet_PSA} (for points
  with error-bars as reported in~\cite{phillips}); dot at $k=0$: measured
  scattering length~\cite{doublet_sca}.}
\label{fig:quartetSPW}
\end{center}
\end{figure}

\begin{figure}[!htbp]
  \begin{minipage}{0.495\linewidth}
    \begin{flushright}
      \includegraphics*[height=0.58\textwidth]
      {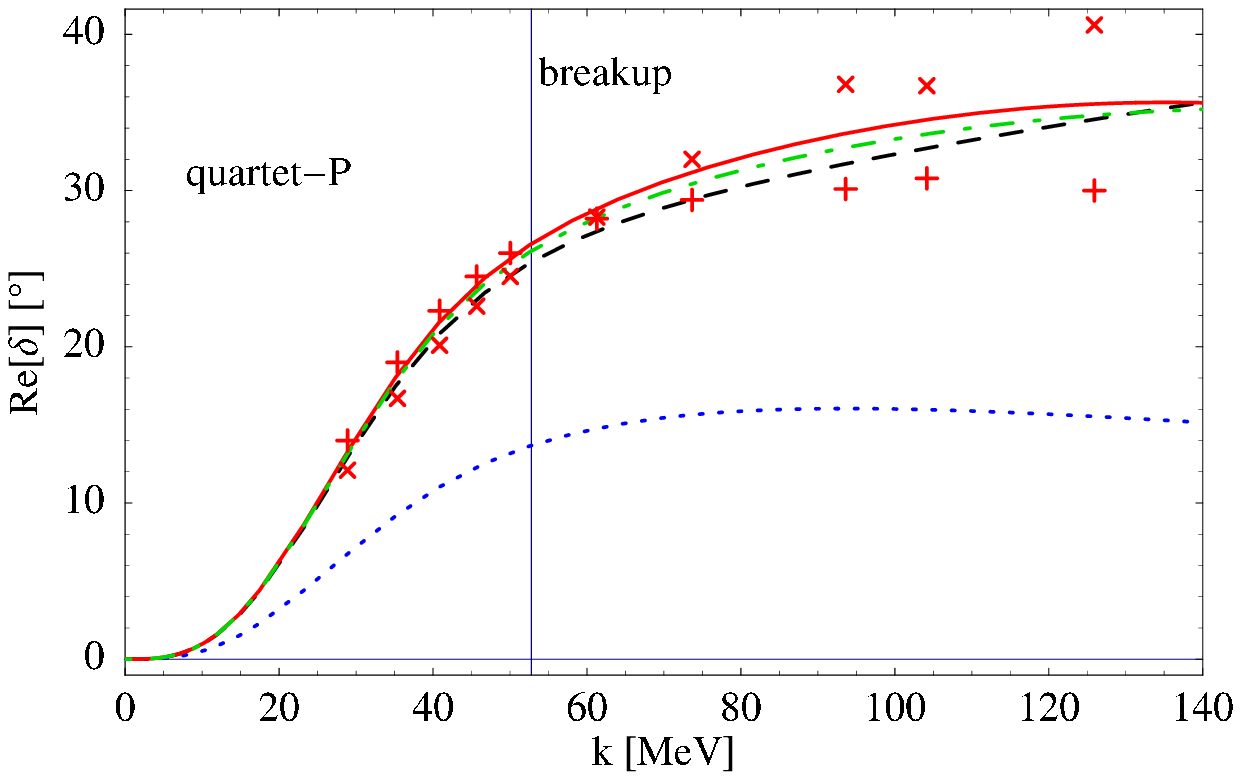}\\[1ex]
      \includegraphics*[height=0.58\textwidth]
      {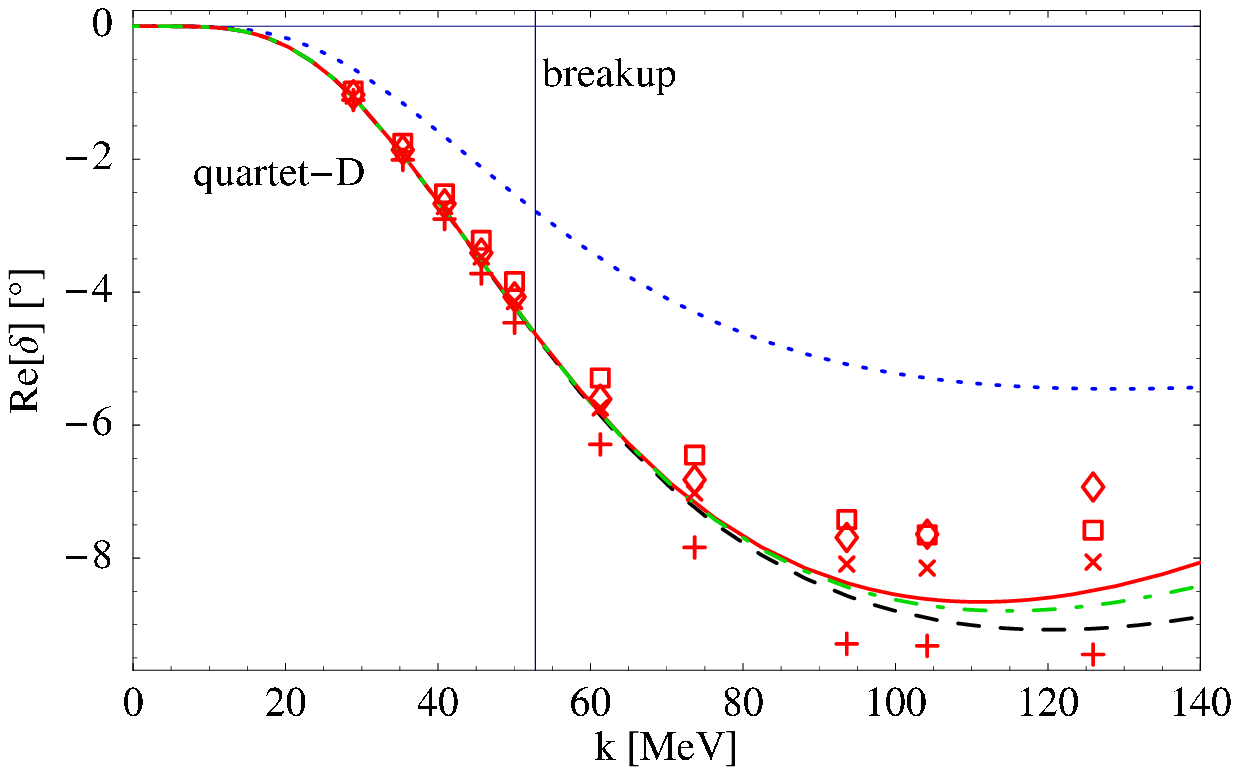}\\[1ex]
      \includegraphics*[height=0.58\textwidth]
      {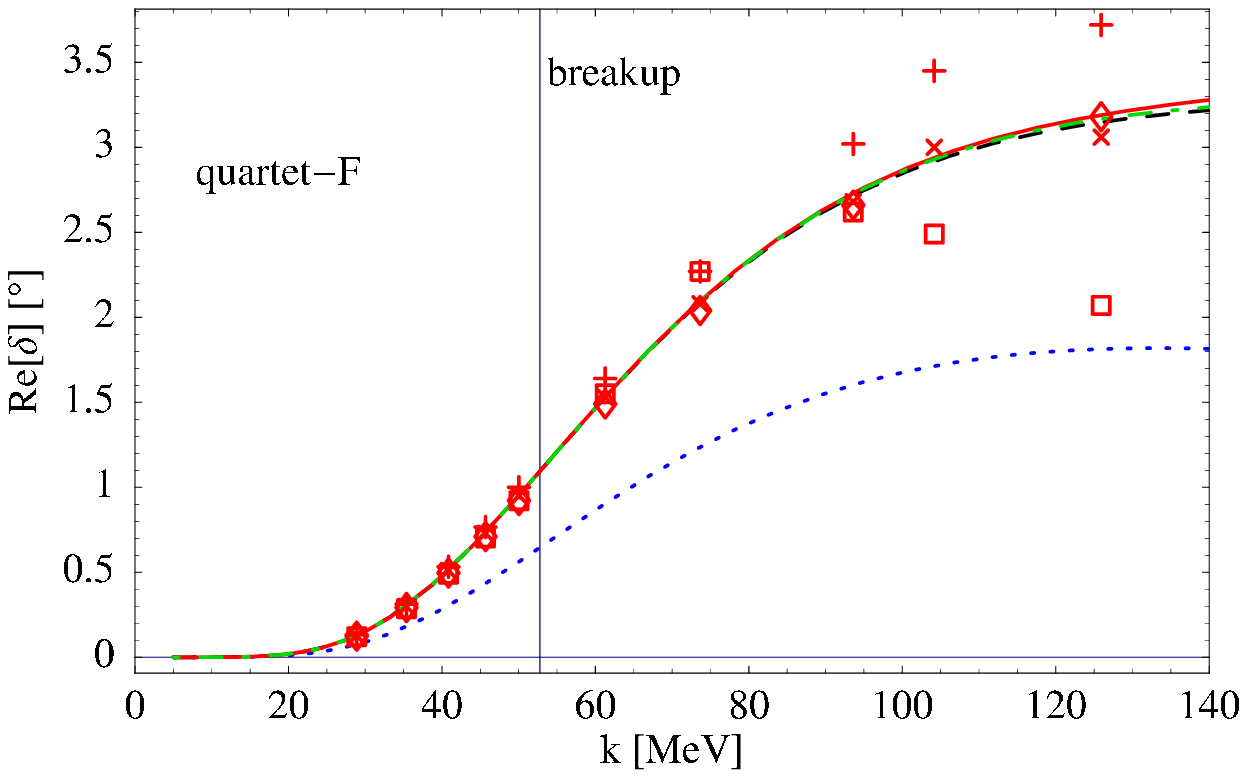}\\[1ex]
      \includegraphics*[height=0.58\textwidth]
      {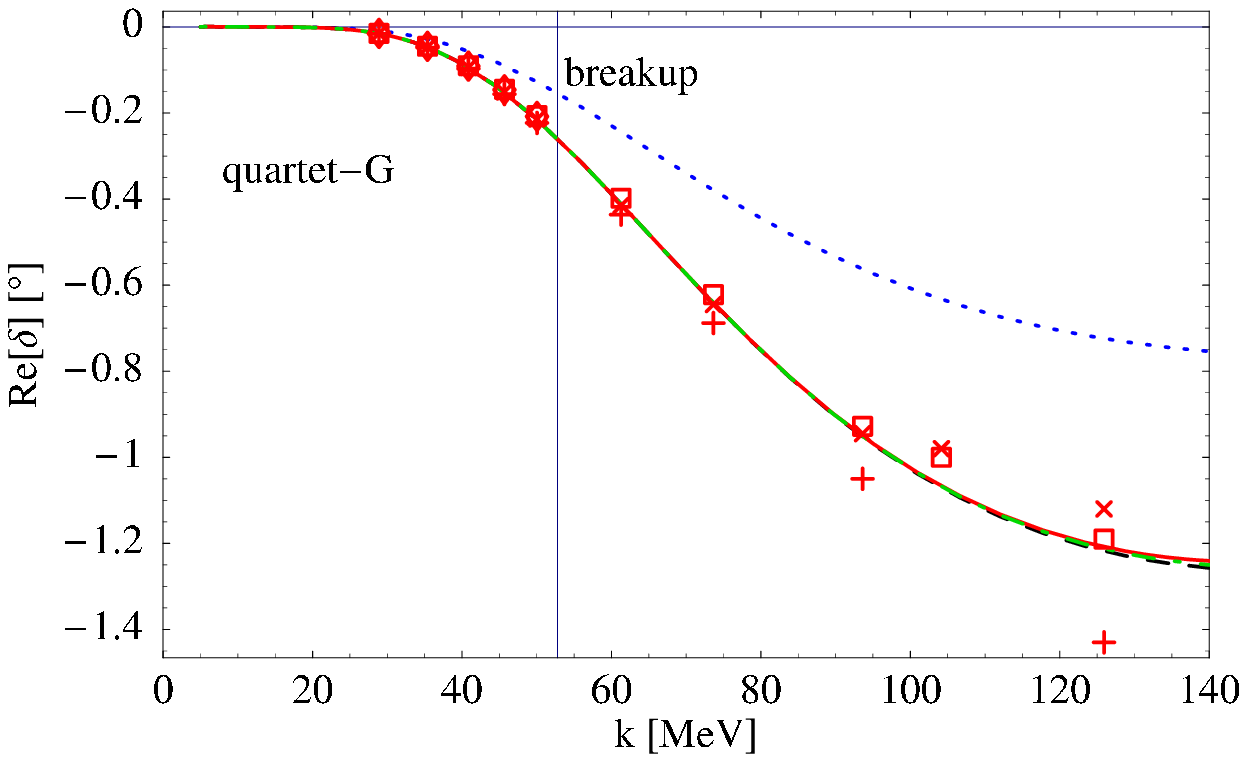}
    \end{flushright}
  \end{minipage}
  \begin{minipage}{0.495\linewidth}
    \begin{flushright}
      \includegraphics*[height=0.58\textwidth]
      {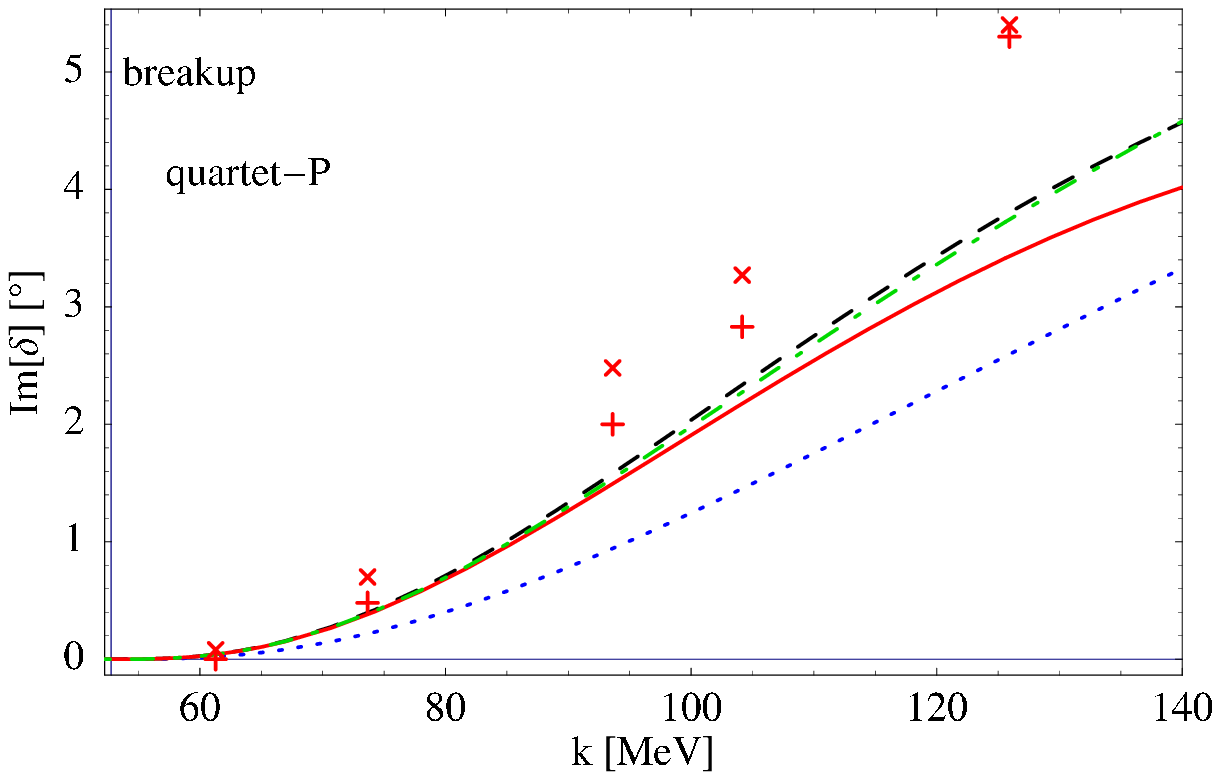}\\[1ex]
      \includegraphics*[height=0.58\textwidth]
      {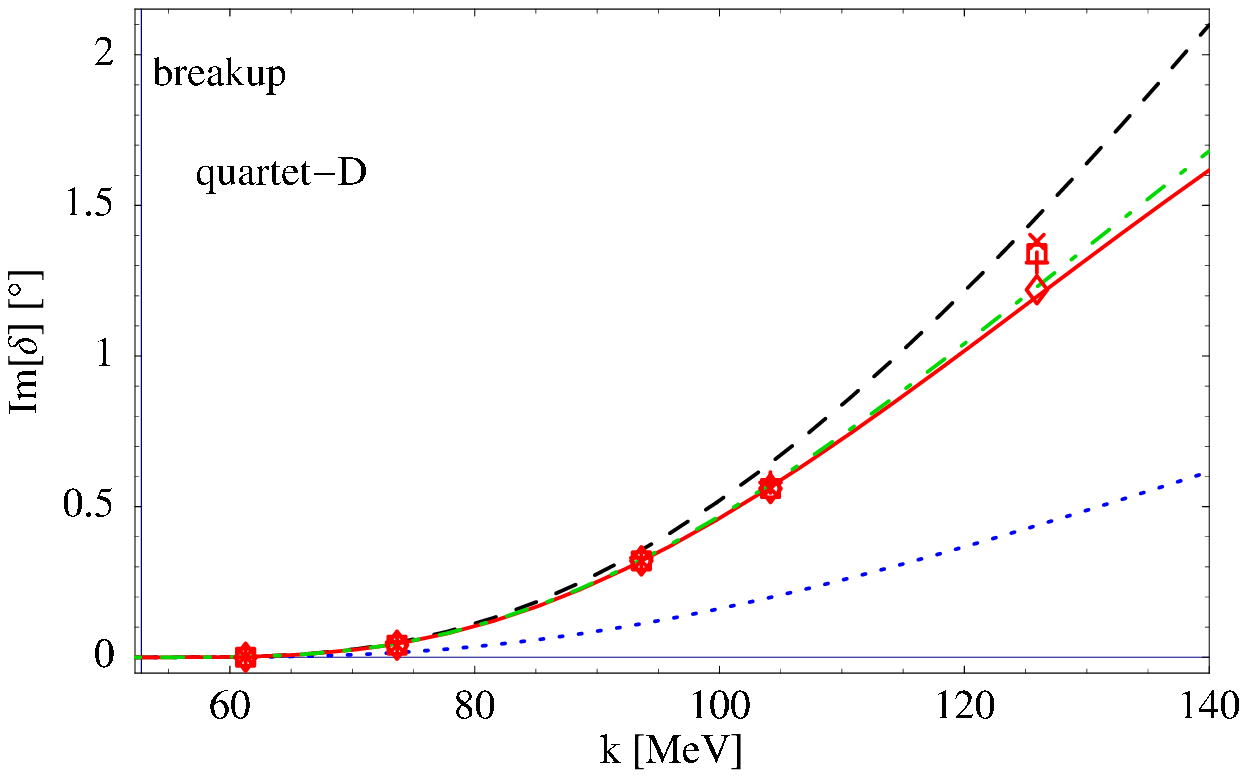}\\[1ex]
      \includegraphics*[height=0.58\textwidth]
      {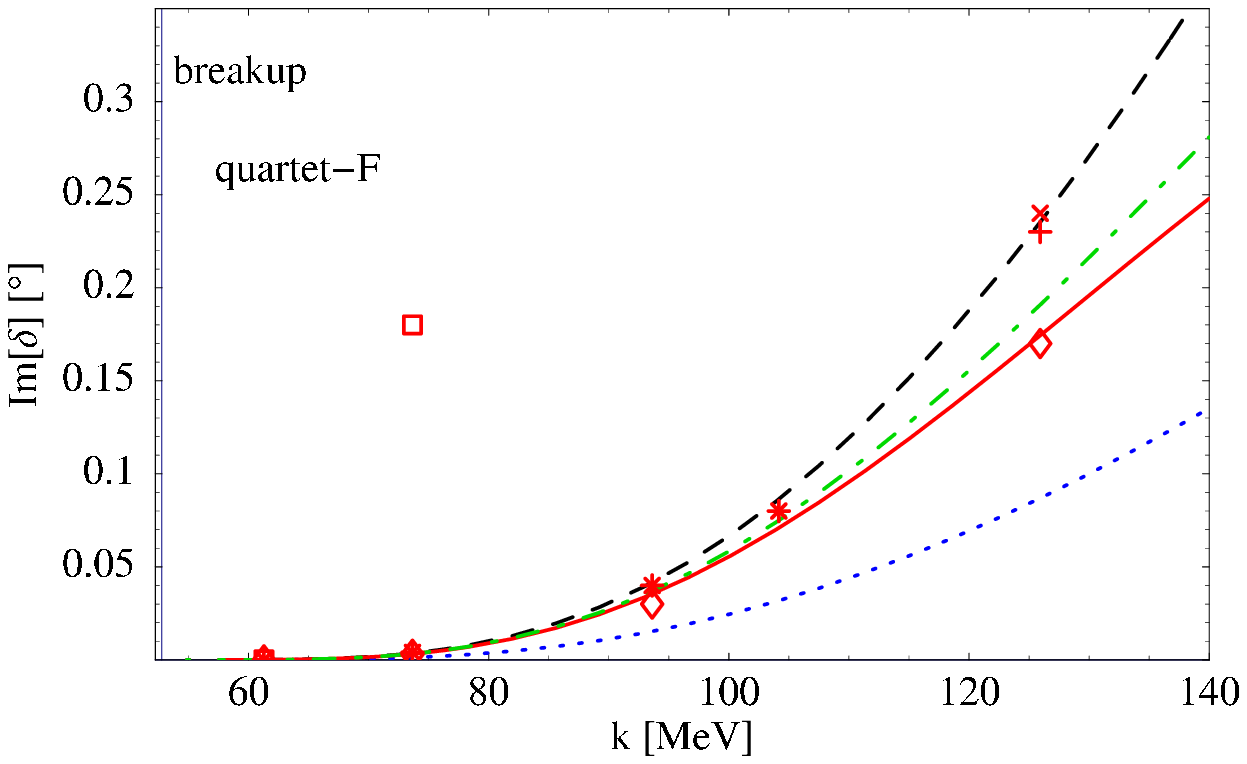}\\[1ex]
      \includegraphics*[height=0.58\textwidth]
      {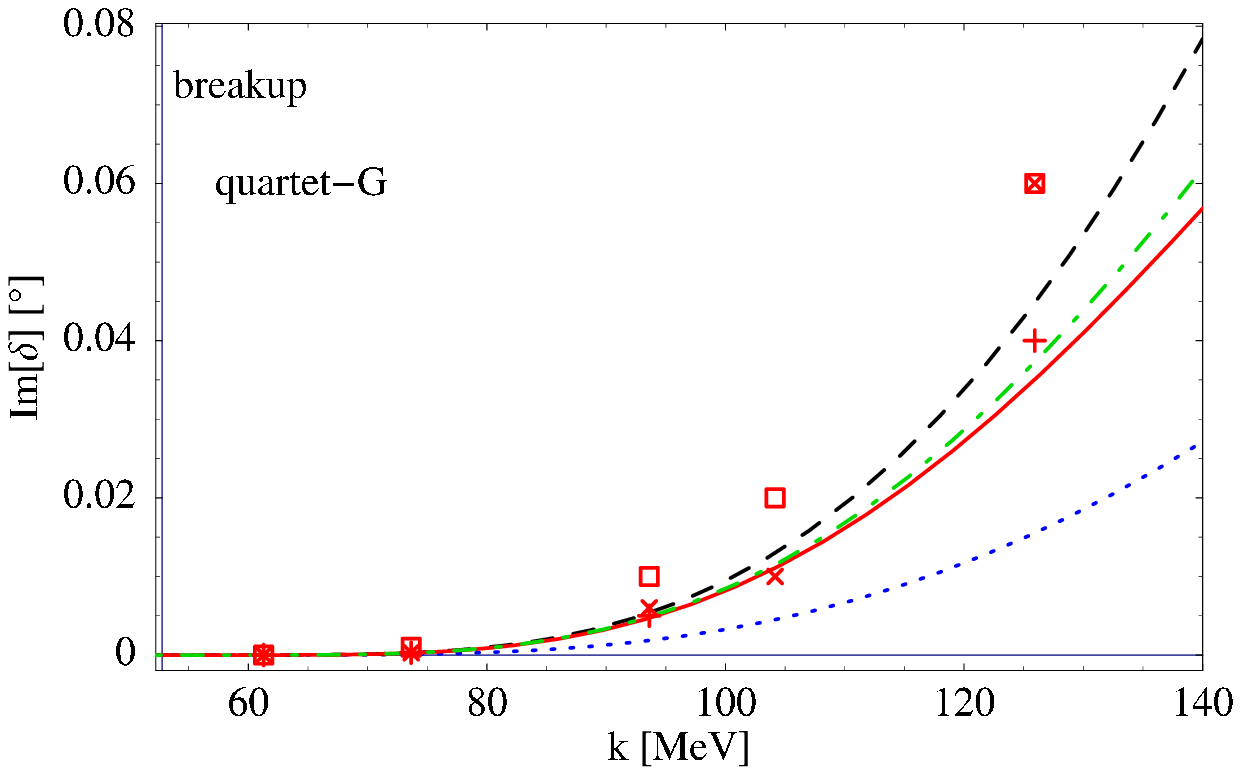}
    \end{flushright}
  \end{minipage}
\begin{center}
\caption{Real (left) and imaginary (right) part of the phase-shifts of the
  higher partial waves in the quartet channel in Z-parameterisation as
  function of the cm momentum $k$, $\Lambda=1000\;\MeV$. Notation as in
  Fig.~\ref{fig:doubletPWs}, and $\Box$: $j=l-\frac{3}{2}$; $\Diamond$:
  $j=l+\frac{3}{2}$ for the ${}^4l_j$-wave.}
\label{fig:quartetPWs}
\end{center}
\end{figure}

Like in the discussion of the \twoS-wave in Sect.~\ref{sec:tritonresults}, the
\NXLO{2}-results are compared to phase-shifts of a modern high-precision
$NN$-potential. Numbers for $nd$-scattering using the AV18-potential with the
Urbana IX-three-body-force were reported by Kievsky et
al.~\cite{kievsky1996,kievskyprivcommun} and H\"uber et
al.~\cite{Hueberetal,Gloeckle:1995jg}.  Within the drawing accuracy, they
agree with the phase-shifts reported for the Nijmegen potentials 93, I and II,
and for CDBonn~\cite{Gloeckle:1995jg}. As the partial waves do not mix in EFT
at \NXLO{2}, the potential-model results for the ${}^{2s+1}l_j$-partial waves
were grouped into bins of fixed spin $s$ and angular momentum $l$, with total
angular momenta $j=l\pm\half$ in the doublet channel, and
$j=l\pm\{\half;\frac{3}{2}\}$ in the quartet.

The higher the partial wave, the better the agreement with the potential model
calculations. This comes as no surprise, since the strong centrifugal barrier
for large angular momenta eliminates sensitivity on short-distance physics and
the solution to the integral equation approaches the result of the Born
approximation.

\begin{table}[!htbp]
  \centering
  \begin{tabular}{|l||c|c||c|}
    \hline
   \rule[-1.5ex]{0ex}{4ex}
   \hspace*{10ex}quartet-S&
            \multicolumn{2}{c||}{$a\;[\fm]$}&
            {$r_0\;[\fm]$}\\
   \hline
   order \rule[-1.5ex]{0ex}{4ex}& Z-param.&ERE-param.   & Z-param.     
   \\
   \hline
   \hline
   LO\rule[-1.5ex]{0ex}{4ex}&\multicolumn{2}{c||}{$5.091 $}&$-0.1 $\\
   \hline
   NLO\rule[-1.5ex]{0ex}{4ex}&$6.410 $&$5.938 $&$1.95 $\\
   \hline
   \NXLO{2}\rule[-1.5ex]{0ex}{4ex}&$6.354\pm0.020 $&$6.204 $&$1.8\pm0.1 $\\
   \hline
   \hline
   experiment~\cite{doublet_sca}
   \rule[-1.5ex]{0ex}{4ex}&\multicolumn{2}{c||}{$6.35\pm0.02$}&
   $$\\ 
   \hline
   \hline
   modern $NN$-potentials~\cite{Witala:2003mr,Friar:1999jd}
   \rule[-1.5ex]{0ex}{4ex}&\multicolumn{2}{c||}{$6.34\dots6.35$}&
   $$\\
   \hline
   \hline
   \NXLO{2} re-summed, cf.~\cite{2stooges_quartet,3stooges_quartet,pbhg}
   \rule[-1.5ex]{0ex}{4ex}&\multicolumn{2}{c||}{$6.365$}&
   $1.8$\\
   \hline
  \end{tabular}
  \\[4ex]
  \begin{tabular}{|l||c|c||c||c|c||c|}
    \hline
    \rule[-1.5ex]{0ex}{4ex}&\multicolumn{2}{c||}{doublet-P}&
   doublet-D&
   \multicolumn{2}{c||}{quartet-P}&
   quartet-D\\
   \hline
   order \rule[-1.5ex]{0ex}{4ex}&
            $\kappa_0\;[\fm^3]$&$\kappa_1\;[\fm^{-1}]$&
            $\kappa_0\;[\fm^5]$&
            $\kappa_0\;[\fm^3]$&$\kappa_1\;[\fm^{-1}]$&
            $\kappa_0\;[\fm^5]$
   \\
   \hline
   \hline
   LO\rule[-1.5ex]{0ex}{4ex}&$33.1 $&$-3.2$&$ -268$&$-76.7 $&$1.0 $&$531 $\\
   \hline
   NLO\rule[-1.5ex]{0ex}{4ex}&$53.9 $&$-2.1$&$ -456$&$-139 $&$0.45$&$894 $\\
   \hline
   \NXLO{2}\rule[-1.5ex]{0ex}{4ex}&$53.3\pm0.2 $&$-2.1$&$-456 $&$-140 $&
      $0.43\pm0.01$&$894 $\\
   \hline
   \hline
   \NXLO{2} re-summed
   \rule[-1.5ex]{0ex}{4ex}&&&&$-140$&$0.43$&$894$
   \\
   \hline
  \end{tabular}
  \caption{The effective-range parameters (\ref{eq:effrangePWs}) of the low
    partial waves in the $Nd$-system from Z-parameterisation in the first
    three orders; theoretical accuracy discussed in the
    text.}
  \label{tab:ERparameters3N}
\end{table}

\absatz The EFT-calculation reproduces of course the well-known behaviour of
the phase shifts at small momenta, e.g.~\cite{GoldbergerWatson},
\begin{equation}
  \label{eq:effrangePWs}
  k^{2l+1}\cot\delta^{(l)}\to
  -\frac{1}{\kappa_0^{(l)}}+
  \frac{\kappa_1^{(l)}}{2}\;k^2+\calO(k^4)\;\;\forall k\to0\;\;,
\end{equation}
where the $\mathrm{S}$-wave parameters $\kappa_0^{(0)}$ and $\kappa_1^{(0)}$
are the scattering length $a$ and effective range $r_0$, the $\mathrm{P}$-wave
parameter $\kappa_0^{(1)}$ the scattering volume, etc. Table
\ref{tab:ERparameters3N} lists parameters for low partial waves of the
$Nd$-system. The theoretical accuracy may for example be estimated
conservatively by $Q\sim\frac{\gamma_t}{\mpi}\approx\frac{1}{3}$ of the
difference between the NLO- and \NXLO{2}-result, or by the difference to the
\NXLO{2}-result with re-summed effective-range corrections as in
(\ref{eq:EREpropagain}), since all \NXLO{2}-calculations must agree within the
\NXLO{2}-accuracy. The numerical uncertainties are easily covered by these
error-bars. In the doublet channel, no re-summed results are given due to the
non-trivial cut-structure induced by the spurious poles of the two-nucleon
propagators, see Sect.~\ref{sec:allorders}. Furthermore, no numbers are
presented for the \twoS-channel as the measured scattering length
(\ref{eq:threebodyexpvalues}) is used to determine the strength of the
three-body force, and the effective range is very large, $\sim 500\;\fm$,
rendering an effective-range expansion useless for this partial wave. In the
other partial waves, the parameters involve the typical length scale $\sim
2\dots5\;\fm$, so that an effective-range expansion is often useful for
momenta up to the deuteron break-up, $k_\mathrm{max}\lesssim 20\dots52\;\MeV$.
More effective-range parameters are obtained without difficulty, but they are
less likely to be measurable in the upcoming partial wave analysis of the
$nd$-system~\cite{black}.

The quartet-$\mathrm{S}$ wave scattering length has drawn substantial interest
recently as its knowledge sets at present the experimental uncertainty in a
recent, indirect determination of the doublet scattering length
(\ref{eq:threebodyexpvalues})~\cite{boundcoherentexp}. At \NXLO{2}, the result
of Z-parameterisation in \EFTNoPion (using a perturbatively expanded kernel
which is the inserted into the integral equation), $[6.35\pm0.02]\;\fm$,
agrees very well with experiment~\cite{doublet_sca} and modern high-precision
potential-model calculations (which give
$a_q=[6.34\dots6.35]\;\fm$)~\cite{Witala:2003mr,Friar:1999jd}, albeit
partial-wave mixing, iso-spin breaking and electro-magnetic effects are not
present in \EFTNoPion at \NXLO{2}. $a_q$ is obviously to a very high degree
sensitive only to the correct asymptotic tail of the deuteron wave function,
as comparison between the NLO result with its correct deuteron residue, the
\NXLO{2}-result and the result of the re-summed deuteron propagator shows. As
the amplitude decays at large off-shell momenta as
$1/p^{3.17\dots}$~\cite{3stooges_quartet}, this is not surprising.

Comparing $k\cot\delta$ in this partial wave at low energies in
Fig.~\ref{fig:quartetSPW}, the ERE-version converges much slower than the
result of Z-parameterisation.  For example, the \NXLO{2}-scattering length
differs from experiment and the Z-parameterisation result of the same order
still by $2\%$, and the correction from NLO to \NXLO{2} is with $5\%$ much
larger than the correction in Z-parametrisation.  Indeed, partial \NXLO{3}-
and \NXLO{4}-calculations in ERE-parameterisation which include only
effective-range effects bring one slowly closer to the experimental number:
\begin{equation}
  a({}^4\mathrm{S})_\mathrm{ERE}=(
  \underbrace{\rule[-1ex]{0ex}{0ex}5.09}_\mathrm{LO}
  +\underbrace{\rule[-1ex]{0ex}{0ex}0.84}_\mathrm{NLO}
  +\underbrace{\rule[-1ex]{0ex}{0ex}0.27}_\text{\NXLO{2}}
  +\underbrace{\rule[-1ex]{0ex}{0ex}0.10}_\text{``\NXLO{3}"}
  +\underbrace{\rule[-1ex]{0ex}{0ex}0.04}_\text{``\NXLO{4}"}+\dots)\;\fm
  \approx 6.34\;\fm
\end{equation}
The re-summed effective-range result of \EFTNoPion at \NXLO{2} was first
reported in~\cite{2stooges_quartet,3stooges_quartet}. The ERE-result at
NLO in~\cite{pbhg}, $6.7\;\fm$, was found not by re-summing the expanded
kernel as here, but by inserting the NLO-correction only once between the LO
amplitudes.  Thus, convergence is also sped up by iterating the perturbed
kernel.

\absatz EFT shares also the rather complex structure found even at low $k$,
e.g.~in the doublet- and quartet-D and -F waves. Some disagreement with the
potential-model results is seen in some imaginary parts at rather high momenta
$k\gtrsim 120\;\MeV$, close to the breakdown scale of \EFTNoPion, where the
expansions parameter $Q$ is approaching $1$. Given that the imaginary parts
are very small compared to the real parts and also most sensitive to details
of the break-up reaction $nd\to nnp$, the overall agreement is not
dis-encouraging. The most notable deviation occurs in the doublet-P wave,
where internal convergence of the EFT-result is achieved up to more that
$k\approx 100\;\MeV$, while significant deviations from potential-model
calculations start around breakup. In this as in all other channels, cut-off
dependence is negligible above $\Lambda\sim900\;\MeV$ (indicating also that a
three-body force is not needed for convergence, see~\cite{suppressed3bfs}),
and numerics is stable. The discrepancy -- already found in~\cite{chickenpaper}
-- has therefore a physical origin. We may expect that it is cured by
cancellation with the well-known, rather strong $\mathrm{P}$-wave interaction
in the $NN$-system. It enters at \NXLO{3}, where the splitting between the
partial waves will also be reproduced more realistically. Clearly, more work
is needed here.

While the NLO-corrections to the LO EFT-result are as expected rather large in
Z-parameterisation, the \NXLO{2}-modifications are in the real parts of the
phase-shifts tiny, and do not exceed $40\%$ of the NLO corrections in the
imaginary parts even as $k\to\LambdaNoPion\sim\mpi$, the ${}^2$P-wave being
again the only notable exception.  The main difference between the strictly
perturbative NLO-calculation of~\cite{chickenpaper} which already utilised
Z-parameterisation and the partially re-summed version used here is that now
the NLO and \NXLO{2}-results lie practically on top of each other, while they
noticeably differ in strictly perturbative Z-parameterisation above break-up.
The difference does however never exceed $5\%$, and hence lies within the
accuracy of the NLO calculation. The \NXLO{2}-results
of~\cite{3stooges_quartet,pbhg,chickenpaper} using the fully re-summed
deuteron and \oneS-propagators (\ref{eq:EREpropagain}) are nearly
in-discernible from the partially re-summed version (\ref{eq:dpropexpanded})
in the real parts even at hight $k$. In the imaginary parts, they do usually
not deviate by more than $10\%$ of the correction from NLO to \NXLO{2} at
$k\sim 140\;\MeV$. This lies again well within the power-counting prediction
that partial inclusion of \NXLO{3}-effects will not increase the accuracy of
the \NXLO{2}-result. It gives however a band within which one expects a full
\NXLO{3}-calculation to lie.

\absatz Finally, an unsolved technical issue has to be mentioned. The problem
that the spurious poles of the re-summed two-body propagators induce
additional cuts in the solution of the integral equation in the doublet
channel, was already discussed in Sect.~\ref{sec:allorders}. Such breaches of
unitarity below breakup, induced by na\"ive integration over a cut at high
off-shell momenta, are the reason why the phase-shifts of the doublet-P and
doublet-D-wave have a small non-zero value at the breakup point.  As the
half-off-shell amplitudes of higher partial waves converge substantially
faster than of the doublet-S-wave, these violations become however less and
less noticeable. For the higher partial waves, they disappear in the numerical
noise.

\absatz To summarise, the overall convergence of the phase-shifts is good in
all channels, both internally and to the available potential-model
calculations. The only noticeable exception is the doublet-P-wave, in which
the radius of convergence is limited to momenta $k$ below break-up. Neglecting
partial-wave splitting, simple observables like the differential cross-section
do therefore agree with potential-model results and direct experimental
measurements, within the level of accuracy of the EFT-prediction. As an
illustration, the elastic differential cross-section is shown in
Fig.~\ref{fig:diffcrosssection} for three neutron energies: just below the
$nd$-breakup ($k=50\;\MeV$), and at two momenta close to the breakdown scale
of \EFTNoPion, $k\approx 90;120\;\MeV$. The accuracy of the calculation is
estimated by varying the \NXLO{2}-correction to $k\cot\delta$ by
$Q\sim\gamma_t/\mpi\approx1/3$ around its central value. The agreement with
experiments~\cite{schwarz,sagara} and potential-model
calculations~\cite{kievsky1996,Kievsky:2001fq} is quite satisfactory even
close to the breakdown scale $k\sim\mpi$. The main source of deviation is of
course the insufficient description of the doublet-$\mathrm{P}$-wave.

\begin{figure}[!htbp]
\begin{center}
  \begin{minipage}[c]{0.71\linewidth}
    \includegraphics*[width=\linewidth]
    {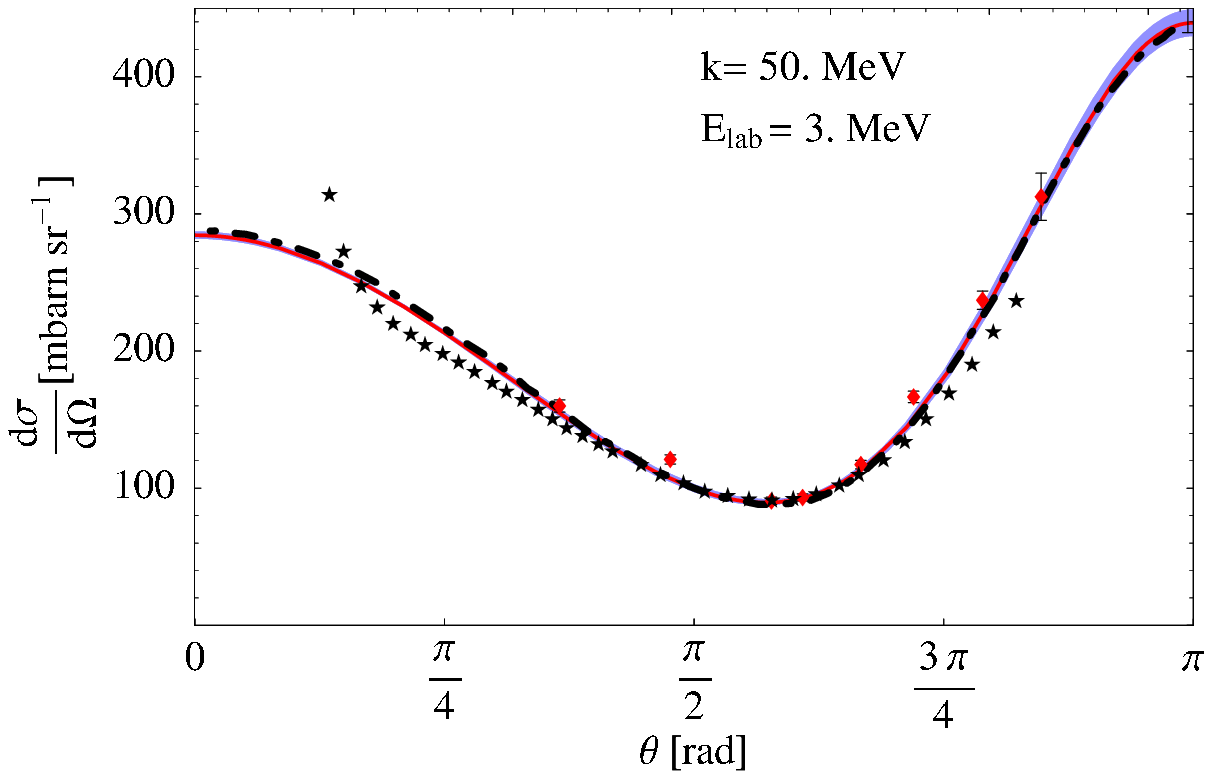}\\[-7.3ex]
    \includegraphics*[width=\linewidth]
    {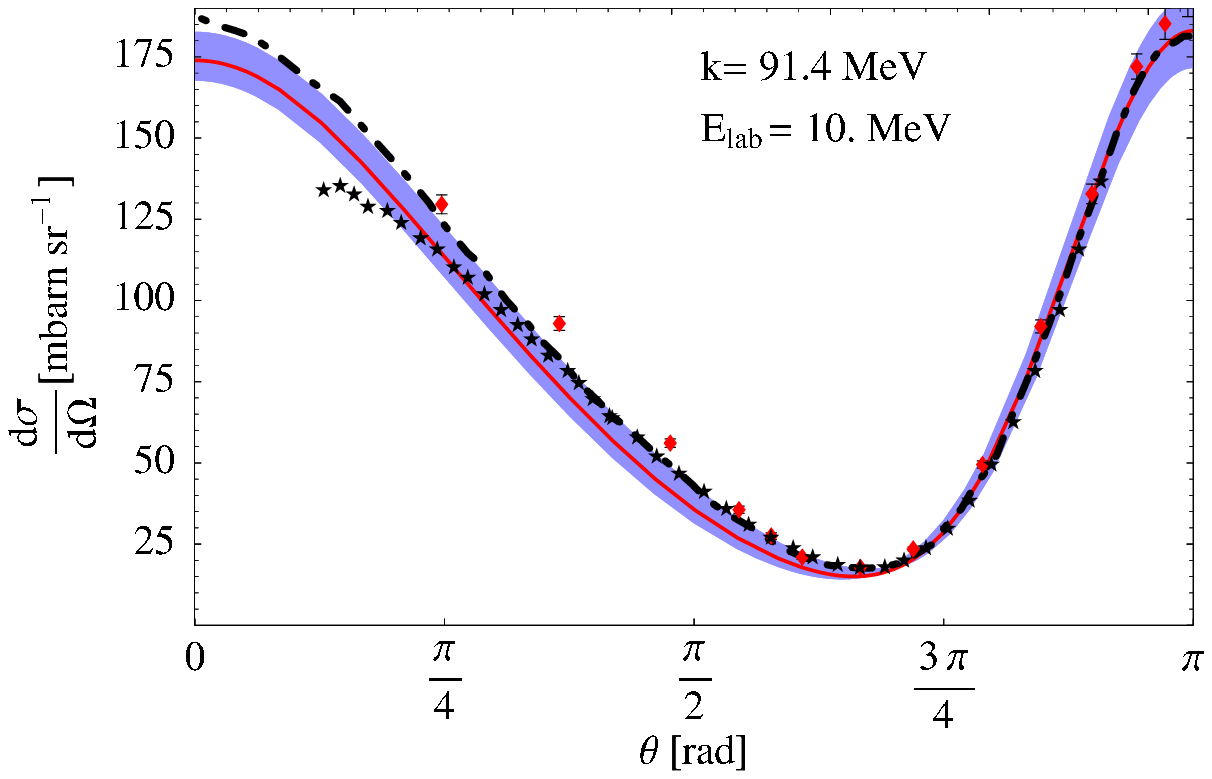}\\[-7.7ex]
    \hspace*{-0.9ex} \includegraphics*[width=1.005\linewidth]
    {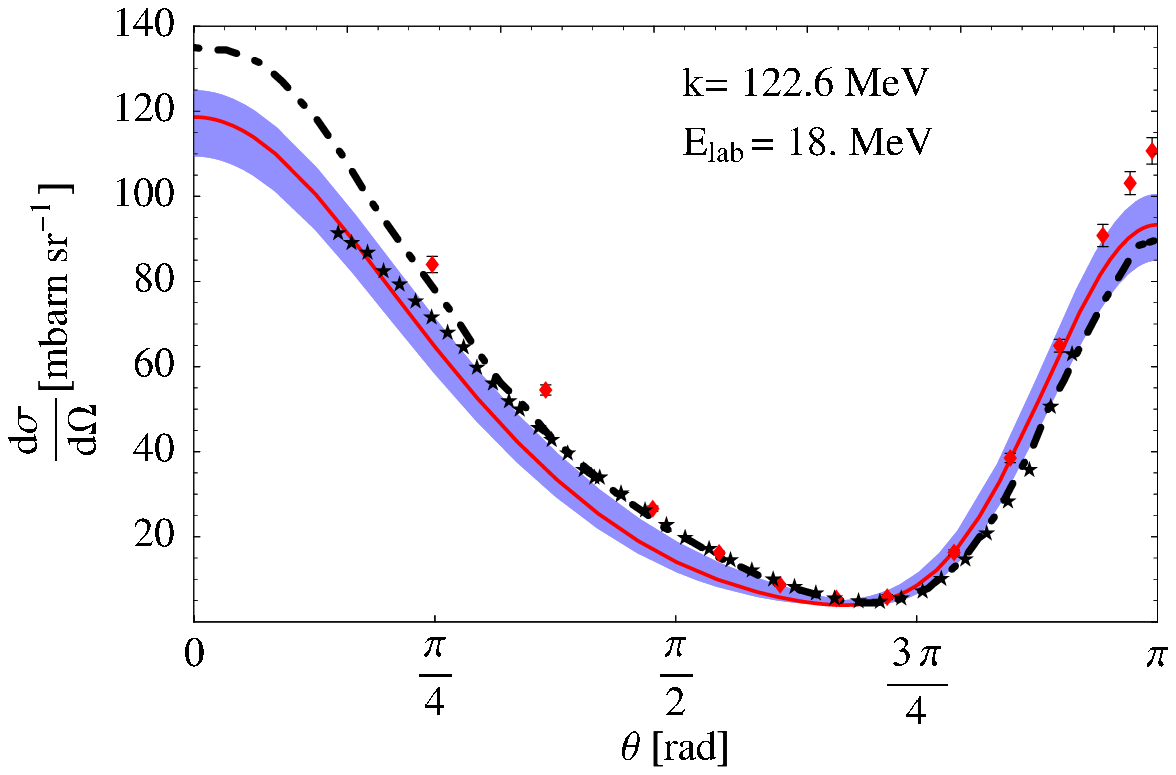}
  \end{minipage}   
\caption{Elastic differential cross-sections of $nd$-scattering at \NXLO{2} in
  the cm-frame in Z-parameterisation (solid) with estimated accuracy (band).
  Dot-dashed: AV18 $+$ Urbana IX from~\cite{kievsky1996} at neutron kinetic
  energy $E_\mathrm{lab}=3\;\MeV$, AV18 from~\cite{Kievsky:2001fq} at
  $E_\mathrm{lab}=10;18\;\MeV$; diamonds: $nd$-data~\cite{schwarz} (taken at
  $3;10.25;18\;\MeV$); stars: $pd$-data~\cite{sagara}.}
\label{fig:diffcrosssection}
\end{center}
\end{figure}

More complex variables need of course more refinement. For example, one
obtains a zero-result for the nucleon-deuteron vector analysing power $A_y$ in
\EFTNoPion at \NXLO{2} because partial-wave splitting is absent. The
well-known under-prediction of this observable in all modern, high-precision
$NN$-potentials~\cite{Huber:1998hu} -- even when supplemented with three-body
forces -- is a challenge a higher-order EFT-calculation has to meet.

%%%%%%%%%%%%%%%%%%%%%%%%%%%%%%%%%%%%%%%%%%%%%%%%%%%%%%%%%%%%%%%%%%%%
\section{Conclusions and Outlook}
\setcounter{equation}{0}
\label{sec:conclusions}

Z-parameterisation~\cite{Phillips:1999hh} fixes the parameters of
``pion-less'' Effective Field Theory at very low energies up to \NXLO{2},
i.e.~up to an estimated theoretical accuracy of $\lesssim 3\%$, to the correct
deuteron pole position and residue, instead of the effective-range parameters.
As the asymptotic fall-off and normalisation of the deuteron wave-function at
large distances is correctly reproduced, this improves drastically convergence
in all processes in which the deuteron is either found as in- or out-state, or
as sub-cluster of a more complex few-body system. The reason for the
improvement is found in a rather large expansion parameter of
$\gamma_t\rho_{0t}\approx0.4$ for effective-range corrections, so that the
deuteron residue is regained only up to $7\%$ at \NXLO{2} when fitting to the
effective range, while it is exact already at NLO in Z-parameterisation. Thus,
the speed of convergence is faster, while the radius of convergence, set by
the first scale on which new Physics enters, is of course not changed.

In this article, Z-parameterisation was extended and applied to
deuteron-nucleon scattering, see Sects.~\ref{sec:Zparam+aux} and
\ref{sec:formalism}. To perform the calculations, a computationally simple and
convenient scheme was again employed which first expands the kernel to the
desired order of accuracy in the power counting and then iterates it to all
orders~\cite{4stooges}.

Numerical problems which origin from additional, spurious poles in the
$NN$-amplitudes when the effective-range corrections are re-summed to all
orders in the deuteron propagator are avoided in Z-parameterisation,
Sect.~\ref{sec:allorders}. Therefore, standard techniques are readily
applicable, and the calculation of scattering-observables also above break-up
becomes computationally trivial, see a Mathematica-code at
\texttt{http://www.physik.tu-muenchen.de/\~{}hgrie}.

In all partial waves, the agreement of the phase-shifts with sophisticated
potential-model calculations is increased in particular in the real and
imaginary parts above the deuteron break-up point, see
Sect.~\ref{sec:higherpws}. What is more, the corrections from one order to the
next in the \EFTNoPion-expansion are smaller in Z-parameterisation, so that a
partial re-summation of effective-range effects increases indeed the
theoretical accuracy of the calculation. This allowed also for high-accuracy
predictions of effective-range parameters in the low partial waves of the
$nd$-system, which can be compared to an upcoming partial-wave analysis of the
$nd$-system~\cite{black}. Most notably, the quartet-$\mathrm{S}$ wave
scattering length is predicted as $a_q=[6.35\pm0.02]\;\fm$ at \NXLO{2}, using
only four observables from the $NN$-system as input, namely the asymptotic
form and normalisation of the deuteron and virtual \oneS-bound state. This
result compares both in magnitude and uncertainty favourably with the most
advanced potential-model calculations
($[6.34\dots6.35]\;\fm$~\cite{Witala:2003mr,Friar:1999jd}), and with the
experimental value ($[6.35\pm0.02]\;\fm$~\cite{doublet_sca}). Simple
observables like differential cross-sections are well reproduced as high as
$E_\mathrm{lab}\sim15\;\MeV$.

The most striking success lies however in the \twoS-channel, which contains
the triton as real bound state, Sect.~\ref{sec:tritonresults}. The numerical
analysis of the cut-off dependence of its phase-shift in Z-parameterisation
clearly supports the analytic finding in Ref.~\cite{4stooges} that one and
only one momentum-dependent three-body force enters at \NXLO{2}, namely the
Wigner-$SU(4)$-symmetric one with two derivatives. Fixing its strength to the
triton binding energy, the accuracy of the calculation -- deduced from the
cut-off dependence of the answer -- is improved at \NXLO{2} from $\sim 4\%$ to
$\sim 1\%$ at a momentum scale of $\sim 90\;\MeV$ typical for the three-body
system. The cut-off dependence follows in Z-parameterisation the pattern
predicted by the power-counting.  The difference to potential-model
calculations is also substantially diminished.

The major source of uncertainty in the triton channel stems thus not from the
theoretical accuracy of the \EFTNoPion-calculation, but from the error in the
determination of its $nd$-scattering length, which is known only to $\sim7\%$
accuracy. To reduce this uncomfortably large error bar to an accuracy of
$\sim0.7\%$ is the goal of an ongoing direct measurement of the incoherent
scattering length at the Paul-Scherer-Institute~\cite{vandenBrandt:2004rb}. A
recent measurement of the bound coherent scattering
length~\cite{boundcoherentexp} relied for the extraction of
$a_d=[0.645\pm0.003(\mathrm{exp})\pm 0.007(\mathrm{theor})]\;\fm$ on
theoretical input for the \fourS-wave scattering length~\cite{Friar:1999jd}.

\absatz Future work includes a complete \NXLO{3}-calculation, including
iso-spin breaking effects and partial-wave splitting and mixing (which should
improve in particular the doublet-$\mathrm{P}$ waves), the extension to
include Coulomb interactions for the ${}^3$He- and $pd$-system, and the
coupling of electro-weak probes to the three-nucleon system. Not only will
this allow for direct comparison with a cornucopia of data and help to shed
light on long-standing puzzles like the $A_y$-problem. It will also lead to
predictions with an accuracy relevant for nuclear astro-physics and neutrino
physics at very low energies, contributing to the ongoing efforts to improve
our knowledge about big-bang nucleo-synthesis, stellar evolution,
neutrino-mass determinations and other fundamental processes of the Standard
Model.

%%%%%%%%%%%%%%%%%%%%%%%%%%%%%%%%%%%%%%%%%%%%%%%%%%%%%%%%%%%%%%%%%%%%
%\begin{equation}
%  \label{eq:}
%\end{equation}

%\begin{eqnarray}
%  \label{eq:}
%\end{eqnarray}

%\begin{figure}[!htb]
%\begin{center}
%  \ifx\feynVersion\AnswerYes
%      %%% if Feynman graphs explicit
%      \feyngraph{100}{100}{}{}{}{}{
%         ******}
%  \else
%      %%% if Feynman graphs as .eps file
%      \includegraphics*[width=0.82\textwidth]{.eps}
%  \fi
%\caption{}
%\label{fig:}
%\end{center}
%\end{figure}

%%%%%%%%%%%%%%%%%%%%%%%%%%%%%%%%%%%%%%%%%%%%%%%%%%%%%%%%%%%%%%%%%%%%%%%%%%%%%%%
%%%%%%%%%%%%%%%%%%%%%%%%%%%%%%%%%%%%%%%%%%%%%%%%%%%%%%%%%%%%%%%%%%%%%%%%%%%%%%%
%%%%%%%%%%%%%%%%%%%%%%%%%%%%%%%%%%%%%%%%%%%%%%%%%%%%%%%%%%%%%%%%%%%%%%%%%%%%%%%

\section*{Acknowledgements}
It is my pleasure to thank P.~F.~Bedaque, H.-W.~Hammer, D.~R.~Phillips and
G.~Rupak for intense discussions and encouragement, and A.~Kievsky and
T.~C.~Black for communications about their results. N.~Kaiser and W.~Weise
provided critical companionship. The warm hospitality and financial support
for stays at the Nuclear Theory Group of Lawrence Berkeley National
Laboratory, at the INT in Seattle and at the ECT* in Trento was instrumental
for this research. In particular, I am grateful to the organisers and
participants of the ``Berkeley Visitors Programme on Effective Field Theories
2003'' and of the ``INT Programme 03-3: Theories of Nuclear Forces and Nuclear
Systems''. This work was supported in part by the Bundesministerium f\"ur
Forschung und Technologie, and by the Deutsche Forschungsgemeinschaft under
contracts GR1887/2-1, 2-2 and 3-1.

\newpage

%%%%%%%%%%%%%%%%%%%%%%%%%%%%%%%%%%%%%%%%%%%%%%%%%%%%%%%%%%%%%%%%%%%%
\appendix
%%%%%%%%%%%%%%% Intro %%%%%%%%%%%%%%%%%%%
\section{Deriving the Faddeev Equation}
\setcounter{equation}{0}
\label{app:appendix}

The Faddeev integral equation in the kinematics defined by
Fig.~\ref{fig:faddeeveq2} is
\begin{eqnarray}
  \label{eq:faddeevunprojected}
  \lefteqn{
    \left(t^{jB}{}_{iA}\right)^{b\beta}{}_{a\alpha}(E; \kv,\pv)=
    \left(K^{jB}{}_{iA}\right)^{b\beta}{}_{a\alpha}(E; \kv,\pv)}\\
    \nonumber
    &&-\int \deintdim{3}{q} \left(K^{jB}{}_{lC}\right)^{b\beta}{}_{c\gamma}(E;
    \qv,\pv)\;\calD(E-\frac{\qv^2}{2M},\qv)\;
    \left(t^{lC}{}_{iA}\right)^{c\gamma}{}_{a\alpha}(E; \qv,\pv)\;\;,
\end{eqnarray}
where $E=\frac{3\kv^2}{4M}-\frac{\gamma_t^2}{M}$ is the total non-relativistic
energy in the cm-frame, $\kv$ ($\pv$) the momentum of the incoming (outgoing)
deuteron/spin-zero field with spin-isospin indices $i/A$ ($j/B$), and the
incoming (outgoing) nucleon has spin-isospin $\alpha a$ ($\beta b$). The
integration over the loop-energy was already performed, setting
$q_0=\frac{\qv^2}{2M}$ under the assumption that no additional poles above the
real $q_0$-axis are hidden in the interaction $K$.  Sub-scripts
(super-scripts) denote quantum numbers for incoming (outgoing) particles.

One decomposes now the integral equation into the pertinent spin-isospin and
angular-momentum channels.

\begin{figure}[!htb]
\begin{center}
  \ifx\feynVersion\AnswerYes
      %%% if Feynman graphs explicit
  
  \vspace*{4ex}
  
  \setlength{\unitlength}{1pt} \feyngraph{75}{64}{18}{0}{0}{0}{
    \fmfleft{i2,i1} \fmfright{o2,o1} \fmf{vanilla,width=2thick}{i1,v1,o1}
    \fmf{vanilla,width=thin}{i2,v5,o2}
    \fmfv{label=$\fs(E-\frac{\kv^2}{2M},,\kv)\;(iA)$,label.angle=90}{i1}
    \fmfv{label=$\fs(\frac{\kv^2}{2M},,-\kv)\;(\alpha a)$,label.angle=-90}{i2}
    \fmfv{label=$\fs(E-\frac{\pv^2}{2M},,\pv)\;(jB)$,label.angle=90}{o1}
    \fmfv{label=$\fs(\frac{\pv^2}{2M},,-\pv)\;(\beta b)$,label.angle=-90}{o2}
    \fmffreeze \fmf{ellipse,foreground=red,rubout=1}{v1,v5} } \hq\hq\hq$=$\hq
  \feyngraph{100}{64}{0}{0}{0}{0}{ \fmfleft{i2,i1} \fmfright{o2,o1}
    \fmf{phantom}{i1,v1} \fmf{phantom}{v1,v2} \fmf{phantom}{v2,o1}
    \fmf{phantom}{i2,v5} \fmf{phantom}{v5,v6} \fmf{phantom}{v6,o2} \fmffreeze
    \fmf{vanilla,width=2thick}{i1,v1} \fmf{vanilla,width=thin}{v6,o2}
    \fmf{vanilla,width=2thick}{v2,o1} \fmf{vanilla,width=thin}{i2,v5}
    \fmffreeze
    \begin{fmfsubgraph}(0.3w,-0.1h)(0.4w,1.2h)
      \fmfleft{ii2,ii1} \fmfright{oo2,oo1}
      \fmf{vanilla,width=thick,foreground=red}{ii1,ii2,oo2,oo1,ii1}
    \end{fmfsubgraph}  
  } \hq$+$\hq \feyngraph{160}{64}{0}{0}{0}{0}{ \fmfleft{i2,i1}
    \fmfright{o2,o1} \fmf{phantom}{i1,v1} \fmf{phantom}{i2,v5}
    \fmf{phantom,tension=0.5}{v1,v2} \fmf{phantom,tension=0.5}{v5,v6}
    \fmf{phantom,tension=2}{v3,o1} \fmf{phantom,tension=2}{v7,o2}
    \fmf{phantom}{v2,v3} \fmf{phantom}{v6,v7} \fmffreeze
    \fmf{vanilla,width=2thick}{i1,v1} \fmf{vanilla,width=2thick,
      label=$\fs\calD$,label.side=left}{v1,v2} \fmf{vanilla,width=thin}{v7,o2}
    \fmf{vanilla,width=2thick}{v3,o1} \fmf{vanilla,width=thin}{i2,v5,v6}
    \fmf{ellipse,foreground=red,rubout=1}{v1,v5} \fmffreeze
              \begin{fmfsubgraph}(0.4w,0.38h)(0.2w,0.2h)
                \fmftop{a,b} \fmfbottom{c,d}
                \fmf{fermion,width=thin,left=1,foreground=green}{a,d}
                \fmf{fermion,width=thin,left=1,foreground=green}{d,a}
                \fmffreeze \fmf{phantom}{a,d} \fmffreeze \fmfipath{pk}
                \fmfiset{pk}{vpath(__a,__d)} \fmfiv{label=$\qv$,
                  label.dist=0pt}{point 1/2 length(pk) of pk}
              \end{fmfsubgraph}    
    \begin{fmfsubgraph}(0.65w,-0.1h)(0.25w,1.2h)
      \fmfleft{ii2,ii1} \fmfright{oo2,oo1}
      \fmf{vanilla,width=thick,foreground=red}{ii1,ii2,oo2,oo1,ii1}
    \end{fmfsubgraph}  
  } \setlength{\unitlength}{1pt}
  
  \vspace*{5ex}
  
  \[
  K=K_{N\text{-ex}}+K_\calH:\hq\hq\hq
  \setlength{\unitlength}{0.7pt} \feyngraph{100}{64}{0}{0}{0}{0}{
    \fmfleft{i2,i1} \fmfright{o2,o1} \fmf{phantom}{i1,v1} \fmf{phantom}{v1,v2}
    \fmf{phantom}{v2,o1} \fmf{phantom}{i2,v5} \fmf{phantom}{v5,v6}
    \fmf{phantom}{v6,o2} \fmffreeze \fmf{vanilla,width=2thick}{i1,v1}
    \fmf{vanilla,width=thin}{v6,o2} \fmf{vanilla,width=2thick}{v2,o1}
    \fmf{vanilla,width=thin}{i2,v5} \fmffreeze
    \begin{fmfsubgraph}(0.3w,-0.1h)(0.4w,1.2h)
      \fmfleft{ii2,ii1} \fmfright{oo2,oo1}
      \fmf{vanilla,width=thick,foreground=red}{ii1,ii2,oo2,oo1,ii1}
    \end{fmfsubgraph}  
  } \hq\hq=\hq\hq \feyngraph{75}{64}{0}{0}{0}{0}{ \fmfleft{i2,i1}
    \fmfright{o2,o1} \fmf{phantom}{i1,v1} \fmf{phantom}{v1,v2}
    \fmf{phantom}{v2,o1} \fmf{phantom}{i2,v5} \fmf{phantom}{v5,v6}
    \fmf{phantom}{v6,o2} \fmffreeze \fmf{vanilla,width=2thick}{i1,v1}
    \fmf{vanilla,width=2thick}{v6,o2} \fmf{vanilla,width=thin}{v1,v2,o1}
    \fmf{vanilla,width=thin}{i2,v5,v6} \fmf{vanilla,width=thin}{v1,v6} }
  \hq\hq+\hq\hq \feyngraph{100}{64}{0}{0}{0}{0}{ \fmfleft{i2,i1}
    \fmfright{o2,o1}
    \fmf{vanilla,width=2thick}{i1,v1}\fmf{vanilla,width=2thick}{o1,v2}
    \fmf{vanilla,width=thin}{i2,v1} \fmf{vanilla,width=thin}{o2,v2}
    \fmf{triple}{v1,v2}} \setlength{\unitlength}{1pt}
  \]
  
  \else
      %%% if Feynman graphs as .eps file
  \includegraphics*[width=1.00\textwidth]{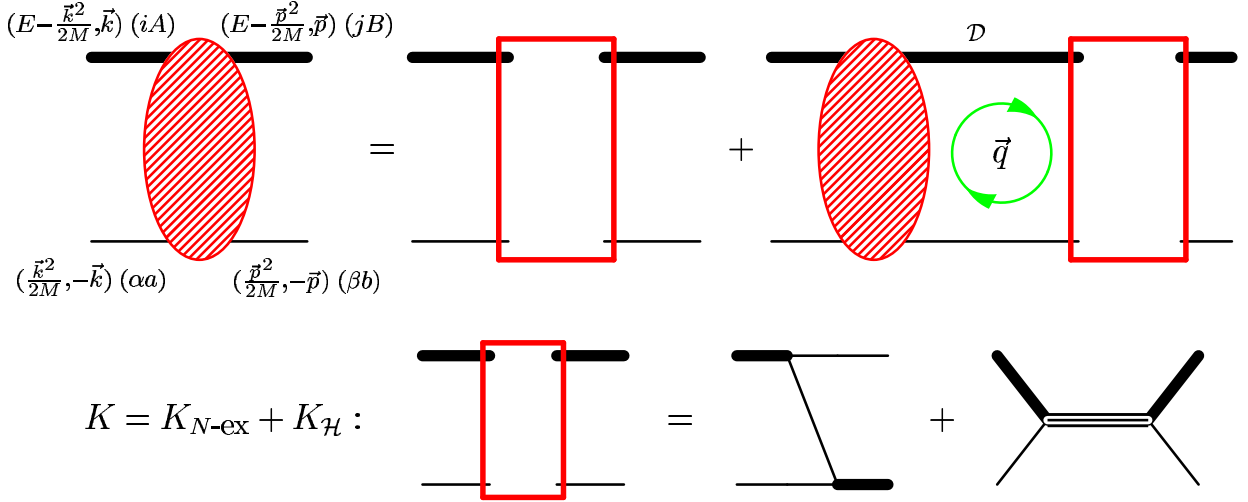} \fi
\caption{The Faddeev equation for $Nd$-scattering to \NXLO{2}. Thick solid
  line: propagator of the two intermediate auxiliary fields $d_s$ and $d_t$,
  denoted by $\calD$, see (\ref{eq:totaltwoparticlepropagator}); triple line:
  propagator of the triton auxiliary field $t$, see
  (\ref{eq:tritonpropagator}); $K$: interaction to \NXLO{2}.}
\label{fig:faddeeveq2}
\end{center}
\end{figure}

%%%%%%%%%%%%%%
\subsection{Projectors}

As two cluster-configurations exist, namely $Nd_t$ and $Nd_s$, it is
convenient to decompose each operator (interactions, amplitudes etc.) as
\begin{equation}
  \calO = N^\dagger_{b\beta} \;\left(d^\dagger_{t,j},\;d^\dagger_{s,B}\right)
  \begin{pmatrix}\calO(Nd_t\to Nd_t)^j_i&\calO(Nd_s\to Nd_t)^j_A\\
    \calO(Nd_t\to Nd_s)^B_i&\calO(Nd_s\to Nd_s)^B_A
  \end{pmatrix}^{b\beta}_{\hq\hq a\alpha}
    \;{d_t^i\choose d_s^A} \;N^{a\alpha}
\end{equation}
In other words, each operator is represented in cluster-configuration space by
a 2x2-matrix which in addition carries spin- and iso-spin indices. It is
understood in the following that the spin-isospin indices on the 2x2-matrix
are applied to each entry separately, and that Kronecker-$\delta$'s in spin or
iso-spin are not displayed. Thus, for example, an entry $\calO(Nd_t\to
Nd_t)=\sigma_i\sigma^j$ in the above matrix is written out as
\begin{equation}
  (\calO(Nd_t\to Nd_t)^j_i)^{b\beta}{}_{a\alpha}=
  (\sigma_i\sigma^j)^\beta{}_\alpha\;\delta^b_a
\end{equation}
$\calO$ is usually not symmetric in any pair of spin-isospin indices $(ij)$,
$(AB)$, $(\alpha\beta)$, $(ab)$.

The projection onto a state with angular momentum $l$, connecting momenta
$\qv,\pv$ with $\qv\cdot\pv=p q \cos\theta$ and $p=|\pv|$ is as usual given by
\begin{equation}
  \label{eq:projectiononl}
  \calO^{(l)}(q,p)=\frac{1}{2}\;\int\limits_{-1}^1\deint{}{\cos\theta}
  \calO(\qv,\pv)\;\;.
\end{equation}

Finally, the projectors onto the possible spin-isospin states of the
three-nucleon system are: combining the auxiliary fields with spin-index $i$
or iso-spin index $A$ with a nucleon of spin-isospin $\alpha,a$ into the
spin-doublet channel with spin $\mu$ and iso-spin $m$
\begin{equation}
  (\calP_{d,iA})^{m\mu}{}_{a\alpha}=\frac{1}{\sqrt{3}}\;\begin{pmatrix}
  \sigma_i&0\\0&\tau_A\end{pmatrix}^{m\mu}_{\hq\hq a\alpha}\;\;,
\end{equation}  
and into the spin-quartet channel with spin $(\mu j)$ and iso-spin $m$
\begin{equation}
  (\calP_{q,i}^j)^{m\mu}{}_{a\alpha}=\begin{pmatrix}
  \delta^j_i-\frac{1}{3}\sigma^j\sigma_i&
  0\\0&0\end{pmatrix}^{m\mu}_{\hq\hq a\alpha}\;\;.
\end{equation}
The projectors are related to their Hermitean conjugates by
\begin{equation}
  \calP_d^{iA}=\left(\calP_{d,iA}\right)^\dagger=
  \mbox{``}\calP_{d,iA}\mbox{''}\;\;,\;\;
  \left(\calP_{q,i}^j\right)^\dagger=\calP_{q,j}^i\;\;,
\end{equation}
and are ortho-normalised:
\begin{equation}
  \calP_{d,iA}\calP_{d}^{iA}=1\;\;,\;\;\calP_{q,j}^k\calP_{q,i}^j=\calP_{q,i}^k
  \;\;,\;\;\calP_{d,iA}\calP_{q,j}^i=0
\end{equation}
For a complete set of spin-isospin projectors, supplement these by the
(ortho-normalised) projector onto the iso-spin quartet channel with iso-spin
$(mB)$ and spin $\mu$, which is not found in nucleon-deuteron scattering:
\begin{equation}
  (\calP_{\text{iso-}q,A}^B)^{m\mu}{}_{a\alpha}=\begin{pmatrix}0&0\\0&
  \delta^B_A-\frac{1}{3}\tau^B\tau_A\end{pmatrix}^{m\mu}_{\hq\hq a\alpha}\;\;,
  \;\;\left(\calP_{\text{iso-}q,A}^B\right)^\dagger=\calP_{\text{iso-}q,B}^A
\end{equation}

%%%%%%%%%%%%%%
\subsection{Projecting the Interaction Terms}

The exchange of a nucleon is from the Lagrangeans
(\ref{eq:threeSlagrangean}/\ref{eq:oneSlagrangean}/\ref{eq:Nlagrangean}) up to
\NXLO{2}:
\begin{equation}
 \left(K_{N\text{-ex}}^{jB}
   {}_{iA}\right)^{b\beta}_{\hq\hq a\alpha} (E; \qv,\pv)=\frac{-M y^2}{2}\;
    \frac{1}{p^2+q^2+\pv\cdot\qv-ME-\ii\epsilon}\;
  \begin{pmatrix}\sigma_i\sigma^j&\sigma^j\;\tau_A\\[1ex]
    \sigma_i\;\tau^B& \tau_A\tau^B\end{pmatrix}^{b\beta}_{\hq\hq a\alpha}
\end{equation}
The projection of the propagator of the exchanged nucleon onto angular
momentum $l$ is obtained by combining (\ref{eq:projectiononl}) with
(\ref{eq:projectedNpropagator}). There is no mixture between the doublet- and
quartet-channels, and one finds in the doublet-channel
\begin{equation}
   \calP_{d,jB}\begin{pmatrix}\sigma_i\sigma^j&\sigma^j\;\tau_A\\[1ex]
    \sigma_i\;\tau^B& \tau_A\tau^B\end{pmatrix}\calP_{d}^{iA} =
  \begin{pmatrix}-1&3\\3&-1\end{pmatrix}
\end{equation}
and in the quartet
\begin{equation}
   \calP_{q,j}^l\begin{pmatrix}\sigma_i\sigma^j&\sigma^j\;\tau_A\\[1ex]
    \sigma_i\;\tau^B& \tau_A\tau^B\end{pmatrix}\calP_{q,k}^{i} =
  2\;\calP_{q,k}^{l}\;\;.
\end{equation}

In cluster-configuration space, one obtains for the three-body interaction
from (\ref{eq:threeNlagrangean})
\begin{equation}
  \label{eq:calHinteraction}
  \begin{split}
 &\left(K_\calH^{jB}{}_{iA}\right)^{b\beta}_{\hq\hq a\alpha}
    (E; \qv,\pv)=
    \\
    \frac{-M y_3^2(\Lambda)}{3M\Omega}\;
%    \frac{1}{M\Omega]-\sum\limits_{n=1}^{\infty}\dis
%      \frac{h_{2n}(\Lambda)}{M^{n-1}}(ME+\gamma_t^2)^{n}}
   & \left[
     1+\;\sum\limits_{m,n=1}^\infty\left[
     \frac{h_{2n}(\Lambda)}{\Omega\;M^{n}}\;(ME+\gamma_t^2)^{n}\right]^m\right]
 \;
  \begin{pmatrix}\sigma^j\sigma_i&-\sigma^j\;\tau_A\\[1ex]
    -\sigma_i\;\tau^B& \tau^B\tau_A\end{pmatrix}^{b\beta}_{\hq\hq a\alpha}
  \;\;.
  \end{split}
\end{equation}
By construction, the only non-zero contribution is in the doublet-$\mathrm{S}$
wave:
\begin{equation}
  \calP_{d,jB}\begin{pmatrix}\sigma^j\sigma_i&-\sigma^j\;\tau_A\\[1ex]
   -\sigma_i\;\tau^B& \tau^B\tau_A\end{pmatrix}\calP_{d}^{iA} =
  3\;\begin{pmatrix}1&-1\\-1&1\end{pmatrix}
\end{equation}

%%%%%%%%%%%%%%%%
\subsection{Result}

Putting these results together and multiplying in the doublet-case the
$2\times 2$-matrix $t^{(l)}$ in cluster-configuration space from the right
with the column vector ${1\choose 0}$, one projects finally onto the
nucleon-deuteron system.  There is no mixture or breaking between in- or
out-states of different individual nucleon spin and iso-spin.  One arrives
thus finally at the integral equations for the quartet (\ref{eq:quartetpw})
and doublet (\ref{eq:doubletpw}) channels quoted in the main text.

%%%%%%%%%%%%%%%%%%%%%%%%%%%%%%%%%%%%%%%%%%%%%%%%%%%%%%%%%%%%%%%%%%%%%%%%%%%%%%%
%%%%%%%%%%%%%%%%%%%%%%%%%%%%%%%%%%%%%%%%%%%%%%%%%%%%%%%%%%%%%%%%%%%%%%%%%%%%%%%

\newpage
%%%%%%%%%%%%%%%%%%%%%%%%%%%%%%%%%%%%%%%%%%%%%%%%%%%%%%%%%%%%%%%%%%%%%%%%%%%%%%%

% The following is used only if the Feynman diagrams are to be generated
% explicitly:
\ifx\feynVersion\AnswerYes
   \end{fmffile}
\fi
\end{document}